\documentclass[preprint,12pt]{elsarticle}

\usepackage{amssymb}
\usepackage{amsmath}

\usepackage{bm}
\usepackage{amsfonts}
\usepackage{algorithm}
\usepackage{booktabs}
\usepackage{multirow,diagbox}
\usepackage[caption=false,font=normalsize,labelfont=sf,textfont=sf]{subfig}
\usepackage{algpseudocode}
\usepackage{float}
\usepackage{adjustbox}
\usepackage{graphicx} 
\usepackage[hidelinks]{hyperref}
\begin{document}

\begin{frontmatter}



\title{Approximating Signed Distance Fields With Sparse Ellipsoidal Radial Basis Function Networks: A Dynamic Multi-Objective Optimization Strategy}

\author[1]{Bobo Lian}
\author[2]{Zidong Wang}
\author[1]{Dandan Wang}
\author[3]{Chenjian Wu\corref{mycorrespondingauthor}}
\ead{cjwu@suda.edu.cn}
\author[1]{Minxin Chen\corref{mycorrespondingauthor}}
\cortext[mycorrespondingauthor]{Corresponding author}
\ead{chenminxin@suda.edu.cn}
\affiliation[1]{organization={the School of Mathematical Sciences},
            addressline={Soochow University}, 
            city={Suzhou},
            postcode={215006}, 
            state={Jiangsu Province},
            country={China}}
\affiliation[2]{organization={the Department of Computer Science},
            addressline={Brunel University London}, 
            city={London},
            postcode={UB8 3PH}, 
            country={United Kingdom}}
\affiliation[3]{organization={the School of Electronic and Information Engineering},
            addressline={Soochow University}, 
            city={Suzhou},
            postcode={215006}, 
            state={Jiangsu Province},
            country={China}}

\begin{abstract}
Accurate and compact representation of signed distance functions (SDFs) of implicit surfaces is crucial for efficient storage, computation, and downstream processing of 3D geometry.
In this work, we propose a general learning method for approximating precomputed SDF fields of implicit surfaces by a relatively small number of ellipsoidal radial basis functions (ERBFs). 
The SDF values could be computed from various sources, including point clouds, triangle meshes, analytical expressions, pretrained neural networks, etc.
Given SDF values on spatial grid points, our method approximates the SDF using as few ERBFs as possible, achieving a compact representation while preserving the geometric shape of the corresponding implicit surface.
To balance sparsity and approximation precision, we introduce a dynamic multi-objective optimization strategy, which adaptively incorporates regularization to enforce sparsity and jointly optimizes the weights, centers, shapes, and orientations of the ERBFs.
For computational efficiency, a nearest-neighbor-based data structure restricts computations to points near each kernel center, and CUDA-based parallelism further accelerates the optimization.
Furthermore, a hierarchical refinement strategy based on SDF spatial grid points progressively incorporates coarse-to-fine samples for parameter initialization and optimization, improving convergence and training efficiency.
Extensive experiments on multiple benchmark datasets demonstrate that our method can represent SDF fields with significantly fewer parameters than existing sparse implicit representation approaches, achieving better accuracy, robustness, and computational efficiency. 
The corresponding executable program is publicly available at \url{https://github.com/lianbobo/SE-RBFNet.git}.
\end{abstract}

\begin{graphicalabstract}
\includegraphics[width=\textwidth]{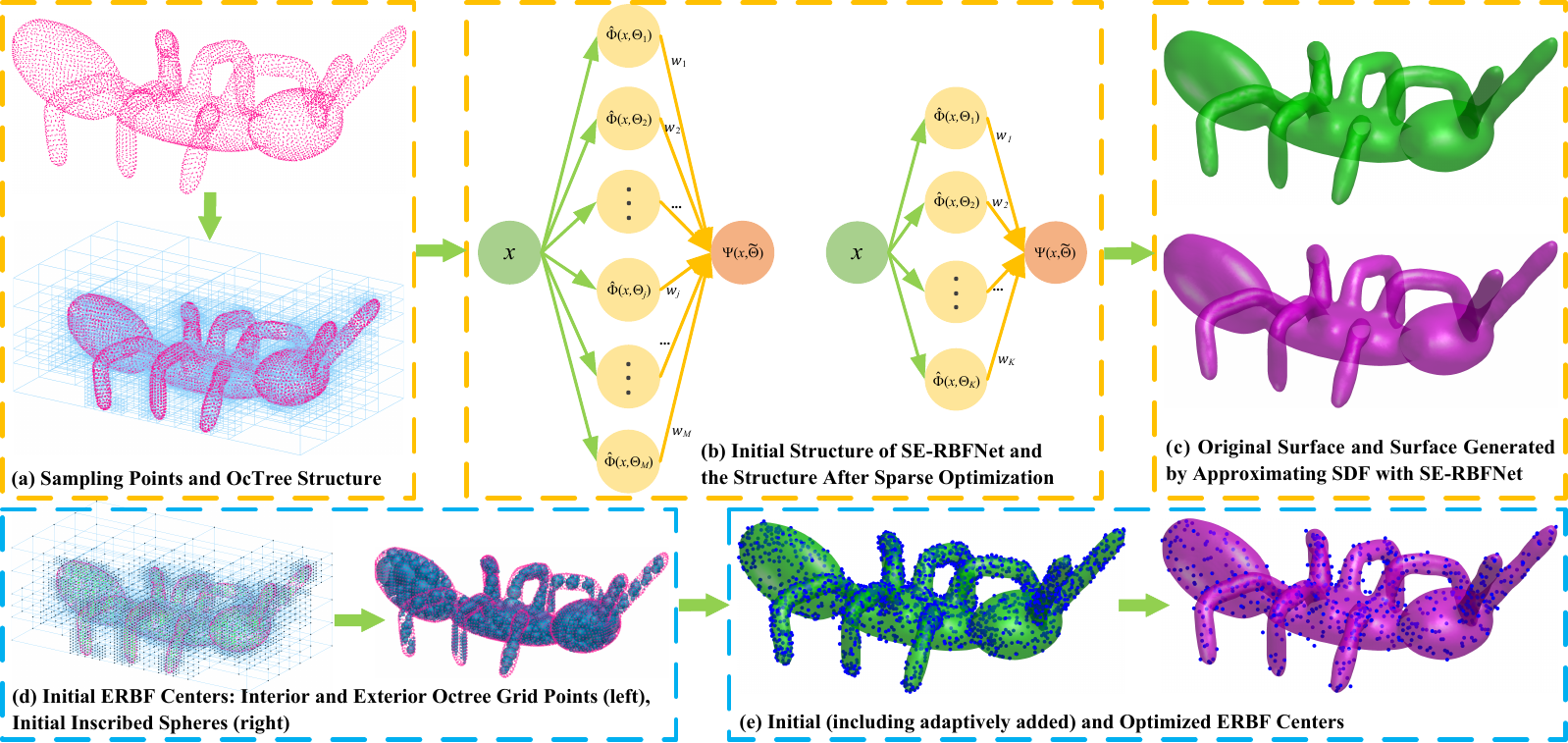}
\end{graphicalabstract}

\begin{highlights}
\item SE-RBFNet is proposed, where the sparse representation of signed distance functions is reformulated as a nonlinear sparse optimization task over the precomputed SDF samples.
\item A dynamic multi-objective optimization strategy is introduced to balance approximation accuracy and sparsity.
\item A coarse-to-fine hierarchical optimization strategy based on SDF grid points is developed, and a nearest-neighbor filtering mechanism is incorporated to improve efficiency and reduce model complexity. 
\item An adaptive basis function addition mechanism iteratively selects new kernel centers from high-error regions and dynamically updates ellipsoidal parameters, enabling improved accuracy and faster convergence.
\end{highlights}

\begin{keyword}
Signed distance function, implicit surface representation, sparse optimization, radial basis function, multi-objective optimization.
\end{keyword}

\end{frontmatter}


\section{Introduction}
Accurate and compact representation of signed distance functions (SDFs) is fundamental for efficient storage, computation, and downstream processing of 3D geometry \cite{li2016sparse}. 
SDFs have gained widespread adoption due to their ability to implicitly encode complex surface geometries as the zero-level set of a continuous function \cite{calakli2011ssd, park2019deepsdf}. This implicit representation facilitates tasks such as surface reconstruction, storage, editing, and physical simulation.

Most implicit surface representation methods, including radial basis function (RBF) interpolation \cite{carr2001reconstruction, dinh2002reconstructing, turk2002modelling, ohtake2006sparse} and neural implicit methods \cite{guerrero2018pcpnet, Erler_2020, gropp2020implicit, wang2023neural} operate directly on point clouds or meshes, learning SDFs from raw geometric data. While these methods can achieve high-fidelity implicit surfaces, they often require a large number of parameters to capture fine-scale or anisotropic features.
Motivated by the success of ellipsoidal radial basis functions (ERBFs) in molecular surface representation \cite{gui2020molecular} and their use in Gaussian splatting \cite{kerbl20233d}, we propose the Sparse Ellipsoidal Radial Basis Function Network (SE-RBFNet), a fast and efficient machine learning approach for sparse representation of precomputed SDFs.
Unlike previous methods, SE-RBFNet does not directly rely on raw point clouds. Instead, it takes precomputed SDF samples—which can be generated by existing point-cloud-based or neural implicit methods, analytical SDFs, or triangle mesh evaluation—as input. By approximating these SDF values using a small set of ERBFs, our method produces a sparser representation of the same implicit surface, achieving further compression in parameter count while preserving geometric accuracy near the zero-level set. Our formulation is thus complementary to existing SDF generation techniques, as it allows their outputs to be efficiently sparsified through subsequent approximation.

Our method shares similarities with SparseRBF \cite{li2016sparse} in its use of RBFs for sparse representation but differs in two key aspects.
First, SparseRBF typically utilizes spherical basis functions, optimizing only the weight coefficients. This design aims to transform the problem into a linear Least Absolute Shrinkage and Selection Operator (LASSO) \cite{tibshirani1996regression} problem, thereby lacking the flexibility to accurately fit complex surfaces. 
In contrast, SE-RBFNet adopts ERBFs and jointly optimizes basis centers, the rotation angles along the principal axes, axis lengths, and the weight coefficients. This is a high-dimensional nonlinear nonconvex optimization problem. This richer parameterization enables high-accuracy approximation of SDF values, allowing fine-scale and elongated surface regions to be represented using far fewer basis functions. 
Furthermore, instead of initializing centers along a predefined axis, we employ an inscribed-sphere initialization strategy, improving adaptability to diverse surface shapes.

In addition, we initially applied radial basis networks for sparse representation in the image domain \cite{chen2020sparse}, and later extended this approach to 3D surface domains \cite{wang2021point}. 
Compared with our earlier work in \cite{wang2021point}, SE-RBFNet incorporates a multi-objective optimization strategy  \cite{sener2018multi},  which adaptively balances L2 error term and L1 regularization term of the basis coefficients to achieve an optimal trade-off between approximation accuracy and sparsity. 
To further improve computational efficiency and robustness, a nearest-neighbor search method is used to identify points near the center of each Gaussian kernel during both the forward and backward steps of optimization, effectively reducing computational complexity. The optimization process is further accelerated via CUDA-based parallel computation.
Moreover, we design a hierarchical optimization strategy based on SDF grid points.
Specifically, ERBF parameters are first initialized and optimized using coarse grid points. As the optimization progresses, finer grid points are progressively incorporated into the training set. This strategy applies to both uniform grids and adaptive grids such as octrees \cite{wilhelms1992octrees}. In addition, extreme error points—identified based on the L2 error at grid points—are introduced as new kernel centers. This coarse-to-fine, iterative process enhances SDF approximation accuracy while accelerating convergence.

We summarize our contributions as follows. 
\begin{itemize} 
    \item SE-RBFNet is proposed, where the sparse representation of SDFs is reformulated as a nonlinear sparse optimization task over the precomputed SDF samples (see Section \ref{subsection:Ellipsoid RBF Neural Network}). 
    \item A dynamic multi-objective optimization strategy is introduced to balance approximation accuracy and sparsity (see Section \ref{subsection:loss_function}). 
    \item A coarse-to-fine hierarchical optimization strategy based on SDF grid points is developed, and nearest-neighbor filtering is incorporated to improve efficiency and reduce model complexity (see Section \ref{subsubsection:Hierarchical Optimization}). 
    \item An adaptive basis function addition mechanism is designed, in which new kernel centers are iteratively selected from high-error regions and ellipsoidal parameters are dynamically updated, enabling improved accuracy and faster convergence (see Section \ref{subsubsection:Adaptive Basis Function Adjustment}). 
\end{itemize}

\section{Related Work}
Implicit surface representation has been a longstanding topic of research. Many methods aim to directly infer implicit functions—such as SDFs or occupancy fields—from raw geometric data. 
In contrast, our work focuses on a different problem: efficiently representing a precomputed set of SDF samples in a compact and memory-efficient manner, independent of how these SDF values were obtained.
This separation between SDF estimation and representation enables us to focus on efficient encoding while fully leveraging high-quality SDF inputs.
In the following, we briefly review classical and learning-based implicit surface representation methods.

\subsection{Classical Implicit Representations}
\label{subsection:Classical Implicit Methods}
Classical methods typically represent a surface as the zero-level set of a continuous scalar field, often an SDF. 
This strategy enables flexible shape representation and efficient surface extraction using algorithms such as Marching Cubes \cite{lorensen1998marching}.
Early work by Curless and Levoy \cite{curless1996volumetric} introduced volumetric SDF fusion, laying the foundation for subsequent approaches. 
Calakli and Taubin \cite{calakli2011ssd} further emphasized that the target implicit function should approximate the true SDF, rather than merely filling volume. 
Pan and Skala \cite{pan2011continuous} proposed a continuous global optimization framework that minimizes a variational energy combining data fidelity and smoothness, enabling robust surface representation from noisy or incomplete oriented point clouds.
In a subsequent work, Pan and Skala \cite{pan2012surface} improved upon their previous approach by incorporating higher-order derivative regularization into the optimization, further enhancing surface smoothness.
Poisson surface reconstruction (PSR) and its variants \cite{kazhdan2006Poisson, kazhdan2013screened, kazhdan2020poisson, sellan2022stochastic} transform reconstruction into solving a spatial Poisson equation, generating smooth and globally consistent surfaces. Building upon the PSR, Hou et al. \cite{hou2022iterative} proposed an iterative PSR method that estimates normals from reconstructed surfaces in each iteration, progressively enhancing surface quality. 

RBFs are another popular choice for implicit surface modeling. These methods place an RBF at each point and optimize the corresponding weights to fit the surface. 
Carr et al. \cite{carr2001reconstruction} proposed multi-harmonic RBFs for smooth surface interpolation and introduced a greedy algorithm, which iteratively added centers associated with large residuals, reducing the total number of basis functions.
Ohtake et al. \cite{ohtake2006sparse} decomposed the global approximation problem into overlapping local subproblems, each solved via least-squares RBF fitting. This significantly reduces the number of required basis functions, enabling sparse yet efficient surface representation.
Samozino et al. \cite{samozino2006reconstruction} proposed selecting RBF centers directly from the Voronoi vertices computed from the input point cloud, rather than placing them on the surface or its offset. This strategy yields a more uniform spatial distribution of centers. However, since the fitting process treats both surface and off-surface points equally, approximation errors near the true surface can increase. 
Further developments on RBF-based surface modeling have explored anisotropic kernels \cite{casciola2006shape}, compactly supported RBFs \cite{walder2006implicit, pan2011two}, orthogonal least squares (OLS) center selection \cite{xia2006orthogonal}, and Hermite RBFs \cite{brazil2010sketching, macedo2011hermite, liu2016closed}, which directly incorporate normal information. 
In large-scale contexts such as geographic surface modeling, space-partitioned and comparative studies \cite{majdisova2017big} have demonstrated both the potential and the limitations of conventional RBFs in handling massive point sets efficiently.
More recently, RBF-based surface representation has been further extended through sparse center selection \cite{wang2021point}, partitioned formulations \cite{drake2022implicit}, and novel interpolation schemes \cite{zeng2022implicit}, broadening its applicability across different surface fitting scenarios.

Beyond RBFs, alternative implicit formulations include Implicit Moving Least Squares \cite{shen2004interpolating, fuhrmann2014floating, liu2021deep}, Fourier bases \cite{Kazhdan_2005}, and Gaussian-based formulations \cite{lu2018surface, lin2022surface}, further illustrating the versatility of implicit representations.

\subsection{Neural Implicit Representations}
\label{subsection:Neural Implicit Representations}
Recent learning-based approaches parameterize implicit functions using neural networks. Mescheder et al. \cite{mescheder2019occupancy} introduced Occupancy Networks, parameterizing implicit functions via neural networks and latent codes. Park et al. \cite{park2019deepsdf} proposed DeepSDF, which represents shapes as continuous SDFs modeled by multilayer perceptrons (MLPs). 
Extensions include convolutional neural networks (CNNs) for capturing shape priors \cite{peng2020convolutional}, geometric regularization \cite{gropp2020implicit}, second-order constraints \cite{ben2022digs}, and Hessian-based smoothness \cite{wang2023neural}.
In contrast to global shape models, Points2Surf \cite{Erler_2020} and PPSurf \cite{erler2024ppsurf} focus on local, patch-based learning with self-supervision, enhancing robustness to noise, sparsity, and varying sampling density. Ma et al. \cite{ma2020neural} further relaxed supervision requirements by proposing an unsupervised learning framework that estimates surface geometry directly from point clouds without relying on ground-truth SDFs.

These studies highlight the importance of SDFs as a fundamental surface representation. Most existing methods—whether classical implicit models or neural implicit models—focus on estimating or learning SDFs directly from raw data, typically requiring a large number of function or network parameters to represent the SDF. Complementary to these approaches, our work targets the sparse representation of precomputed SDF samples, approximating the implicit surface as a combination of ellipsoids. This design enables our method to operate in a data-agnostic manner, leveraging SDFs from any source, including oriented point clouds, triangle meshes, analytically generated SDFs, or SDFs produced by the previously discussed methods. As a general SDF representation framework, SE-RBFNet can efficiently encode a given set of SDF samples with high approximation accuracy and sparsity.

\section{Method}
\label{section:Method}
\begin{figure}[!t]
\centering
\includegraphics[width=\textwidth]{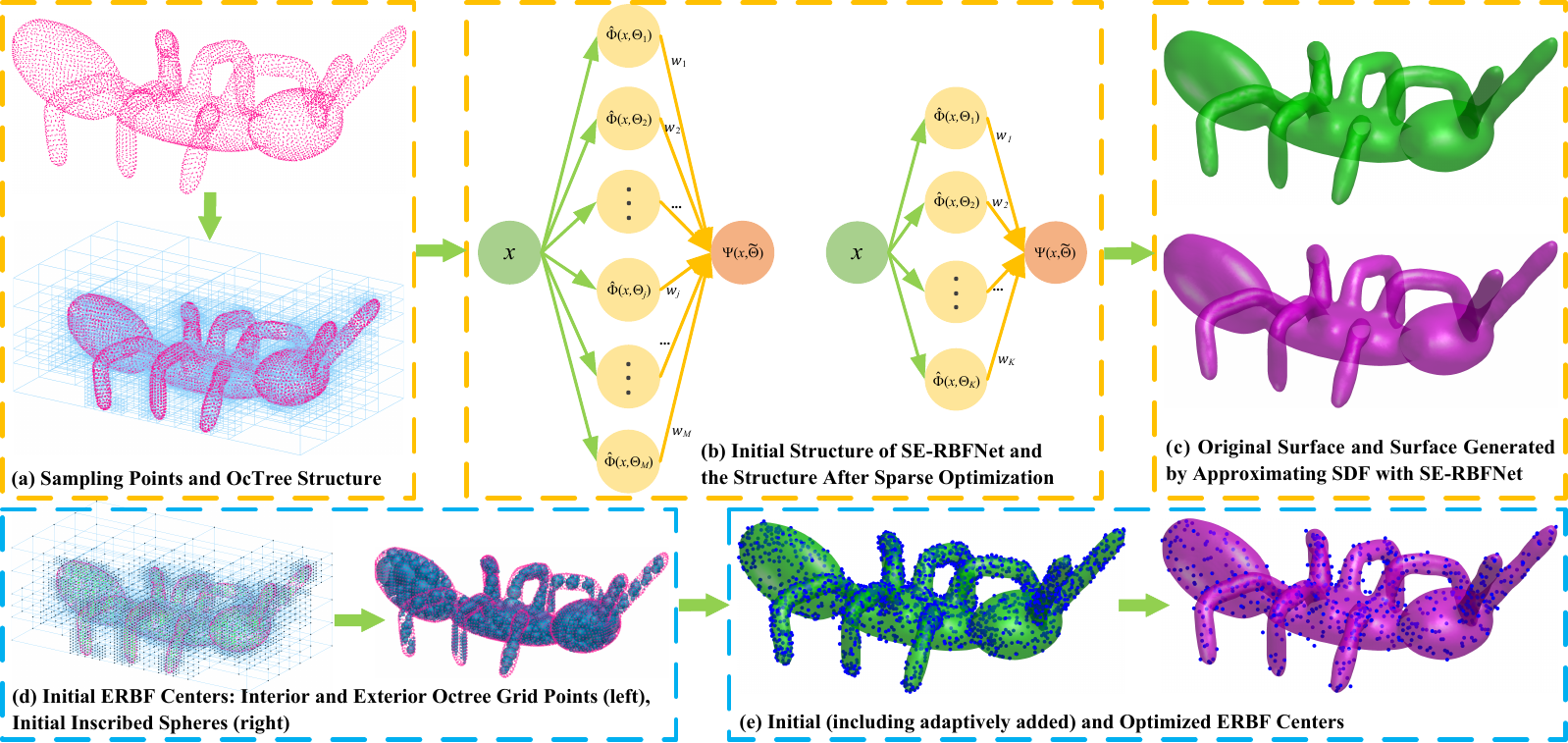}
\caption{The workflow and results of the SE-RBFNet on SDFs of implicit surfaces. (a) takes the octree structure as an example, where the signed distances of all grid points can be obtained using arbitrary SDF-generation methods and used as training data; (b) shows the initial structure of SE-RBFNet and the structure after sparse optimization. SE-RBFNet takes the SDF values on the octree grids and the sampling points on the implicit surface as input and outputs the optimized parameters of ERBFs; (c) shows the explicit surfaces extracted from the original SDF values on octree grids (in green) and from the SE-RBFNet approximated SDF values on octree grids (in purple); (d) shows the process of extracting the initial ERBF centers using maximum inscribed spheres. The left part illustrates the interior (in green) and exterior (in black) octree grid points. The right part shows the computed maximum inscribed spheres (in light blue); (e) shows that the number of optimized ERBF bases is dramatically reduced while the surface shape is preserved.}
\label{fig:total}
\end{figure}

In this section, we describe the SE-RBFNet framework for approximating precomputed SDFs to achieve a compact and sparse representation of implicit surfaces. ERBFs are used as nodes in each hidden layer of the neural network. Sparsity is introduced to represent the surface using fewer RBFs. 
The overall workflow of SE-RBFNet is illustrated in Figure \ref{fig:total}. SDFs derived from an implicit surface are used here as an example. For clarity, we describe the process using an octree grid to provide spatial sampling points for training and subsequent surface extraction, though other sampling strategies, such as uniform grids, could be used as well.

\subsection{RBF Networks}
\label{subsection:Theoretical Foundation}
RBF networks have been extensively studied and are well-known for their universal approximation property, which guarantees that they can approximate any continuous function on a compact subset of $\mathbb{R}^{n}$ to arbitrary precision, given a sufficient number of basis functions. This theoretical property has been rigorously established in foundational works \cite{Park-rbf1993, majdisova2017radial, ismayilova2024universal}, forming the basis for the widespread application of RBFs in function approximation and geometric modeling tasks.

Despite their theoretical expressiveness, traditional RBF networks typically employ isotropic Gaussian basis functions with fixed shape, making them less effective in capturing anisotropic structures or highly detailed variations in complex surfaces. In the context of 3D implicit surface representation, this limitation becomes especially pronounced when the input data exhibits uneven curvature, fine-grained features, or elongated geometries that cannot be efficiently modeled by spherical support regions. Moreover, the approximation quality of RBFs is highly sensitive to the choice of kernel shape parameters. Skala et al. \cite{skala2020radial} showed that assigning individual shape parameters to each basis improves flexibility but also introduces many local optima, making robust estimation challenging.

To overcome these limitations, SE-RBFNet extends the classical RBF framework by replacing spherical kernels with learnable ERBFs \cite{zwicker2001ewa}. This allows each kernel to adapt its shape, orientation, and scale to better align with the underlying geometry. The resulting anisotropic flexibility enhances approximation fidelity and parameter efficiency, as fewer ellipsoidal kernels are needed to represent complex structures compared with isotropic ones.
Furthermore, by integrating ERBFs into a sparse optimization framework, SE-RBFNet retains the universal approximation property of RBF networks while enabling efficient and compact representation of precomputed SDF samples, improving adaptability to geometric variations in 3D surfaces.
 
\subsection{Ellipsoid RBF Neural Network}
\label{subsection:Ellipsoid RBF Neural Network}
 
A function of the form $\Phi(\boldsymbol{x, c})=\Phi(\|\boldsymbol{x}-\boldsymbol{c}\|)$ is called an RBF, whose value depends only on the distance from $\boldsymbol{x}$ to the center $\boldsymbol{c}$. There are many RBF expressions. One example is given by:
\begin{equation}
\label{equ:3-1}
\Phi(\boldsymbol{x, c})=e^{-\|\boldsymbol{x}-\boldsymbol{c}\|^2},
\end{equation}
where $\boldsymbol{x}=\left(x_1, x_2, x_3\right)^\mathrm{T} \in \mathbb{R}^3$, $\boldsymbol{c}=\left(c_1, c_2, c_3\right)^\mathrm{T} \in \mathbb{R}^3$. 

In this work, the ERBF \cite{zwicker2001ewa} extends the classical RBF by adopting a more flexible ellipsoidal form. The ERBF in $\mathbb{R}^3$ is defined as follows:
\begin{equation}
\label{equ:3-7}
\begin{aligned}
\hat{\Phi}(\boldsymbol{x}, \Theta)=e^{-\|\boldsymbol{D} \boldsymbol{R}(\boldsymbol{x}-\boldsymbol{c})\|^2},
\end{aligned}
\end{equation}
where $\boldsymbol{D}=\operatorname{diag}\left(d_1, d_2, d_3\right)$, $d_1, d_2, d_3 \in \mathbb{R}$ indicate the lengths of the ellipsoid along its
principal axes. $\boldsymbol{R}$ represents the rotation matrix, defined as follows:
\begin{equation}
\label{equ:3-6}
\boldsymbol{R}\left(\theta_x, \theta_y, \theta_z\right)=\boldsymbol{R}\left(\theta_z\right) \cdot \boldsymbol{R}\left(\theta_y\right) \cdot \boldsymbol{R}\left(\theta_x\right),
\end{equation}
and
\begin{equation*}
\begin{aligned}
& \boldsymbol{R}\left(\theta_x\right)=\left[\begin{array}{ccc}
1 & 0 & 0 \\
0 & \cos \theta_x & -\sin \theta_x \\
0 & \sin \theta_x & \cos \theta_x
\end{array}\right], \\
& \boldsymbol{R}\left(\theta_y\right)=\left[\begin{array}{ccc}
\cos \theta_y & 0 & -\sin \theta_y \\
0 & 1 & 0 \\
\sin \theta_y & 0 & \cos \theta_y
\end{array}\right], \\
& \boldsymbol{R}\left(\theta_z\right)=\left[\begin{array}{ccc}
\cos \theta_z & -\sin \theta_z & 0 \\
\sin \theta_z & \cos \theta_z & 0 \\
0 & 0 & 1
\end{array}\right],
\end{aligned}
\end{equation*}
$\boldsymbol{R}(\theta_x)$, $\boldsymbol{R}(\theta_y)$, $\boldsymbol{R}(\theta_z)$ are the rotation matrices of the three principal axes, $\theta_x$, $\theta_y$, $\theta_z$ are the rotation angles of the ellipsoid along the principal axis.  $\Theta=\left[\boldsymbol{c}, \boldsymbol{D}, \theta_{x}, \theta_{y}, \theta_{z}\right]^\mathrm{T}$, represents the parameters of the ERBF, including the centers, the rotation angles, and the lengths.

\begin{figure}[!t]
\centering
\includegraphics[width=0.5\textwidth]{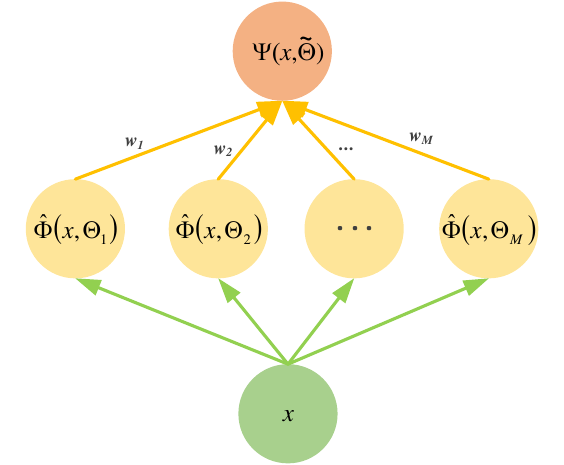}
\caption{The neural network structure of SE-RBFNet.}
\label{fig:se-rbfnet}
\end{figure}

Figure \ref{fig:se-rbfnet} shows the structure of the SE-RBFNet. When ERBF is chosen as the hidden layer activation function, given an input $\boldsymbol{x_i}$, the output of the EBBF network is as follows:
\begin{equation}
\label{equ:3-8}
\begin{aligned}
\Psi(\boldsymbol{x}_i, \boldsymbol{\widetilde{\Theta}}) = \sum_{j=1}^M w_j^2 \hat{\Phi}(\boldsymbol{x}_i, \Theta_j)
= \sum_{j=1}^M w_j^2 e^{-\left\|\boldsymbol{D}_j \boldsymbol{R}_j\left(\boldsymbol{x}_i-\boldsymbol{c}_j\right)\right\|_2^2},
\end{aligned}
\end{equation}
where $\boldsymbol{x}_i$ is an arbitrary point in three-dimensional space, $\boldsymbol{x}_i=(x_{i 1}, x_{i 2}, x_{i 3})^\mathrm{T}$, $i=1,2, \ldots, N$, $N$ is the total number of input points, $M$ is the number of ERBFs in the hidden layer. $\boldsymbol{\widetilde{\Theta}}=\{\boldsymbol{C}\textrm{,}\bar{\boldsymbol{D}}\textrm{,} \boldsymbol{A}, \boldsymbol{W}\}$ indicates all the parameters of the ERBFs in the hidden layer, which are also all the parameters to be optimized in SE-RBFNet, where $\boldsymbol{C} \in \mathbb{R}^{M \times 3}$ is the center of all ERBFs and $\boldsymbol{c}_j=(c_{j1}, c_{j2}, c_{j3})$ is the $j$-th row of $\boldsymbol{C}$, indicating the center of the $j$-th ERBF.
Similarly, $\bar{\boldsymbol{D}} \in \mathbb{R}^{M \times 3 \times 3}$, $\boldsymbol{D}_j=\operatorname{diag}\left(d_{j 1}, d_{j 2}, d_{j 3}\right)$, where the diagonal entries describe the lengths of the $j$-th ellipsoid along its principal axes. $\boldsymbol{A} \in \mathbb{R}^{M \times 3}$ is the rotation angles of all ERBFs, $\boldsymbol{a}_j=(\theta_{xj}, \theta_{yj}, \theta_{zj})$. $\boldsymbol{R}_j=\boldsymbol{R}_j(\theta_{xj}, \theta_{yj}, \theta_{zj})$ is the rotation matrix of the $j$-th ERBF. $\boldsymbol{W} \in \mathbb{R}^{M \times 1}$ is the coefficient matrix from the hidden layer to the output layer, $w_j^2$ is the weight of the $j$-th ERBF from the hidden layer to the output layer. The weight is nonnegative to prevent different ERBFs from canceling each other out through opposing weighting and to make the output of Eq. (\ref{equ:3-8}) nonnegative.

Suppose the set of grid points used for surface extraction is denoted by $\boldsymbol{X}$, consisting of $N$ points: $\boldsymbol{X}=\displaystyle\bigcup_{i=1}^{N}\boldsymbol{x}_i$, where $\boldsymbol{x}_i$ denotes the $i$-th point. Next, according to Eq. (\ref{equ:3-9}), each point $\boldsymbol{x}_i$ is fed into SE-RBFNet to obtain the corresponding predicted SDF value. The output of SE-RBFNet is denoted as $\boldsymbol{O}$, as follows:
\begin{equation}
\label{equ:3-9}
\begin{aligned}
\boldsymbol{O} &= \begin{bmatrix}
\hat{\Phi}(\boldsymbol{x}_1, \Theta_1) & \hat{\Phi}(\boldsymbol{x_1}, \Theta_2) & \cdots & \hat{\Phi}(\boldsymbol{x}_1, \Theta_M) \\
\hat{\Phi}(\boldsymbol{x}_2, \Theta_1) & \hat{\Phi}(\boldsymbol{x_2}, \Theta_2) & \cdots & \hat{\Phi}(\boldsymbol{x}_2, \Theta_M) \\
\vdots & \vdots & & \vdots \\
\hat{\Phi}(\boldsymbol{x}_N, \Theta_1) & \hat{\Phi}(\boldsymbol{x_N}, \Theta_2) & \cdots & \hat{\Phi}(\boldsymbol{x}_N, \Theta_M)
\end{bmatrix} 
\begin{bmatrix}
w^2_1 \\ w^2_2 \\ \vdots \\ w^2_M
\end{bmatrix} \\
&= \begin{bmatrix}
\Psi(\boldsymbol{x}_1, \boldsymbol{\widetilde{\Theta}}) \\
\Psi(\boldsymbol{x}_2, \boldsymbol{\widetilde{\Theta}}) \\
\vdots \\
\Psi(\boldsymbol{x}_N, \boldsymbol{\widetilde{\Theta}})
\end{bmatrix}
= \boldsymbol{F} \cdot \boldsymbol{W}.
\end{aligned}
\end{equation}
For convenience of notation, we introduce the feature matrix $\boldsymbol{F} \in \mathbb{R}^{N \times M}$ in Eq. (\ref{equ:3-9}), where each element of $\boldsymbol{F}$ corresponds to the response value of the $i$-th point $\boldsymbol{x}_i$ under the influence of the $j$-th ERBF $\hat{\Phi}(\boldsymbol{x}_i, \Theta_j)$.  
Subsequently, the marching cubes algorithm \cite{lorensen1998marching} is applied to the output $\boldsymbol{O}$ to extract the explicit surface.

\subsection{Loss Function Design}
\label{subsection:loss_function}
Given a set of sampling points on an implicit surface $P=\left\{\boldsymbol{p}_i\right\}_{i=1}^{K_p}$, $K_p$ is the number of points, we assume that the corresponding octree grid points $G=\left\{\boldsymbol{g}_j\right\}_{j=1}^{K_g}$ and the signed distance value $S(\boldsymbol{g}_j)$ at each grid point are provided as input, where $K_g$ is the number of grid points. 
The surface itself is characterized by the zero-level set of the signed distance: $\{\boldsymbol{g}_j \in \mathbb{R}^3, S(\boldsymbol{g}_j) = 0\}$. For points inside (outside) the surface, $S(\boldsymbol{g}_j)$ is negative (positive). As can be seen from Eq. (\ref{equ:3-8}), the output of $\Psi$ is nonnegative. Therefore, to fit the signed distance using the SE-RBFNet, we scale all signed distances $S$ to [0, 2] using the following nonlinear transformation.
\begin{equation}
\label{equ:3-10}
\begin{aligned}
\hat{S}&=2 \cdot e^{-h(S-m)^2}\textrm{,}
\end{aligned}
\end{equation}
where $m=min(S)$, $h=\frac{\ln(2)}{m^2}$.
Eq. (\ref{equ:3-10}) normalizes $S$ values less than 0 (inside the surface) to the interval [1, 2] and values greater than 0 (outside the surface) to the interval [0, 1]. Notably, the zero-level set of $S=0$ (i.e., the reconstructed surface) is mapped to $\hat{S}=1$. 

For the network model, we merge the grid points $G$ and the sampling points $P$ as the total training points, denoted as $V$, $V=\{G, P\}=\left\{\boldsymbol{v}_i\right\}_{i=1}^{N}$, $N=K_p+K_g$ is the total number of training points, $\boldsymbol{v}_{i}$ denotes the $i$-th training point. $\hat{S}$ is merged with the all-1 vector $I$ as the label of $P$, denoted as $T$, $T=\{\hat{S}, I\}=\left\{t_i\right\}_{i=1}^{N}$, $t_{i}$ is the $i$-th SDF value. Here, the length of $I$ is equal to the number of sampling points.
The SE-RBFNet model is used to approximate the given SDF values of $V$ as follows:
\begin{equation}
\label{equ:3-13}
\begin{aligned}
&\boldsymbol{O} = \boldsymbol{F} \cdot \boldsymbol{W} = \begin{bmatrix}
\Psi(\boldsymbol{v}_1, \boldsymbol{\widetilde{\Theta}}) \\
\Psi(\boldsymbol{v}_2, \boldsymbol{\widetilde{\Theta}}) \\
\vdots \\
\Psi(\boldsymbol{v}_N, \boldsymbol{\widetilde{\Theta}})
\end{bmatrix}
\approx
\begin{bmatrix}
t_1 \\
t_2 \\
\vdots \\
t_N
\end{bmatrix}
\end{aligned}
\end{equation}
We set the loss function as:
\begin{equation}
\label{equ:3-14}
\mathcal{L}(\boldsymbol{\widetilde{\Theta}})=\alpha \cdot \sum_{i=1}^N\left(\Psi\left(\boldsymbol{v}_i, \boldsymbol{\widetilde{\Theta}}\right)-t_i\right)^2+\beta \cdot\|\boldsymbol{W}\|_1 .
\end{equation}
The loss function consists of two primary components: 
\begin{equation}
\label{equ:3-14_2}
\begin{aligned}
& \mathcal{L}_2(\boldsymbol{\widetilde{\Theta}})=\sum_{i=1}^N\left(\Psi\left(\boldsymbol{v}_i, \boldsymbol{\widetilde{\Theta}}\right)-t_i\right)^2 , \\
& \mathcal{L}_1(\boldsymbol{W})=\|\boldsymbol{W}\|_1 = \sum_{i=1}^M |w_{i}|.
\end{aligned}
\end{equation}
Based on Eq. (\ref{equ:3-14_2}), Eq. (\ref{equ:3-14}) can be rewritten concisely as follows: $\mathcal{L}=\alpha \mathcal{L}_2+\beta \mathcal{L}_1$, where $\mathcal{L}_2$ ensures approximate accuracy by minimizing the squared error between predicted and ground-truth values on training points $V$.
$\mathcal{L}_1$ enforces sparsity in the network parameters by penalizing the $\mathcal{L}_1$ norm of the weights, encouraging a sparse solution for $\boldsymbol{W}$. Weights in $\boldsymbol{W}$ that are close to zero correspond to basis elements with negligible contributions to the network, which can safely be removed.
$\alpha\ge0$ and $\beta\ge0$ are dynamic weighting coefficients that balance the two loss terms. Our aim is to minimize the loss function $\mathcal{L}$ and find the corresponding optimal network parameters, $\boldsymbol{\widetilde{\Theta}}$.  

However, in surface representation, more attention is paid to the points close to the surface, where $\left\{ \boldsymbol{x} \in \mathbb{R}^3 \,\middle|\, \Psi(\boldsymbol{x}) = 1 \right\}$. To further reduce the computational cost, two thresholds, $\tau_1$ (with $\tau_1<1$) and $\tau_2$ (with $\tau_2>1$), each close to 1, are introduced to filter the training points for loss computation at each iteration, as defined by the following formula:
\begin{equation}
\label{equ:3-18}
\begin{aligned}
& id_0=\{i \mid \tau_1<t_i<\tau_2\}, \\
& id_1=\{i \mid t_i \leq \tau_1 \quad \text {and} \quad \Psi(\boldsymbol{v}_i, \boldsymbol{\widetilde{\Theta}})>\tau_1\}, \\
& id_2=\{i \mid t_i \geq \tau_2 \quad \text {and} \quad \Psi(\boldsymbol{v}_i, \boldsymbol{\widetilde{\Theta}})<\tau_2\}, \\
& i d_{a l l}=\displaystyle\bigcup_{j\in \{0\textrm{,} 1\textrm{,} 2\}}id_{j}.
\end{aligned}
\end{equation}
In Eq. (\ref{equ:3-18}), $id_0$ selects from both interior and exterior regions close to the surface, which are crucial for accurately approximating the SDF near the surface and ensuring effective surface representation through ERBFs. $id_1$ ($id_2$) selects points with erroneous predictions in the exterior (interior) region to help the network learn and optimize further, avoiding artificial bulges or dents on the surface represented by the predicted SDF. For instance, $id_1$ ($id_2$) prevents misclassification of points that should properly reside outside (inside) the surface as being erroneously close to or crossing the surface boundary.
Only the points corresponding to $id_{all}$ are subject to loss calculation, so Eq. (\ref{equ:3-14}) can be rewritten as:
\begin{equation}
\label{equ:3-19}
\begin{aligned}
\mathcal{L}(\boldsymbol{\widetilde{\Theta}})=\alpha \cdot \sum_{i \in id_{all}}\left(\Psi\left(\boldsymbol{v}_i, \boldsymbol{\widetilde{\Theta}}\right)-t_i\right)^2+\beta \cdot\|\boldsymbol{W}\|_1.
\end{aligned}
\end{equation}

During the optimization of the loss function, the $\mathcal{L}_1$ regularization term is applied only when two conditions are simultaneously satisfied. First, starting from epoch $e_p \geq k^{l_2}$, the $\mathcal{L}_2$ loss is considered converged, meaning that its standard deviation over the last $k^{l_2}$ epochs falls below a threshold $\tau^{l_2}$:
\begin{equation}
\label{equ:l1_activate}
\begin{aligned}
\operatorname{std}\left(\left\{\mathcal{L}_2^{(e)}(\boldsymbol{\widetilde{\Theta}})\right\}_{e=e_p-k^{l_2}+1}^{e_p}\right)< \tau^{l_2},
\end{aligned}
\end{equation}
where $\operatorname{std}(\cdot)$ denotes the standard deviation operation, $\mathcal{L}_2^{(e)}$ represents the $\mathcal{L}_2$ loss at epoch $e$, $e_p$ indicates the current training epoch. 
Second, the maximum absolute error between the predicted and the ground-truth SDFs must be below a threshold $\tau_m$:
\begin{equation} 
\label{equ:l1_activate2} 
\max_{i \in id_{all}}  \left|\Psi (\boldsymbol{v}_i, \boldsymbol{\widetilde{\Theta}})-t_i \right|  < \tau_m, 
\end{equation} 
If either of these conditions is not satisfied, the optimization proceeds using only the $\mathcal{L}_2$ loss, i.e., $\alpha=1 \text{}, \beta=0$.

When performing sparse optimization with the $\mathcal{L}_1$ regularization term, we adopt a dynamic weighting mechanism \cite{sener2018multi} instead of using fixed coefficients $\alpha$ and $\beta$ in Eq. (\ref{equ:3-19}). 
This mechanism is inspired by the Frank-Wolfe method for Pareto multi-task learning \cite{sener2018multi} and dynamically adjusts the trade-off between accuracy and sparsity during the optimization process. 
Specifically, at each optimization step, the coefficient $\alpha$ is computed based on the gradients of the two loss terms, reflecting their current relative contributions, and $\beta$ is set as $1 - \alpha$. 
This dynamic approach is formulated as:
\begin{equation}
\label{equ:3-21}
\begin{aligned}
& \alpha=\left[\frac{\left(\frac{\partial \mathcal{L}_1}{\partial \boldsymbol{W}}-\frac{\partial \mathcal{L}_2}{\partial \boldsymbol{W}}\right)^\mathrm{T} \frac{\partial \mathcal{L}_1}{\partial \boldsymbol{W}}}{\left\|\frac{\partial \mathcal{L}_2}{\partial \boldsymbol{W}}-\frac{\partial \mathcal{L}_1}{\partial \boldsymbol{W}}\right\|_2^2}\right]_{+,{ }_{\tau}^1}, \\
& \beta = 1 - \alpha,
\end{aligned}
\end{equation}
where $[\cdot]_{+,{ }_\tau^1}$ represents clipping to $[0, 1]$ as $[\sigma]_{+,{ }_\tau^1}=\max (\min (\sigma, 1), 0)$.
This dynamic multi-objective optimization strategy (step 19 in Algorithm \ref{alg:se-rbfnet}) automatically adjusts the coefficients during training, balancing accuracy and sparsity according to the model’s state, which improves generalization and robustness.

\subsection{Optimization Algorithm}
\label{subsection:Sparse Optimization}

\subsubsection{Inscribed Sphere-Based Initialization for ERBF}
\label{subsubsection:Inscribed Sphere-Based Initialization for Radial Basis Functions}

The hidden layer of SE-RBFNet consists of multiple ERBFs. To effectively initialize the network parameters of SE-RBFNet, we use the largest inscribed sphere method \cite{cornea2024curve} to calculate the initialization parameters of each ERBF. This method ensures that the initialization process can adapt to the geometry of the target surface, facilitating the learning of the SDF used for surface representation.

For interior points $G_{\text {in}}=\left\{g_i \mid 1<\hat{S}_i<2\right\}$ and their corresponding signed distance $\hat{S}_{\text {in}}$, we iteratively detect the maximal inscribed spheres using Algorithm \ref{alg:misc} to obtain the initial centers $C=\left\{ \boldsymbol{c}_j\right\}_{j=1}^M$ and the initial weights $W=\left\{ w_j\right\}_{j=1}^M$. To visually demonstrate the calculation process of the inscribed sphere, Figure \ref{fig:total}.(d) illustrates the inner and outer points (left), as well as the distribution of the maximum inscribed spheres (right). As depicted, the inscribed spheres densely cover the internal space of the implicit surface, providing an effective approximation of its geometric structure.

\begin{algorithm}[ht]
\caption{Inscribed Sphere-Based Initialization for ERBF}
\label{alg:misc}
\hspace*{0.02in} {\bf Input:}
interior gird points $G_{\text {in}}$, the corresponding signed distance $\hat{S}_{\text {in}}$, sampling points $P$.
\begin{algorithmic}
\State Compute the closest distance of each interior point to $P$, denoted as $\boldsymbol{d}^{in}$. 
\State Initialize empty matrices for centers $\boldsymbol{C}$ and weights $\boldsymbol{W}$.
\While{$G_{\text{in}}$ is not empty}
    \State Find the index of the maximum $\boldsymbol{d}^{in}$ value:
        \[ i = \arg\max (\boldsymbol{d}^{in}), \quad r = \boldsymbol{d}^{in}[i] \]
    \State Append the center and weight to the matrices:
        \[
        \boldsymbol{C} \gets G_{\text{in}}[i], \quad \boldsymbol{W} \gets \hat{S}_{\text{in}}[i]
        \]
    \State Compute squared Euclidean distances from $G_{\text{in}}[i]$:
        \[
        d_k = \| G_{\text{in}}[k] - G_{\text{in}}[i] \|^2, \quad \forall k
        \]
    \State Remove points within the inscribed sphere:
        \[
        G_{\text{in}} \gets G_{\text{in}}[d_k > r]
        \]
    \State Update $\boldsymbol{d}^{in}$ values $\boldsymbol{d}^{in} \gets \boldsymbol{d}^{in}[d_k > r]$
    \State Update $\hat{S}_{\text{in}}$ values $\hat{S}_{\text{in}} \gets \hat{S}_{\text{in}}[d_k > r]$
\EndWhile
\end{algorithmic}
\hspace*{0.02in} {\bf Output:}
 Centers $\boldsymbol{C}$, Weights $\boldsymbol{W}$.
\end{algorithm}

From Algorithm \ref{alg:misc}, the weights and centers are initialized. Then, the shape parameters $\boldsymbol{D}_j, j=1,2, \ldots, M$ are initialized with identical diagonal elements, i.e., $d_{j1}=d_{j2}=d_{j3}$.
Assume that the Gaussian distribution is satisfied between each Gaussian kernel center and its nearest neighboring Gaussian kernel center.
To minimize the mutual influence between Gaussian kernels, we ensure that the function value of each Gaussian kernel at the position of its nearest neighboring kernel is sufficiently small. This effectively reduces the interference between different kernels as well.
Under this assumption, $d_{j1}$ satisfies the following equation:
\begin{equation}
\label{equ:3-11}
w_j^2 e^{-d_{j1}^2\left(\hat{d}_{i} / 2\right)^2}=\gamma .
\end{equation}
Here, $\gamma$ is a threshold value chosen to minimize the mutual influence between kernels. $\hat{d}_{i}=\min \left\|c_j-c_i\right\|_2, i=1,2, \ldots, M$ and $i \neq j$. This means that $\hat{d}_{i}$ is the shortest distance from the current kernel center $\boldsymbol{c}_j$ to any other kernel center. After simplification, the $d_{j1}$ can be derived from Eq. (\ref{equ:3-11}) as:
\begin{equation}
\label{equ:3-12}
d_{j1}=d_{j2}=d_{j3}=\frac{2 \sqrt{-\ln \left(\gamma / w_j^2\right)}} {\hat{d}_i}.
\end{equation}
Initialize $\boldsymbol{D}_j=\operatorname{diag}\left(d_{j 1}, d_{j 2}, d_{j 3}\right)$. The rotation angle of each ellipsoid along the principal axis is set to zero, so that each corresponding rotation matrix $\boldsymbol{R}_{j}$ is initialized as the identity matrix.

\subsubsection{Hierarchical Optimization}
\label{subsubsection:Hierarchical Optimization}

In this section, we describe a hierarchical optimization method for training the aforementioned SE-RBFNet on multi-level grid points. Although the method is applicable to any hierarchical grid structure, we use an octree as an example for illustration. Specifically, we perform the training process on the octree grid points layer by layer. Assuming that $G^{i}$ represents the grid point of the $i$-th layer of the octree, $G$ can be represented as follows:
\begin{equation}
\label{equ:3-16}
\begin{aligned}
G&=\displaystyle\bigcup_{i\in \{1\textrm{,} 2\textrm{,}...\textrm{,} l\}}G^i ,
\end{aligned}
\end{equation}
and $G^i=\{G^{i}_1\textrm{,} G^{i}_2\textrm{,} ...\textrm{,} G^{i}_{k_i}\}$, 
where ${k_i}$ indicates the number of grid points in the $i$-th layer, and the total number of layers in the octree is $l$. Similarly, the signed distance corresponding to each grid point is also layered, that is,
\begin{equation}
\label{equ:3-17}
\begin{aligned}
\hat{S}&=\displaystyle\bigcup_{i\in \{1\textrm{,} 2\textrm{,}...\textrm{,} l\}}\hat{S}^i ,
\end{aligned}
\end{equation}
and $\hat{S}^i=\{\hat{S}^{i}_1\textrm{,} \hat{S}^{i}_2\textrm{,} ...\textrm{,} \hat{S}^{i}_{k_i}\}$.
During optimization, the training process begins with the grid points of the coarse layer, and data from finer layers is progressively incorporated. The specific steps are as follows:
\begin{enumerate} 
    \item Set the starting layer to $l_s$, and calculate the inscribed sphere using the points $\hat{G}^{l_s}=\{G^{1}\textrm{,} G^{2}\textrm{,} ...\textrm{,} G^{l_s}\}$ to initialize the ERBF parameters (Section \ref{subsubsection:Inscribed Sphere-Based Initialization for Radial Basis Functions}).
    \item Combine $\hat{G}^{l_s}$ with the sampling points $P$ to form $V^{l_s}=\{\hat{G}^{l_s}\textrm{,} P\}$ as the initial training set, send it to the model along with the corresponding label $T^{l_s}=\{\hat{S}^{1}\textrm{,} \hat{S}^{2}\textrm{,} ...\textrm{,} \hat{S}^{l_s}\textrm{,} I\}$ for supervised optimization.
    \item Using Eq. (\ref{equ:3-19}), compute the loss and optimize until convergence, then add the points of the next finer layer to the training set.
    \item Repeat the third step until all grid points of the octree have been included in the optimization.
\end{enumerate}
This hierarchical method effectively reduces the difficulty of model learning. In addition, trained parameters in the current layer are used as initialization parameters for learning in the next layer, similar to a transfer learning model, thereby improving convergence and stability. \par

Note that the formulations presented here, as well as those introduced below, are derived with respect to the initial layer $l_s$, while the same optimization strategy is applied to subsequent layers. In practice, the initial layer $l_s$ is typically set to the third-to-last layer.

\subsubsection{Adaptive Basis Function Addition}
\label{subsubsection:Adaptive Basis Function Adjustment}

We introduce an adaptive basis function addition mechanism that automatically adds suitable basis functions to areas with large errors during the optimization of the loss function.
Before adding new basis functions, two prerequisites must be satisfied: (1) the condition in Eq. (\ref{equ:l1_activate}) has been fulfilled, meaning the $\mathcal{L}_2$ loss has stabilized; (2) the number of effective basis functions remains stable over a predefined number of optimization steps, formulated as:
\begin{equation}
\label{equ:add_activate}
\begin{aligned}
\resizebox{0.65\textwidth}{!}{$\max\left(\left\{b_f^{(e)}\right\}_{e=e_p-k^{l_1}+1}^{e_p}\right) - \min\left(\left\{b_f^{(e)}\right\}_{e=e_p-k^{l_1}+1}^{e_p}\right) < \tau^{l_1},$}
\end{aligned}
\end{equation}
where $b_f^{(e)}$ indicates the number of effective basis functions at epoch $e$, and is computed as: $b_f^{(e)}=\sum_{j=1}^M \mathbb{I}\left(\left|w_j\right| \geq \tau_d \right)$, $\mathbb{I}(\cdot)$ is the indicator function, which returns 1 if the condition is true and 0 otherwise. The threshold $\tau_d$ is used to determine whether a basis function is effective. $k^{l_1}$ and $\tau^{l_1}$ represent the number of iterations used to determine the stability of the number of basis functions and the corresponding threshold, respectively.
Once the condition of Eq. (\ref{equ:add_activate}) is triggered, the basis function addition process is automatically performed.
After adding the new basis functions, the $\mathcal{L}_1$ regularization term is temporarily deactivated and the network switches to pure $\mathcal{L}_2$ optimization to fully train the newly added basis functions. 
The $\mathcal{L}_1$ regularization term is reactivated for sparse optimization when Eq. (\ref{equ:l1_activate}) and Eq. (\ref{equ:l1_activate2}) are satisfied again. 
The process repeats as follows:
\begin{enumerate} 
    \item Basis function addition is performed when the condition in Eq. (\ref{equ:add_activate}) is met.
    \item This is followed by pure $\mathcal{L}_2$ optimization to train the newly added functions.
    \item When Eq. (\ref{equ:l1_activate}) and Eq. (\ref{equ:l1_activate2}) are satisfied, $\mathcal{L}_1$ regularization is reactivated for sparse optimization.
    \item If the basis function addition condition is met again, the process returns to step 1 and repeats the steps.
\end{enumerate} \par

\begin{algorithm}[ht]
\caption{Adaptive ERBF Addition Algorithm}
\label{alg:add_gaussian_basis}
\hspace*{0.02in} {\bf Input:}
Error vector $E$, the training points $V^{l_s}$, $id_{all}$ and $\tau_{m}$.
\begin{algorithmic}[1]
\State Identify high-error points: $\mathcal{I}_{add} = \{i \mid |E_i| > \frac{\tau_m}{2} \}$.
\State Construct a KD-tree on $V^{l_s}(id_{all})$ and find neighbors within radius $r$ for each selected point.
\State Initialize an empty list for extreme points $\mathcal{I}_{ext}$.
\For{each $i \in \mathcal{I}_{add}$}
    \State Retrieve neighbor indices $\mathcal{N}_i$ from the KD-tree.
    \State Extract absolute errors $E_{\mathcal{N}_i}$ in the neighborhood.
    \If{$|E_i| \geq \max(|E_{\mathcal{N}_i}|)$}
        \[
        \mathcal{I}_{ext} \gets i
        \]
    \EndIf
\EndFor
\If{$|\mathcal{I}_{ext}| = 0$} 
    \State \Return $\emptyset, \emptyset, \emptyset, \emptyset$
\EndIf
\State Compute new ERBF parameters using Eq. (\ref{equ:3-24}):
\end{algorithmic}
\hspace*{0.02in} {\bf Output:}
 $\boldsymbol{C}^a, \boldsymbol{W}^a, \boldsymbol{d}^a, \boldsymbol{A}^a$.
\end{algorithm}

The process alternates cyclically until the predefined number of training epochs $T^e$ is reached.
The detailed process for adding basis functions is given in Algorithm \ref{alg:add_gaussian_basis}.
First, calculate the error vector for each training point in $V^{ls}$, and select the subset indexed by $id_{all}$, which is computed using Eq. (\ref{equ:3-18}):
\begin{equation}
\label{equ:3-22}
\begin{aligned}
E=\Psi(V^{l_s}, \boldsymbol{\widetilde{\Theta}})\left[id_{all}\right]-T^{l_s}\left[id_{all}\right].
\end{aligned}
\end{equation}
Then, we select the points with larger absolute errors based on $\tau_m$. Among these candidates, points whose absolute error is larger than that of all other points in their local neighborhood are identified as extreme error points.
The parameters of the newly added basis functions are computed as follows:
\begin{equation}
\label{equ:3-24}
\begin{aligned}
& \boldsymbol{C}^a =  V^{l_s}\left[id_{all}\right] \left[\mathcal{I}_{ext}\right], \\
& \boldsymbol{W}^a = -\operatorname{sign}(E[\mathcal{I}_{ext}]) \cdot {|E[\mathcal{I}_{ext}]|}, \\
& \boldsymbol{d}^a = \sqrt{-\ln(\varepsilon / |E[\mathcal{I}_{ext}]|) / (\boldsymbol{\bar{d}})^2}, \\
& \boldsymbol{A}^a = \mathbf{0}, \\
\end{aligned}
\end{equation}
where $\mathcal{I}_{ext}$ denotes the indices of the extreme error points. The length of $\mathcal{I}_{ext}$ is $k^a$, which indicates the number of newly added basis functions. 
$\boldsymbol{C}^a \in \mathbb{R}^{k^a \times 3}$ represents the centers of the newly added basis functions.
$\boldsymbol{W}^a \in \mathbb{R}^{k^a \times 1}$ represents the corresponding weights.
$\boldsymbol{d}^a \in \mathbb{R}^{k^a}$ represents the axis lengths of the newly added basis functions. Since the axis lengths are initialized to be identical along all three directions, each scalar element in $\boldsymbol{d}^a$ is expanded into a $3 \times 3$ diagonal matrix to construct $\boldsymbol{D}^a \in \mathbb{R}^{k^a \times 3 \times 3}$.
$\boldsymbol{A}^a \in \mathbb{R}^{k^a \times 3}$ denotes the rotation angles, which are initialized to zero.
$\bar{d}_i$ refers to the $i$-th element of $\boldsymbol{\bar{d}}$, which is defined as the minimum distance from the center $\boldsymbol{c}^a(i)$ to the sampling points $P$, $\bar{d}_{i}=\min \left\|\boldsymbol{c}^a(i)-P\right\|_2, i=1,2, \ldots, k^a$.
$\operatorname{sign}()$ is the sign function:
\begin{equation*}
\operatorname{sign}(x)= \begin{cases}-1, & x<0 \\ 0, & x=0 \\ 1, & x>0\end{cases}. 
\end{equation*}
Next, the SE-RBFNet parameters will be updated according to Eq. (\ref{equ:3-25}), and optimization will continue.
\begin{equation}
\label{equ:3-25}
\begin{aligned}
\boldsymbol{C} &\leftarrow [\boldsymbol{C}; \: \boldsymbol{C}^a]; \quad 
\boldsymbol{W} \leftarrow [\boldsymbol{W}; \: \boldsymbol{W}^a] \\
\bar{\boldsymbol{D}} &\leftarrow [\bar{\boldsymbol{D}}; \: \boldsymbol{D}^a]; \quad 
\boldsymbol{A} \leftarrow [\boldsymbol{A}; \: \boldsymbol{A}^a] \\
\end{aligned}
\end{equation}

In Eq. (\ref{equ:3-22}), $E$ is defined as the difference between the predicted and true values. If the error value is positive, the $\boldsymbol{W}^a$ should be negative to reduce the error. However, in Eq. (\ref{equ:3-8}), the $w$ are restricted to squared values ($w^2$), which in themselves ensure non-negativity. To overcome this restriction, we modify Eq. (\ref{equ:3-8}) as follows:
\begin{equation}
\label{equ:3-26}
\begin{aligned}
\Psi(\boldsymbol{v}_i, \boldsymbol{\widetilde{\Theta}}) &= \sum_{j=1}^M w_j \cdot |w_j| \hat{\Phi}(\boldsymbol{v}_i, \Theta_j) \\
&= \sum_{j=1}^M w_j \cdot |w_j| e^{-\left\|\boldsymbol{D}_j \boldsymbol{R}_j\left(\boldsymbol{v}_i-\boldsymbol{c}_j\right)\right\|_2^2}.
\end{aligned}
\end{equation}
By replacing $w^2$ with $w \cdot |w|$, the network can dynamically learn both positive and negative contributions during the addition of ERBFs, thereby enhancing error correction capability.

During SE-RBFNet optimization, some basis functions have coefficients that become so small that their contribution to the overall output can be ignored. We introduce a threshold $\tau_{d}$ and remove the basis functions whose $w$ coefficients are less than $\tau_{d}$ at the specified iteration interval. That is:
\begin{equation}
\label{equ:3-27}
\begin{aligned}
\boldsymbol{\widetilde{W}}& =\left\{w_j \in \boldsymbol{W}| \: |w_j|<\tau_{d}\right\}, \\
\boldsymbol{\widetilde{C}}& =\left\{\boldsymbol{c}_j \in \boldsymbol{C}| \: |w_j|<\tau_{d}\right\}, \\
\boldsymbol{\widetilde{D}}& =\left\{\boldsymbol{D}_j \in \bar{\boldsymbol{D}}| \: |w_j|<\tau_{d}\right\}, \\
\boldsymbol{\widetilde{A}}& =\left\{\boldsymbol{a}_j \in \boldsymbol{A}| \: |w_j|<\tau_{d}\right\}. \\
\end{aligned}
\end{equation}
Deleting these basis functions allows the network to automatically adjust its complexity, increasing sparsity and computational efficiency.
When training reaches $T^e$ epochs, the optimization focuses exclusively on the $\mathcal{L}_2$ loss term to further improve the precision of the SDF approximation, continuing until the maximum iteration count $M^e$ is reached. At this stage, no further basis function additions or deletions occur, and $\mathcal{L}_1$ regularization is no longer applied.

\subsubsection{Additional Techniques for Training Acceleration}
\label{subsubsection:Parallel Computing Acceleration}

In practice, the size of the input set $G$ contains a very large number of points, and the problem of optimization $\mathcal{L}$ is a high-dimensional non-convex nonlinear problem. To reduce computational burden and increase convergence speed, we primarily employ the following techniques for accelerating the training process.

In Eq. (\ref{equ:3-13}), $\boldsymbol{F}$ has a high sparsity due to the fact that the ERBF value of points far from the center of the basis function approaches 0. To improve efficiency, we introduce a point screening strategy based on nearest neighbor. 
Specifically, for each ERBF, we only calculate those points that are relatively close to the center of the basis function, with the distance range defined as follows:
\begin{equation}
\label{equ:3-15}
\|\boldsymbol{c}_{j} - V\|_2 \leq \sqrt{\frac{-\ln (\varepsilon)}{\lambda_{j}}} \text{,}
\end{equation}
where $\boldsymbol{c}_{j}$ is the center of the $j$-th ERBF, $\lambda_{j}$ represents the minimum eigenvalues of the matrix $\boldsymbol{R}_{j}^{T}\boldsymbol{D}_{j}^{T}\boldsymbol{D}_{j}\boldsymbol{R}_{j}$. Since $\boldsymbol{R}_{j}$ is orthogonal and $\boldsymbol{D}_{j}$ is a diagonal matrix, $\lambda_{j} = \min(d^2_{j1}, d^2_{j2}, d^2_{j3})$.
The entries of $\boldsymbol{F}$ can be re-expressed as:
\begin{equation}
\label{equ:3-15-1}
\boldsymbol{F}_{i j}= \begin{cases}\hat{\Phi}\left(\boldsymbol{v}_i, \Theta_j\right), & \text { if }\left\|\boldsymbol{c}_j-\boldsymbol{v}_i\right\|_2<\sqrt{\frac{-\ln (\varepsilon)}{\lambda_{j}}} \\ 0, & \text { otherwise. }\end{cases}
\end{equation}

\begin{algorithm}[htbp]
\caption{SE-RBFNet Sparse Optimization Algorithm}
\label{alg:se-rbfnet}
\begin{adjustbox}{max width=\textwidth}
\begin{minipage}{1.4\textwidth}
\hspace*{0.02in} {\bf Input:}
Sampling points $P$ on an implicit surface, batch size $B_s=10000$, $M^e=2000$, $T^e=1600$, initial learning rate $lr=0.01$, maximum depth of the octree $l_{max}=10$,  $l_s$ is set by default to the third-to-last layer, $\gamma=10^{-3}$, $\varepsilon=10^{-7}$, $\tau_1=0.9$, $\tau_2=1.1$, $\tau_m=0.02$, $\tau_d=0.01$, $\tau^{l_1}=5$, $\tau^{l_2}=0.5$, $k^{l_1}=50$, $k^{l_2}=10$.
\begin{algorithmic}[1]
\State Construct an octree from the sampling points $P$, and compute the SDF at each octree grid point $G$, which is then normalized according to Eq. (\ref{equ:3-10}) to obtain $\hat{S}$ as the target for sparse optimization;
\State According to the $l_s$, obtain the initial training points $V^{l_s}$ and corresponding real labels $T^{ls}$, and then calculate the parameters of the initial ERBF using Algorithm \ref{alg:misc}.
\State Initialize SE-RBFNet parameters: $\boldsymbol{\widetilde{\Theta}}\leftarrow\{\boldsymbol{C}\textrm{,}\bar{\boldsymbol{D}}\textrm{,} \boldsymbol{A}, \boldsymbol{W}\}$.
\State Set $l_{1}\_{optim}$=False, $add\_point$=False, $e_p=1$;
\While {$e_{p} < M^e$}
    \State $e_{p} \leftarrow e_{p} + 1$,
    \If{$add\_point$ and $l_s<l_{max}$}
        \State The grid points from the $(l_s+1)$-th layer are selected and appended to the training set, followed by updating: $V^{l_s} \gets V^{l_s+1}$, $T^{ls} \gets T^{ls+1}$ and $l_s \gets l_s+1$.
        \State Set $add\_point$ = False.
    \EndIf
    \State Shuffle training points and corresponding labels.
    \For{$i = 0$ to $N$ with step $B_{s}$}
        \State Sample batch $V^{l_s}_{b}$ and $T^{l_s}_{b}$
        \State Use Eq. (\ref{equ:3-9}) to calculate the network forward result: $\boldsymbol{O} \leftarrow \text{SE-RBFNet}(V^{l_s}_{b})$
        \State Calculate $id_{all}$ according to Eq. (\ref{equ:3-18})
        \State Compute $\mathcal{L}_2$ and $\mathcal{L}_1$ loss using Eq. (\ref{equ:3-14_2}) 
        \State Explicitly compute the gradient $\nabla{\mathcal{L}(\boldsymbol{\widetilde{\Theta}})}$ using the formulas in \ref{section:gradients}.
        \If{$l_{1}\_{optim}$=True and Eq. (\ref{equ:l1_activate2}) is True}
            \State Use Eq. (\ref{equ:3-21}) to compute adaptive $\alpha, \beta$
            \State Update loss: $\mathcal{L} \gets \alpha \mathcal{L}_{2} + \beta \mathcal{L}_1$
        \Else
            \State Update loss: $\mathcal{L} \gets \mathcal{L}_{2}$
        \EndIf
        \State Update network parameters $\boldsymbol{\widetilde{\Theta}}$ via Adam \cite{kingma2014adam}.
    \EndFor
    \If{$e_{p} \: \% \: k^{l_2}=0 $ and $l_{1}\_{optim}$=True and $e_{p} < T^e$}
        \State Delete invalid basis functions using Eq. (\ref{equ:3-27}).
    \EndIf
    \If {$l_{1}\_{optim}$=False and Eq. (\ref{equ:l1_activate}) is True}
        \State Update $l_{1}\_{optim}$=True.
    \EndIf
    \If{$l_{1}\_{optim}$=True and Eq. (\ref{equ:add_activate}) is True} 
        \State Add new ERBF according to Eq. \ref{equ:3-24}.
        \State Update $l_{1}\_{optim}$=False.
    \EndIf
    \If{$e_{p} = T^e$}
        \State During the iterations from $T^p$ to $M^p$, steps 26 to 35 are skipped, and only $\mathcal{L}_2$ optimization is performed. $l_{1}\_{optim}$=False, $lr=10^{-3}$ and adjusted using a cosine annealing schedule, with a minimum value of $10^{-5}$.
    \EndIf
    \If{$e_{p}$ in $[400, 800]$}
        \State $add\_points = \text{True}$
    \EndIf
\EndWhile
\end{algorithmic}
\hspace*{0.02in} {\bf Output:}
 Network parameters $\boldsymbol{\widetilde{\Theta}}$, predicted SDF $\boldsymbol{O}$, a surface via the marching cubes algorithm  \cite{lorensen1998marching}.
\end{minipage}
\end{adjustbox}
\end{algorithm}

The matrix $\boldsymbol{F}$ is used not only for the forward calculation of the network but also plays a role in the calculation of the gradient of each parameter. To avoid repeated calculations, we explicitly compute the gradient of each parameter based on $\boldsymbol{F}$ and the loss $\mathcal{L}$ (Eq. (\ref{equ:3-14})) according to the chain rule. 
The gradient of each parameter is denoted as:
\begin{equation*}
\label{equ:3-15-2}
\begin{aligned}
\nabla{\mathcal{L}(\boldsymbol{\widetilde{\Theta}})}=\left\{\alpha \cdot \frac{\partial \mathcal{L}_2}{\partial \boldsymbol{C}}, \alpha \cdot \frac{\partial \mathcal{L}_2}{\partial \bar{\boldsymbol{D}}}, \alpha \cdot \frac{\partial \mathcal{L}_2}{\partial \boldsymbol{A}}, \alpha \cdot \frac{\partial \mathcal{L}_2}{\partial \boldsymbol{W}} + \beta \cdot \frac{\partial \mathcal{L}_1}{\partial \boldsymbol{W}}\right\}.
\end{aligned}
\end{equation*}
The detailed derivations of the gradients can be found in \ref{section:gradients}. As shown in the \ref{section:gradients}, when the value of $\boldsymbol{F}$ at a certain location is close to zero, the corresponding gradient also approaches zero. Consequently, to improve the calculation efficiency, we take advantage of the nearest-neighbor strategy as described in Eq. (\ref{equ:3-15}), only the gradient values of the points near the center points of the basis functions are calculated. 

Finally, due to the inherent independence of each ERBF in the network, we leverage CUDA to parallelize the computation of the value of the loss function and the gradient with respect to each parameter in $\Theta$. This independence allows each ERBF to be computed independently and concurrently, leading to a significant reduction in computational time.

SE-RBFNet is optimized using the Adam method \cite{kingma2014adam}, with the complete optimization details provided in Algorithm \ref{alg:se-rbfnet}.

\section{Experiment and analysis}
\label{section:Experiment}
In this section, we evaluate the accuracy, sparsity, and efficiency of SE-RBFNet. It is important to emphasize that SE-RBFNet is designed to approximate the SDF of a general implicit surface, rather than directly reconstruct surfaces from raw point clouds. Therefore, we do not perform direct comparisons with full surface reconstruction methods. For experimental demonstration, we conducted two complementary experiments: one using ground-truth SDFs from triangle meshes, and the other using predicted SDFs from neural implicit methods (or any other SDF-generating methods).


First, to assess sparsity and parameter efficiency, we compare SE-RBFNet with SparseRBF \cite{li2016sparse}, which also employs sparse RBFs for implicit surface representation.
In the original SparseRBF framework, the inputs consist of surface points (with zero SDF), offset points, and center points, where the SDF values are approximated from Voronoi diagram poles \cite{amenta2001power} rather than computed exactly. To ensure a fair comparison, we instead provide both methods with identical input data: area-weighted uniformly sampled points on the mesh surface (SDF = 0) and octree grid points, whose SDF values are defined as the directional point-to-mesh distances and computed directly from the triangle mesh geometry using the GPU-accelerated Kaolin API \cite{jatavallabhula2019kaolin}. 
The octree construction follows the implementation of \cite{lin2022surface}.
This setup eliminates potential inaccuracies introduced by approximate SDF estimation in SparseRBF and ensures that the comparison focuses solely on the sparsity and approximation accuracy of the two methods. 

Second, to demonstrate the generality of SE-RBFNet, we evaluate its ability to sparsely approximate arbitrary SDF data without emphasizing the source and the accuracy of the SDF computation. In this scenario, we start from sampled points on the implicit surface to construct the corresponding octree and obtain SDF values on the octree grid points using neural implicit methods or arbitrary SDF-generation methods. SE-RBFNet further sparsifies the representation by approximating the input SDFs using significantly fewer parameters, while preserving high accuracy of the surface from the approximated SDF by SE-RBFNet.
This highlights the compatibility of SE-RBFNet with existing SDF generation methods and its advantage in reducing storage and transmission costs. 

Together, these two parts of the experiments allow us to evaluate both the sparsity and parameter efficiency of SE-RBFNet, as well as its generality across different SDF sources. All experiments were performed on Ubuntu 18.04 with an Intel Core i9-7940X CPU @ 3.1GHz and an NVIDIA GeForce RTX 2080 GPU. For reproducibility, the executable used in our experiments is available at \url{https://github.com/lianbobo/SE-RBFNet.git}

\subsection{Datasets and Parameter Settings}
\label{subsection: Datasets and Parameter Settings}
To evaluate the algorithm, we mainly used four commonly used datasets: the Famous dataset \cite{Erler_2020}, MeshSegBenchmark (MeshSeg) \cite{chen2009benchmark}, ABC \cite{koch2019abc}, and Thingi10K \cite{Thingi10K}. The Famous dataset, which contains 22 well-known models in geometric processing, and MeshSeg, a 3D mesh segmentation benchmark consisting of 380 meshes from 19 object categories. Since the meshes within each category share similar characteristics, we randomly selected 60 meshes for our experiments.
The ABC dataset contains approximately 1 million CAD models, while Thingi10K includes 10,000 3D printable meshes. Using all available models for evaluation would result in a prohibitively high computational cost. To balance efficiency and representativeness, we follow the experimental setup of Points2Surf \cite{Erler_2020} and select a subset of 100 meshes from each dataset for our experiments.
In addition, we also used several real scanned data \cite{Erler_2020} to demonstrate the effectiveness of SE-RBFNet.

We set the parameters of SE-RBFNet as follows: batch size $B_s=10000$, max epoch $M^e=2000$, learning rate $lr=0.01$, maximum depth of the octree $l_{max}=10$, the two thresholds for filtering training points: $\tau_1=0.9$ and $\tau_2=1.1$, $\tau^{l_1}=5$ and $k^{l_1}=50$ are used to determine whether the number of basis functions has stabilized, while $\tau^{l_2}=0.5$ and $k^{l_2}=10$ are used to judge the stability of the $\mathcal{L}_2$ loss. $\tau_m=0.02$ controls the sparsity optimization process, and $\tau_d=0.01$ is used as the deletion threshold for ERBFs.
The initial training layer $l_s$ is set by default to the third-to-last layer. However, it must also satisfy the following condition: $\hat{S}^{l_s}_{\text{in}} > 100$, otherwise, $l_s$ is incremented by one, i.e., $l_s = l_s + 1$. 
This constraint ensures that the number of inner grid points in the $l_s$ layer must be greater than 100 to avoid having too few initial ERBFs. Experiments have found that $l_s$ is generally 6 or 7. The specific use of all parameters is shown in Algorithm \ref{alg:se-rbfnet}. 
All comparison methods are used with their default parameter settings. 
\par

\subsection{Evaluation Metric}
In our setting, SE-RBFNet aims to approximate a precomputed ground-truth SDF. To ensure a fair and meaningful evaluation, the ground-truth SDF is computed directly from the original triangle mesh geometry, avoiding intermediate errors of SDF values.
It is important to note that our loss function assigns greater emphasis to points close to the surface (see Eq.~\ref{equ:3-18}).
Therefore, accurately approximating SDF values on all grid points is not the purpose of SE-RBFNet. Instead, SE-RBFNet aims to preserve the geometry of the zero-level set (i.e., the implicit surface).
Accordingly, our evaluation focuses on the geometric deviation between the zero-level set surface from the outputs of SE-RBFNet and the isosurface from the ground-truth SDF, as measured by surface-to-surface distances and normal consistency.
Specifically, the network-predicted normalized SDF values are first transformed back to the physical SDF range according to Eq.~(\ref{equ:3-10}). Using the same octree structure, the iso-surface at zero level is then extracted from the transformed ERBF-approximated SDF and compared with the ground truth surface directly obtained from the ground-truth SDF, which represents the true zero-level set. The explicit surface extraction from the SDF values on grids is performed using the method provided in \cite{lin2022surface}.
We use the following metrics—Hausdorff Distance (HD) \cite{lin2022surface}, Chamfer Distance (CD) \cite{Erler_2020}, and Cosine Similarity (CS) \cite{park2019deepsdf}—to compare the zero-level set surface from the outputs of SE-RBFNet with the isosurface of the ground-truth SDF.
Smaller HD/CD values and higher CS values indicate that the surfaces extracted from the approximated SDF closely match the true zero-level set, thus reflecting higher fidelity of the SDF representation of the implicit surface.

Suppose that $S_t$ and $S_r$ are the ground truth surface and the surface extracted from the approximated SDF, respectively. 
Both HD and CD are computed using point-to-point distances, where the distance from a point to a surface is defined as the minimum Euclidean distance to any point on that surface.
Formally, the distance from a point $x$ to a surface $S$ is defined as: 
\begin{equation} 
\label{equ:point2face} 
d(x, S) = \min_{y \in S} \|x - y\|_2, 
\end{equation} 
where $y \in S$ denotes a surface point.

The HD measures the worst-case deviation between the surface $S_t$ and $S_r$, and is defined as: 
\begin{equation} 
\label{equ:haustof}
\begin{aligned}
\mathrm{HD}(S_t, S_r) = \max \left( \max_{x \in S_t} d(x, S_r), \: \max_{y \in S_r} d(y, S_t) \right). 
\end{aligned}
\end{equation}

The Chamfer Distance (CD) measures the bidirectional average deviation between the surface $S_t$ and $S_r$, and is computed as: 
\begin{equation} 
\label{equ:chamfer}
\begin{aligned}
\mathrm{CD}(S_t, S_r) = \frac{1}{2|S_t|} \sum_{x \in S_t} d(x, S_r) + \frac{1}{2|S_r|} \sum_{y \in S_r} d(y, S_t), 
\end{aligned}
\end{equation}
where $|\cdot|$ denotes the number of points sampled uniformly on the surface.

In addition to geometric distances, normal consistency is another critical factor in assessing surface representation quality. To evaluate the similarity between normal vectors, we compute the cosine similarity between the normals of corresponding nearest-neighbor points:
\begin{equation}
\label{equ:cosin sim}
\begin{aligned}
\mathrm{CS}\left(S_t, S_r\right)&=\frac{1}{2\left|S_t\right|} \sum_{x \in S_t } \left| v\left(x\right) \cdot v\left(\text {closest}_{S_r}(x)\right)\right|_1 \\
& +\frac{1}{2\left|S_r\right|} \sum_{y \in S_r} \left|v\left(y\right) \cdot v(\text {closest}_{S_t}(y))\right|_1,
\end{aligned}
\end{equation}
where $v(x)$ is the surface normal at point $x$, $\text {closest}_{S_r}(x)$ is the point in $S_r$ closest to $x$. For all the above metrics, $10^5$ points are uniformly sampled from each surface for evaluation.

\subsection{Sparsity and Parameter Efficiency Evaluation}

\begin{figure}[!t]
\centering
    \subfloat[\small Error Map + Details (Horse)]{
    \label{fig:horse}
        \includegraphics[width=0.45\textwidth]{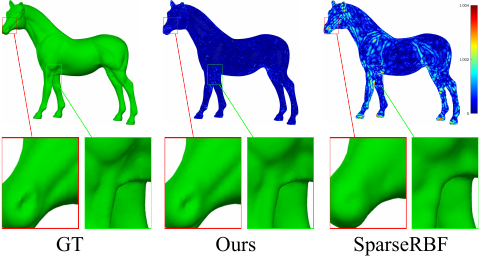}
    } 
    \hspace{0.05\textwidth}
    \subfloat[\small Error Map + Details (Hand)]{
    \label{fig:hand}
        \includegraphics[width=0.45\textwidth]{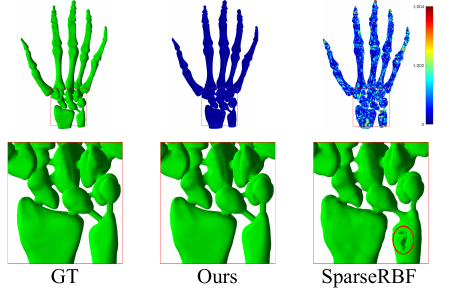}
    }
    \hfil 
    \subfloat[\small  Error Map + Details (Person)]{
    \label{fig:person}
        \includegraphics[width=0.45\textwidth]{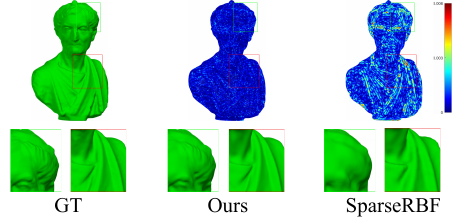}
    }
    \hspace{0.05\textwidth}
    \subfloat[\small Error Map + Details (Dragon)]{
    \label{fig:dragon}
        \includegraphics[width=0.45\textwidth]{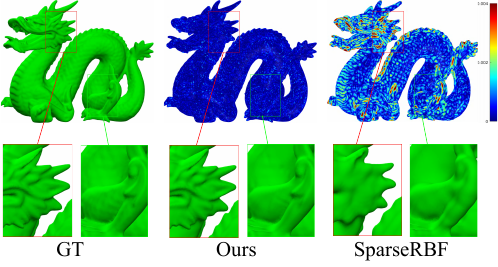}
    }
    \hfil 
    \subfloat[\small  Error Map + Details \scriptsize (statue\_ps\_outliers)]{
    \label{fig:statue-ps-outliers}
        \includegraphics[width=0.45\textwidth]{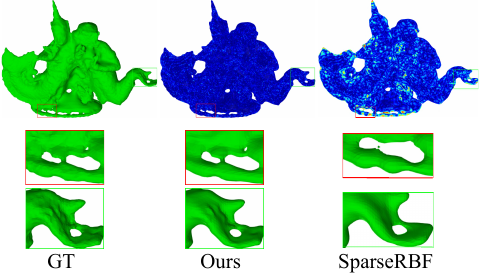}
    }
    \hspace{0.05\textwidth}
    \subfloat[\small Error Map + Details \scriptsize (torch\_ps\_outliers)]{
    \label{fig:torch-ps-outliers}
        \includegraphics[width=0.45\textwidth]{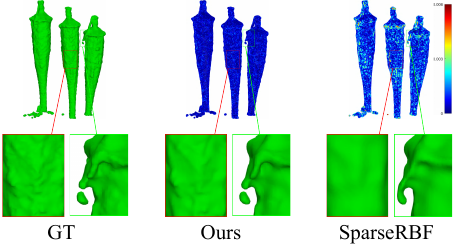}
    }
    \caption{Comparison of implicit surface approximation results obtained by our method and SparseRBF. (a), (b), and (d) are from the Famous dataset; (c) is from the Thingi10k dataset; (e) and (f) correspond to two real-world scanned datasets provided by Erler et al. \cite{Erler_2020}.}
    \label{fig:eval-sim}
\end{figure}

In this experiment, we focus on evaluating the sparsity and parameter efficiency of our method in comparison with SparseRBF~\cite{li2016sparse}, which also employs sparse RBFs for surface representation. 
For each mesh, 40K points are first sampled using the \texttt{sample\_surface\_even} function in the Trimesh~\cite{trimesh} API, which performs area-weighted uniform sampling over the mesh surface. These sampled points are then used to construct an octree, and the ground-truth SDF values at the octree grid points (defined by the directional point-to-mesh distances) are precomputed directly from the mesh geometry using the Kaolin API~\cite{jatavallabhula2019kaolin}, serving as the training samples for our network. 

Figure \ref{fig:eval-sim} shows the extracted surfaces for several representative cases. 
As shown in the figure, the surfaces obtained from SE-RBFNet’s approximated SDF more closely match the ground-truth surface, effectively preserving geometric details while avoiding artifacts. 
Specifically, Figure \ref{fig:horse} presents the error map and detailed comparison of the Horse model. Our method preserves more detailed features, such as the horse’s head and leg contours, whereas SparseRBF appears smoother but loses geometric details. 
For the Hand model (Figure \ref{fig:hand}), the SparseRBF approach produces noticeable holes, resulting in incomplete surface representations. 
Similarly, the Person and Dragon cases further highlight the robustness of our approach in accurately approximating SDFs of the ground-truth implicit surface, with the Dragon model showing substantially reduced errors compared with SparseRBF.
In addition, Figures \ref{fig:statue-ps-outliers} and \ref{fig:torch-ps-outliers} demonstrate results on two real-world scanned datasets. As illustrated by the local views in the second row of these figures, SE-RBFNet achieves visually optimal results, effectively preserving fine holes and thin structural connections, while SparseRBF tends to oversmooth these delicate features. These qualitative results confirm the effectiveness of our approach in accurately approximating the SDF, thereby enabling the extraction of high-fidelity implicit surfaces.

\begin{table}[ht]
\centering
\renewcommand{\arraystretch}{1.2}
\caption{Accuracy, Time (in seconds) , and Sparsity Comparison Between Our Method and SparseRBF for Implicit Surfaces Shown in Figure \ref{fig:eval-sim}.}
\label{table:result_comparison1}
\resizebox{\textwidth}{!}{
\begin{tabular}{c|c|cccc||ccc}
\toprule
\multirow{2}{*}{\vspace{-1ex}\bfseries Implicit Surface\vspace{-1ex}} & 
\multirow{2}{*}{\vspace{-1ex}\bfseries Method\vspace{-1ex}} & 
\multicolumn{4}{c||}{\bfseries Surface Geometric Metrics} & 
\multicolumn{3}{c}{\bfseries Sparsity Comparison} \\
\cmidrule(lr){3-6} \cmidrule(lr){7-9}
 & & 
\bfseries HD  & \bfseries CD  &  \bfseries CS & \bfseries Time  &  \bfseries Init Basis & \bfseries Opt Basis  & \bfseries Param \\
\midrule
\multirow{2}{*}{\bfseries Horse} 
& \bfseries Ours        & \textbf{0.0042}  & \textbf{0.0016}  & \textbf{0.9958}  & \textbf{106.9}  & 3223  & \textbf{612}  & \textbf{6120} \\
& \bfseries SparseRBF   & 0.0081          & 0.0017            & 0.9919          & 809.7  & 17789 & 3490   & 17450   \\
\midrule
\multirow{2}{*}{\bfseries Hand} 
& \bfseries Ours        & \textbf{0.0033}  & \textbf{0.0015}  & \textbf{0.9900}  &  \textbf{90.5}  & 2683  & \textbf{1149} & \textbf{11490} \\
& \bfseries SparseRBF   & 0.0126          & 0.0017           & 0.9773          & 654.2  & 15935 & 4970    & 24850       \\
\midrule
\multirow{2}{*}{\bfseries Person} 
& \bfseries Ours        & \textbf{0.0044}  & \textbf{0.0016}  & \textbf{0.9949}  &  \textbf{132.0}  & 6314  & \textbf{893} & \textbf{8930}  \\
& \bfseries SparseRBF   & 0.0062          & 0.0018           & 0.9886          & 1030.2  & 18127 & 5256  & 26280        \\
\midrule
\multirow{2}{*}{ \bfseries Dragon} 
& \bfseries Ours        & \textbf{0.0053}  & \textbf{0.0016}  & \textbf{0.9914}  &  \textbf{138.3}  & 5076  & \textbf{1859}  & \textbf{18590} \\
& \bfseries SparseRBF   & 0.0206          & 0.0019           & 0.9769          & 1117.3  & 18455 & 6426  & 32130         \\
\midrule
\multirow{2}{*}{\bfseries  \shortstack{\bfseries statue\_ps\_outliers}} 
& \bfseries Ours        & \textbf{0.0092}  & \textbf{0.0015}  & \textbf{0.9861}  & \textbf{136.7}  & 5217  & \textbf{1192} & \textbf{11920} \\
& \bfseries SparseRBF   & 0.0256    & 0.0018     & 0.9700    & 883.2  & 17540 & 5574 & 27870      \\
\midrule
\multirow{2}{*}{\bfseries \shortstack{\bfseries torch\_ps\_outliers}} 
& \bfseries Ours        & \textbf{0.0082}  & \textbf{0.0016}  & \textbf{0.9830}  & \textbf{143.1}  & 6419  & \textbf{1386} & \textbf{13860} \\
& \bfseries SparseRBF   & 0.0506          & 0.0020    & 0.9637    & 934.8  & 17852 & 5105     & 25525   \\
\bottomrule
\end{tabular}}
\end{table}

In Table \ref{table:result_comparison1}, we present a comprehensive quantitative study of the experimental results of the implicit surfaces shown in Figure \ref{fig:eval-sim}. The results show that our method achieves better performance across all metrics, including HD, CD, and CS. 
The superior performance of SE-RBFNet highlights its ability to accurately represent the SDF using a sparse set of ERBFs.

Using as few basis functions as possible to represent the SDFs of implicit surfaces is crucial for reducing storage and computational overhead. This is particularly important when processing SDFs of complex surfaces with rich geometric details.
As demonstrated in \cite{li2016sparse}, SparseRBF achieves substantially better sparsity than prior RBF-based methods \cite{carr2001reconstruction, ohtake2006sparse, samozino2006reconstruction} by combining medial-axis-based center selection with a linear $\mathcal{L}_{1}$ sparsity optimization strategy.
The right side of Table \ref{table:result_comparison1} shows a comparison of the number of effective basis functions obtained from SparseRBF and our method. Here, Init Basis denotes the number of initial basis functions. For our method, Init Basis includes both the initial and the newly added basis functions during optimization, while Opt Basis denotes the number of effective basis functions remaining after the optimization. In addition, Param denotes the number of parameters required to represent an implicit surface. Specifically, SE-RBFNet represents the surface using a sparse set of anisotropic ERBFs. Each ERBF is parameterized by its center, rotation angles, axis lengths, and a weight coefficient, totaling 10 parameters per ellipsoid. In contrast, SparseRBF approximates surfaces using a set of isotropic Gaussian RBFs, where each basis function is defined by its center coordinates, a scalar radius, and a weight, totaling $5$ parameters per RBF.
Compared with SparseRBF, our method significantly reduces the number of basis functions, requiring on average only about 45\% of its parameters while achieving higher accuracy in implicit surface representation. This demonstrates that SE-RBFNet achieves a higher compression ratio while maintaining superior approximation accuracy. 
Moreover, the computation speed of our method is on average approximately seven times faster than that of SparseRBF, fully validating the effectiveness of our sparse optimization strategy. Note that the time values reported in Table \ref{table:result_comparison1} refer only to the optimization/training process and do not include the time required for data preparation, extracting the explicit surface from the SDF, etc.  
In summary, these results show that SE-RBFNet can approximate precomputed SDFs with high fidelity while using fewer parameters and requiring less computation time, making it suitable for efficient and accurate implicit surface representation in practical applications.

\begin{table}[ht]
\centering
\renewcommand{\arraystretch}{1.2}
\caption{Average Geometric Approximation Accuracy, Training Time (in seconds), and Sparsity of Implicit Surface Representations on Four Datasets.}
\label{table:performance_in_four_datasets}
\resizebox{\textwidth}{!}{
\begin{tabular}{c|c|cccc||cccc}
\toprule
\multirow{2}{*}{\vspace{-1ex}\bfseries Dataset\vspace{-1ex}} & 
\multirow{2}{*}{\vspace{-1ex}\bfseries Method\vspace{-1ex}} & 
\multicolumn{4}{c||}{\bfseries Surface Geometric Metrics} & 
\multicolumn{4}{c}{\bfseries Sparsity Comparison} \\
\cmidrule(lr){3-6} \cmidrule(lr){7-10}
 & & 
\bfseries HD  & \bfseries CD  &  \bfseries CS & \bfseries Time  &  \bfseries Init Basis & \bfseries Opt Basis  & \bfseries Param & \bfseries Ratio \\
\midrule
\multirow{2}{*}{\bfseries ABC} 
& \bfseries Ours        & \textbf{0.0071}  & \textbf{0.0021}  & \textbf{0.9910}  & \textbf{119.0}  & 4806  & \textbf{1170}  & \textbf{11700} & \multirow{2}{*}{\bfseries 0.38}\\
& \bfseries SparseRBF   & 0.0529          & 0.0049            & 0.9616          & 923.8  & 16920 & 6079   & 30395 &   \\
\midrule
\multirow{2}{*}{\bfseries Famous} 
& \bfseries Ours        & \textbf{0.0058}  & \textbf{0.0019}  & \textbf{0.9846}  &  \textbf{167.8}  & 6361  & \textbf{2050} & \textbf{20500} & \multirow{2}{*}{\bfseries 0.68}\\
& \bfseries SparseRBF   & 0.0384          & 0.0022          & 0.9643          & 918.6  & 17596 & 5969    & 29845 &       \\
\midrule
\multirow{2}{*}{\bfseries MeshSeg} 
& \bfseries Ours        & \textbf{0.0086}  & \textbf{0.0024}  & \textbf{0.9957}  &  \textbf{103.7}  & 3619  & \textbf{582} & \textbf{5820} & \multirow{2}{*}{\bfseries 0.23}  \\
& \bfseries SparseRBF   & 0.0195          & 0.0026           & 0.9910          & 797.5  & 16980 & 4961  & 24805  &      \\
\midrule
\multirow{2}{*}{ \bfseries Thingi10k} 
& \bfseries Ours        & \textbf{0.0066}  & \textbf{0.0018}  & \textbf{0.9940}  &  \textbf{140.8}  & 6060  & \textbf{1097}  & \textbf{10970} & \multirow{2}{*}{\bfseries 0.43} \\
& \bfseries SparseRBF   & 0.0191          & 0.0021           & 0.9844          & 895.5  & 17009 & 5086  & 25430 &         \\
\bottomrule
\end{tabular}}
\end{table}

\begin{figure}[ht]
\centering
    \includegraphics[width=0.65\textwidth]{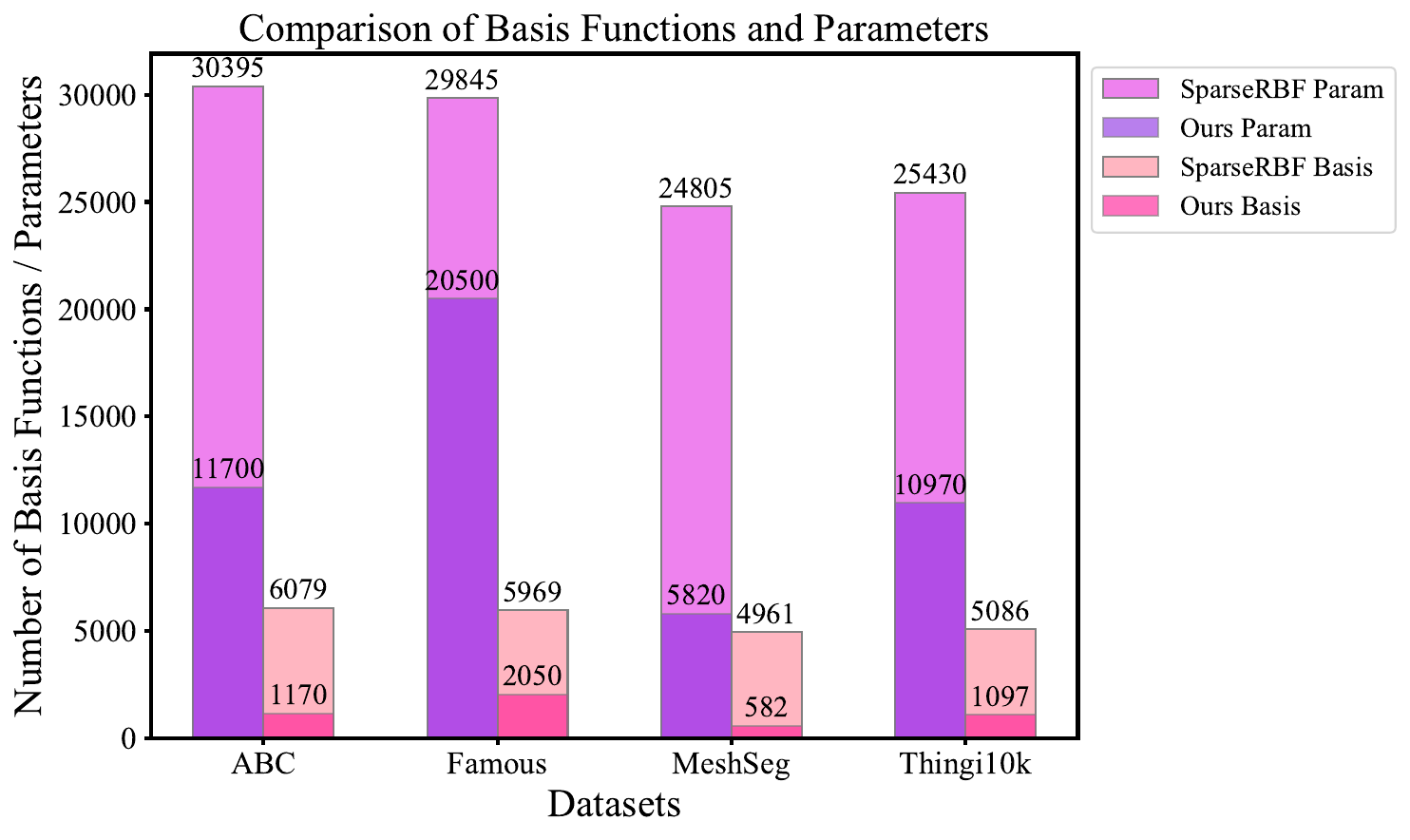}
    \caption{Comparison of average number of basis functions and parameters on four benchmark datasets.}
    \label{fig:basis_num}
\end{figure}

To further validate the generality and robustness of our proposed method, we performed a comprehensive evaluation on four datasets: ABC, Famous, MeshSeg, and Thingi10k. 
Table \ref{table:performance_in_four_datasets} reports a comprehensive comparison between our method and SparseRBF in terms of geometric approximation accuracy, sparsity, and computational efficiency. In terms of surface quality, our method consistently achieves significantly lower HD and CD values and higher CS scores across all datasets, indicating more accurate implicit surface representations and better preservation of fine geometric details. SparseRBF, in contrast, struggles to capture fine details on complex surfaces, often producing overly smoothed surface representations and leading to degraded accuracy. Beyond accuracy, our method also demonstrates strong sparsity advantages. The right columns of Table \ref{table:performance_in_four_datasets} and Figure \ref{fig:basis_num} compare the average numbers of effective basis functions and parameters used by our method with those of SparseRBF. The "Ratio" column indicates the proportion of parameters required by our method relative to those of SparseRBF. The results demonstrate that, compared with SparseRBF, our method reduces the number of basis functions by about 78\% on average, corresponding to only about 44\% of the parameters on average required by SparseRBF, while still maintaining superior accuracy. This reduction not only decreases memory consumption but also accelerates computation. As shown in the Time column of Table \ref{table:performance_in_four_datasets}, our method consistently achieves a 6–7$\times$ speedup over SparseRBF across all datasets. Taken together, SE-RBFNet achieves a favorable balance between approximation accuracy, sparsity, and efficiency.
Its ability to approximate complex SDFs with fewer parameters and significantly lower computation time makes it particularly suitable for real-world applications where both precision and efficiency are crucial.

\subsection{General SDF Sparse Representation}

As a general sparse implicit surface representation method, SE-RBFNet can be trained on any given SDF sample sets and directly implements sparse representation from SDF point values, regardless of their source (e.g., point clouds, triangle meshes, analytical SDFs, or existing neural SDFs).
To further validate the method's universality, we apply it to SDF data generated by the neural implicit method Neural-Singular-Hessian (NSH) \cite{wang2023neural}. Specifically, NSH is first used to predict SDF values at octree nodes of the input sampling points, upon which SE-RBFNet constructs its sparse representation. In addition to comparing with SparseRBF, we further compare against the fast RBF interpolation algorithm RBF-QNN provided by the ALGLIB library \cite{alglib}. RBF-QNN adopts an automatic radius selection mechanism, keeps the centers and shapes of RBFs fixed during computation, and achieves computational complexity close to $O(N \log N)$, where $N$ denotes the number of training points defined previously. For fairness, all three algorithms are evaluated using the same input data to assess approximation quality.

\begin{figure}[!t]
\centering
    \subfloat[\small Surface Represented by the Approximate SDF and Its Error Map (3DBenchy)]{
    \label{fig:3DBenchy_sdf}
        \includegraphics[width=\textwidth]{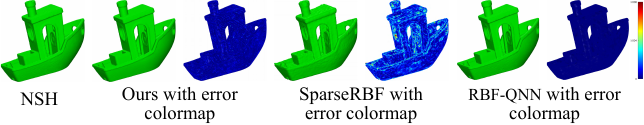}
    }
    \hfil
    \subfloat[\small Surface Represented by the Approximate SDF and Its Error Map (tortuga)]{
    \label{fig:tortuga_sdf}
        \includegraphics[width=\textwidth]{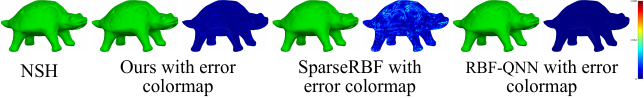}
    }
    \hfil
    \subfloat[\small Surface Represented by the Approximate SDF and Its Error Map (00010218)]{
    \label{fig:00010218_sdf}
        \includegraphics[width=\textwidth]{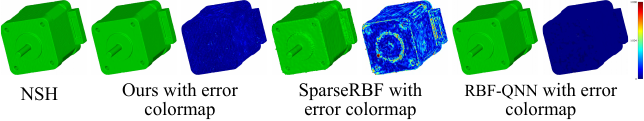}
    }
    \hfil
    \subfloat[\small Surface Represented by the Approximate SDF and Its Error Map (331105)]{
    \label{fig:331105_sdf}
        \includegraphics[width=\textwidth]{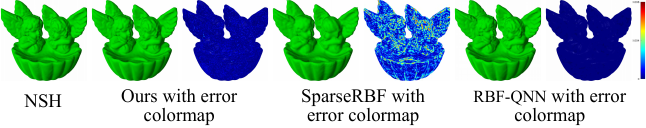}
    }
    \caption{Comparison of the implicit surfaces extracted from the approximated SDFs by our method, SparseRBF, and RBF-QNN. (a) and (b) are from the MeshSeg dataset, (c) is from the ABC dataset, while (d) is from the Thingi10k dataset.}
    \label{fig:eval-interest}
\end{figure}

Experimental results are presented in Figure \ref{fig:eval-interest}. For consistency, explicit surfaces are extracted under the same octree grid using the program provided in \cite{lin2022surface}, and comparisons are made against the target surfaces generated from NSH under identical conditions. As shown in Figure \ref{fig:eval-interest}, both our method and RBF-QNN closely approximate the SDF values predicted by NSH, while SparseRBF exhibits the most pronounced error map, primarily due to oversmoothing of fine surface details that results in significant deviations. Quantitative results are summarized in Table \ref{table:result_comparison_sdf}.
Our method consistently outperforms SparseRBF in geometric accuracy metrics and achieves results on par with RBF-QNN. In terms of sparsity, our approach requires the fewest parameters, delivering superior results with an average of only about $35\%$ of the parameters used by SparseRBF. By contrast, RBF-QNN lacks sparsity, with a parameter count that scales linearly with the number of input points, resulting in the largest parameter size among the three methods. Regarding computational efficiency, our method exhibits the shortest training times on average: it is approximately $8\times$ faster than SparseRBF and even faster than RBF-QNN, while preserving high surface fidelity. 
Moreover, NSH itself requires average $264\,\mathrm{K}$ parameters to present the SDF on grids of the test cases in Table \ref{table:result_comparison_sdf}, whereas our method achieves high-fidelity sparse representation using fewer than $2,000$ basis functions, corresponding to no more than $20\,\mathrm{K}$ parameters. This indicates that, compared with NSH, our method maintains high approximation accuracy while achieving over $10\times$ parameter compression for the test cases in Table \ref{table:result_comparison_sdf}, greatly improving storage and transmission efficiency.
More importantly, this experiment not only validates the superiority of the proposed method over traditional RBF interpolation baselines but also demonstrates its strong compatibility with advanced neural implicit representation models. In other words, SE-RBFNet provides a lightweight and efficient sparse representation for neural SDFs, effectively reducing computational and storage costs without sacrificing geometric fidelity. This highlights its potential for efficient geometric representation and resource-constrained scenarios.

\begin{table}[ht]
\centering
\renewcommand{\arraystretch}{1.2}
\caption{Accuracy, Training Time (in seconds), and Parameter Comparison of Our Method, SparseRBF, and RBF-QNN for Implicit Surfaces Shown in Figure \ref{fig:eval-interest}. The ground-truth SDFs used in this evaluation are obtained from NSH \cite{wang2023neural}.}
\label{table:result_comparison_sdf}
\resizebox{0.95 \textwidth}{!}{
\begin{tabular}{c|c|cccc||cc}
\toprule
\multirow{2}{*}{\vspace{-1ex}\bfseries Implicit Surface\vspace{-1ex}} & 
\multirow{2}{*}{\vspace{-1ex}\bfseries Method\vspace{-1ex}} & 
\multicolumn{4}{c||}{\bfseries Surface Geometric Metrics} & 
\multicolumn{2}{c}{\bfseries Parameter Comparison} \\
\cmidrule(lr){3-6} \cmidrule(lr){7-8}
 & & 
\bfseries HD  & \bfseries CD  &  \bfseries CS & \bfseries Time  &  \bfseries Basis Count & \bfseries Param\\
\midrule 
\multirow{3}{*}{\bfseries  \shortstack{\bfseries 3DBenchy}} 
& \bfseries Ours        & 0.0068  & 0.0022  & \textbf{0.9891}  &  \textbf{130.7}  & 2601  & \textbf{26010} \\
& \bfseries SparseRBF   & 0.0912  & 0.0025  & 0.9725  &  901.6  & 11143  & 55715 \\
& \bfseries RBF-QNN     & \textbf{0.0067}    & \textbf{0.0021}     & 0.9890    & 150.9  & 632023 & 3160116      \\
\midrule
\multirow{3}{*}{\bfseries  \shortstack{\bfseries tortuga}} 
& \bfseries Ours        & 0.0662  & 0.0023  & \textbf{0.9888}   & \textbf{119.1}  & 1688  & \textbf{16880} \\
& \bfseries SparseRBF   & 0.0650  & 0.0024  & 0.9861  & 1188.4  & 8926  & 44630 \\
& \bfseries RBF-QNN     & \textbf{0.0647}    &  \textbf{0.0022}     & 0.9885   & 192.5  & 738927 & 3694635      \\
\midrule
\multirow{3}{*}{\bfseries  \shortstack{\bfseries 00010218}} 
& \bfseries Ours        & 0.0069  & \textbf{0.0021}  & \textbf{0.9947}  & \textbf{120.1}  & 1638  & \textbf{16380} \\
& \bfseries SparseRBF   & 0.2128  & 0.0084  & 0.9767  & 1007.9  & 10485  & 52425 \\
& \bfseries RBF-QNN     & \textbf{0.0066}    & \textbf{0.0021}   & \textbf{0.9947}    & 168.4  & 659887 & 3299435       \\
\midrule
\multirow{3}{*}{\bfseries \shortstack{\bfseries 331105}} 
& \bfseries Ours        & \textbf{0.0065}  & 0.0019  & 0.9962  & \textbf{130.1}  & 1494  & \textbf{14940}  \\
& \bfseries SparseRBF   & 0.0109  & 0.0020  & 0.9918  & 1163.1  & 11314  & 56570 \\
& \bfseries RBF-QNN     & \textbf{0.0065} & \textbf{0.0018}    & \textbf{0.9963}    & 183.1  & 721626 & 3608130     \\
\bottomrule
\end{tabular}}
\end{table}

\subsection{Parameter Analysis}
We analyze the key parameters used in SE-RBFNet, focusing on those that significantly affect model performance. The primary parameters include the threshold $\tau_m$ and the batch size $B_s$, both of which play a critical role in balancing speed, accuracy, and sparsity. The threshold $\tau_m$, introduced in Step 1 of Algorithm \ref{alg:add_gaussian_basis} and in Eq. \ref{equ:l1_activate2}, serves two purposes. First, during the basis function addition process, $\tau_m$ is used to assist in the identification of local extreme error points. Second, during optimization, once the max absolute error drops below $\tau_m$, the $\mathcal{L}_1$ regularization term is incorporated into the loss function to perform sparse optimization.
We also conducted experiments using area-weighted uniform sampling on the mesh surface with varying numbers of sampled points to evaluate the robustness of the implicit surface approximation.
Other parameters not explicitly mentioned follow the settings described in Section \ref{subsection: Datasets and Parameter Settings}.

\subsubsection{\texorpdfstring{Impact of $\tau_m$ on Approximation Results}{Impact of tau\_m on Approximation Results}}

\begin{figure}[!t]
\centering
    \subfloat[\small Armadillo Surface Approximation Results and Error Colormaps under Different $\tau_m$]{
    \label{fig:Armadillo_thres}
        \includegraphics[width=\textwidth]{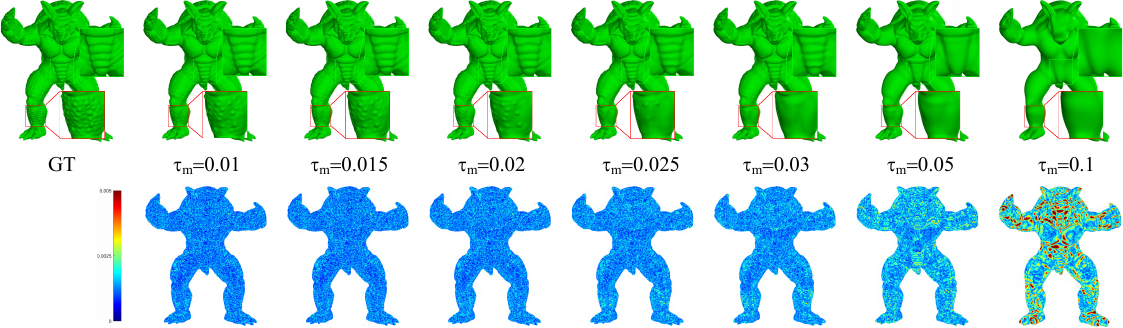}
    }
    \hfil
    \subfloat[\small Variation of CD, Time, $\mathcal{L}_2$ Loss and Basis Function Count under Different $\tau_m$]{
    \label{fig:Armadillo_thres_eval}
        \includegraphics[width=0.24\textwidth]{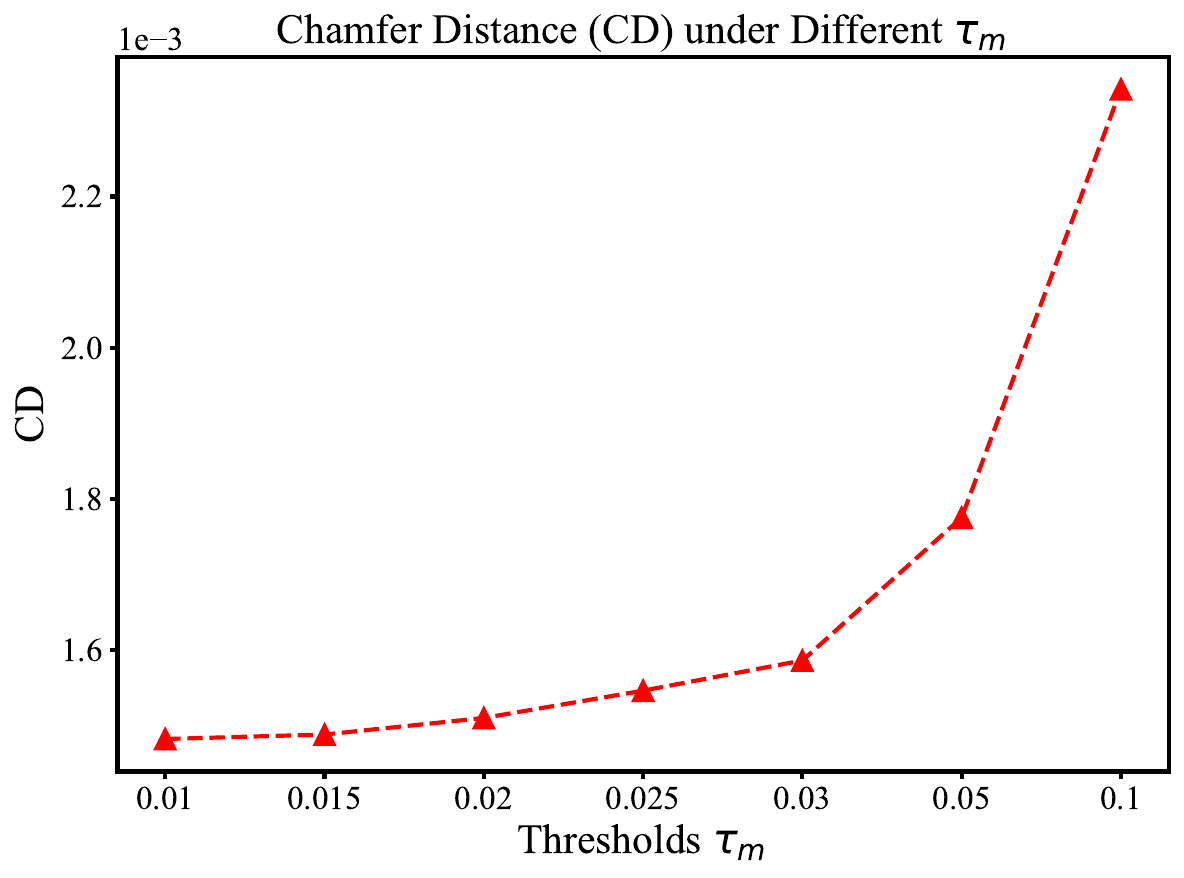}
        \includegraphics[width=0.24\textwidth]{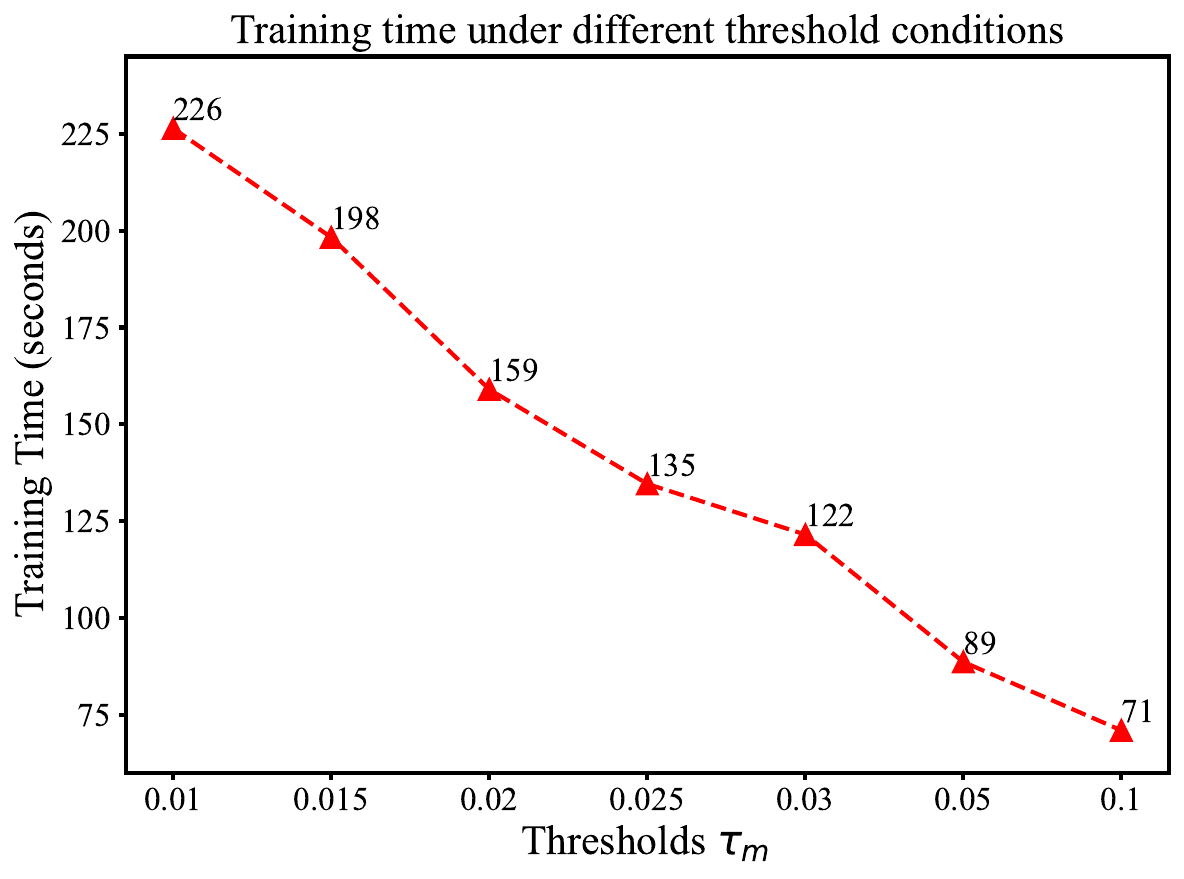}
        \includegraphics[width=0.24\textwidth]{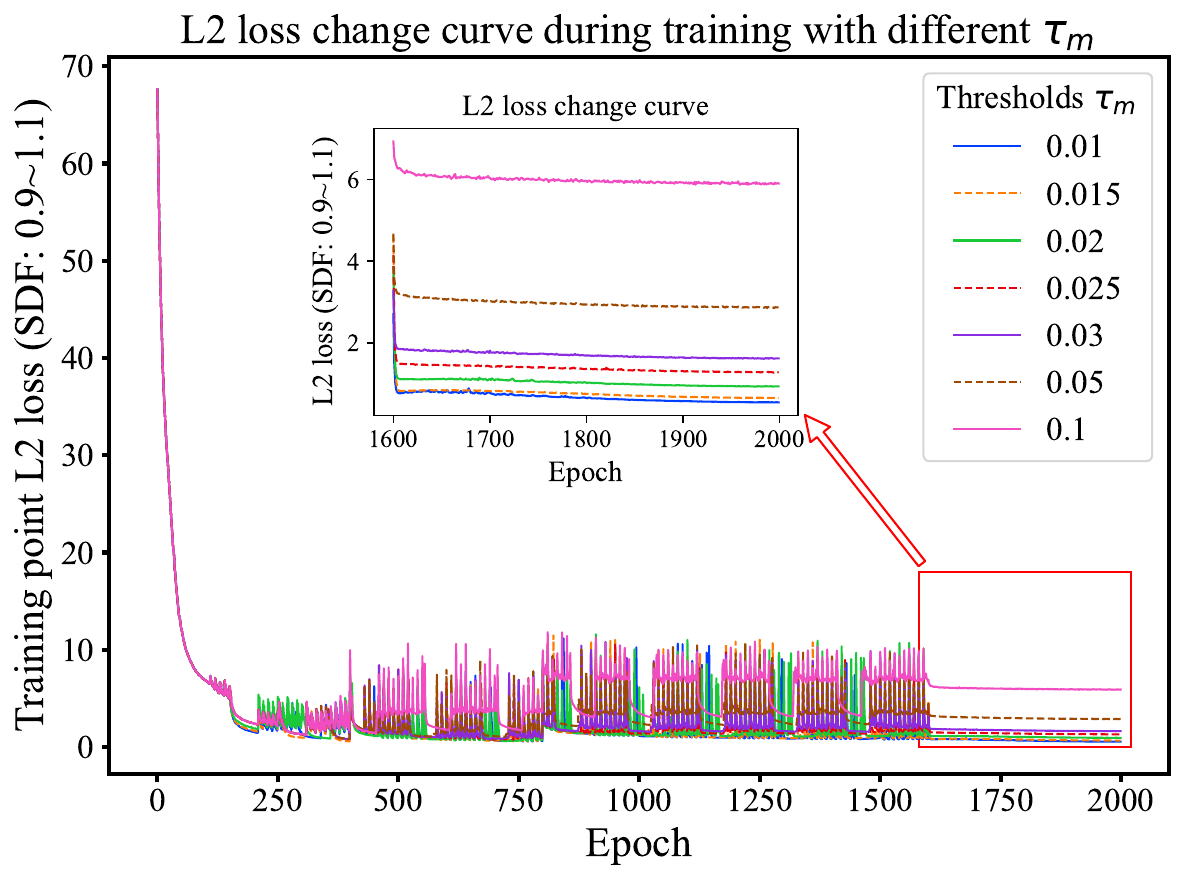}
        \includegraphics[width=0.24\textwidth]{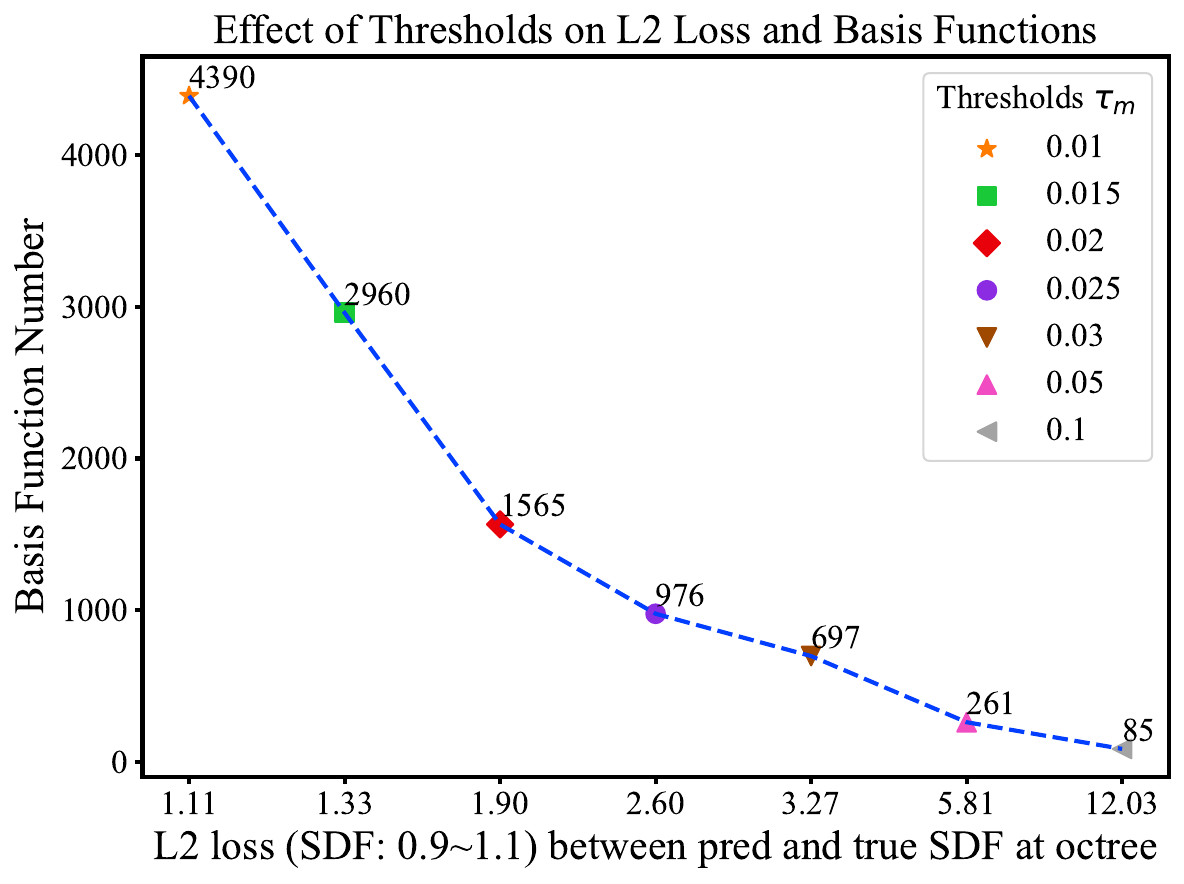}
        
    }
    \caption{Effect of $\tau_m$ on implicit surface approximation results, (a) Armadillo implicit surface approximation results and error colormaps under different $\tau_m$; (b) shows the corresponding metric variation curves. From left to right, these curves represent the variation of the chamfer distance with different $\tau_m$, computational times, loss convergence during training, and the number of effective basis functions.}
    \label{fig:thres_analyse}
\end{figure}

In Figure \ref{fig:thres_analyse}, we analyze the effect of different $\tau_m$ on approximation results. Figure \ref{fig:Armadillo_thres} shows the implicit surfaces extracted from the approximated SDF and corresponding error colormaps under various $\tau_m$. As $\tau_m$ increases, finer surface details tend to be lost. This occurs because $\tau_m$ controls the sparse optimization trigger: when the maximum absolute error falls below $\tau_m$ (see Eq. \ref{equ:l1_activate2}), sparse optimization prunes basis functions, reducing their number. A larger $\tau_m$ thus activates pruning earlier and more frequently, potentially causing insufficient detail capture. For example, small bumps on the leg and stripes on the abdomen become less pronounced at higher $\tau_m$.
Figure \ref{fig:Armadillo_thres_eval} (leftmost) presents the CD between the implicit surfaces extracted from the approximated SDF and the ground-truth surface. Smaller values of $\tau_m$ lead to lower CD, indicating higher approximation accuracy.
The error colormaps in Figure \ref{fig:Armadillo_thres} also confirm this finding. Figure \ref{fig:Armadillo_thres_eval}  (third from left) illustrates the $\mathcal{L}_2$ loss during training under different $\tau_m$ values, where smaller thresholds lead to lower converged losses, reflecting better SDF fitting. Figure \ref{fig:Armadillo_thres_eval} (rightmost) depicts the relationship between $\tau_m$, final $\mathcal{L}_2$ loss, and the number of basis functions after optimization. Decreasing $\tau_m$ results in lower loss but requires more basis functions and longer training time, as shown in Figure \ref{fig:Armadillo_thres_eval} (second from left).
Based on these observations, $\tau_m=0.02$ strikes an optimal balance between approximation accuracy and efficiency: it achieves nearly the same accuracy as $\tau_m=0.015$ while using roughly half the number of basis functions and significantly reducing computation time.

\subsubsection{\texorpdfstring{Impact of $B_s$ on Approximation Results}{Impact of B\_s on Approximation Results}}

\begin{figure}[!t]
\centering
    \subfloat[\small Implicit Surface Approximation Results and Distributions of Basis Function Centers under Different $B_s$ (Utah\_teapot\_solid)]{
    \label{fig:Utah_bs}
        \includegraphics[width=\textwidth]{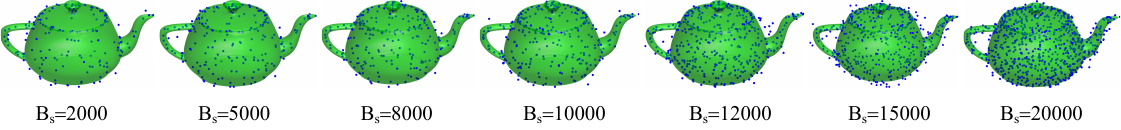}
    }
    \hfil
    \subfloat[\small Variation of CD, Time, $\mathcal{L}_2$ Loss and Basis Function Count under Different $B_s$]{
    \label{fig:Utah_bs_eval}
        \includegraphics[width=0.24\textwidth]{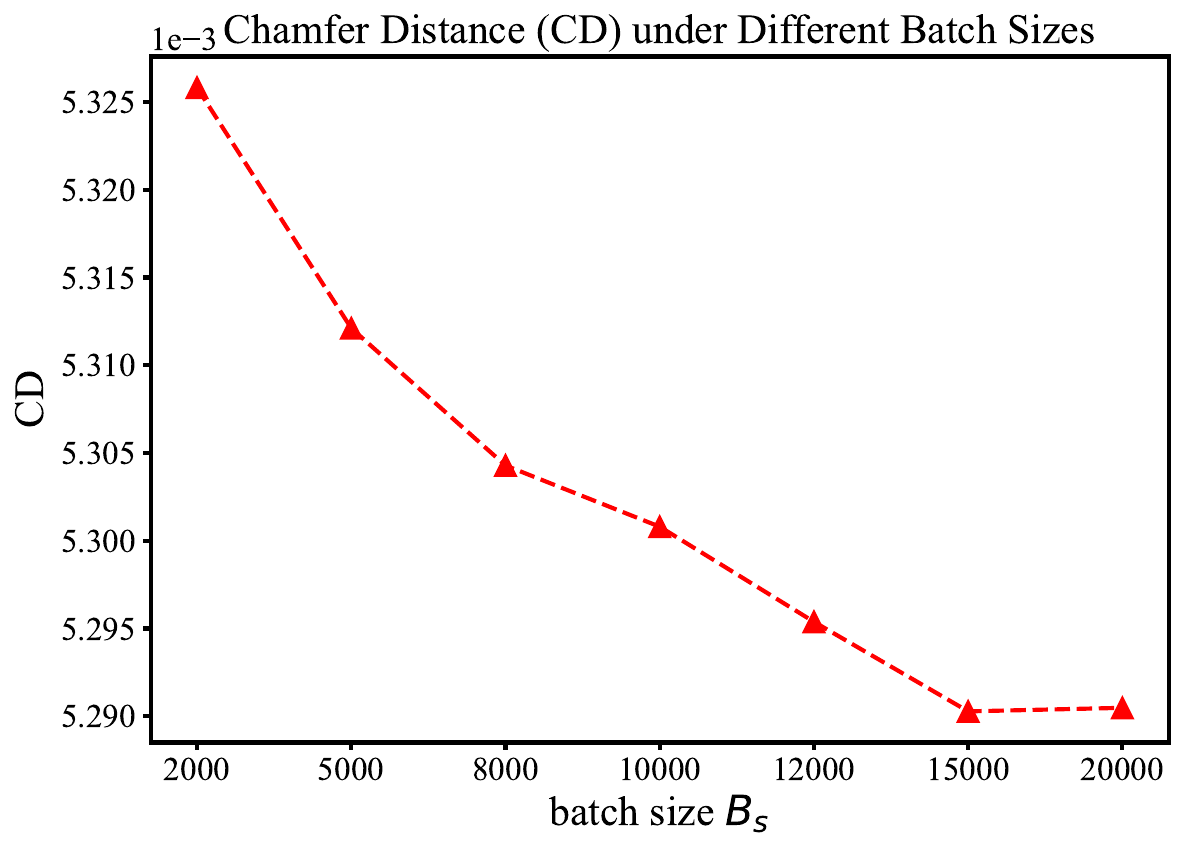}
        \includegraphics[width=0.24\textwidth]{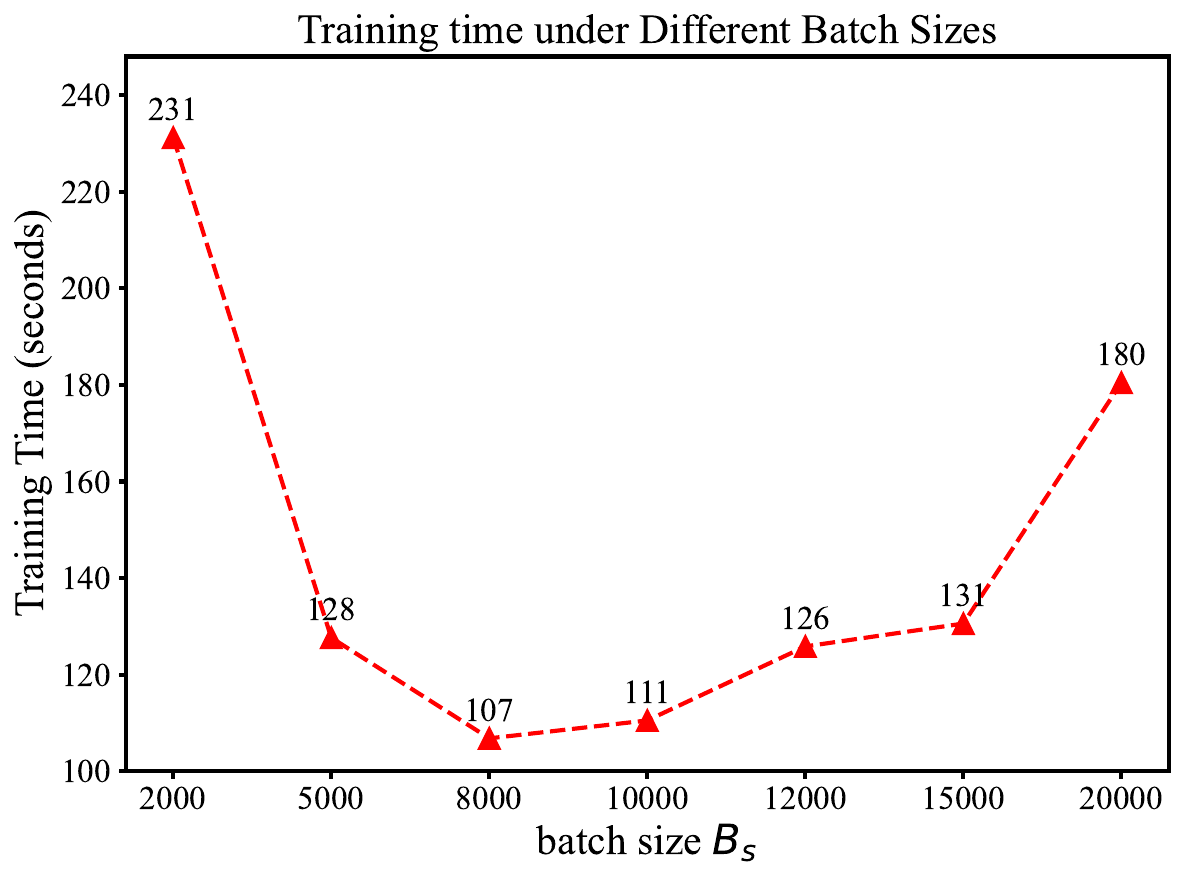}
        \includegraphics[width=0.24\textwidth]{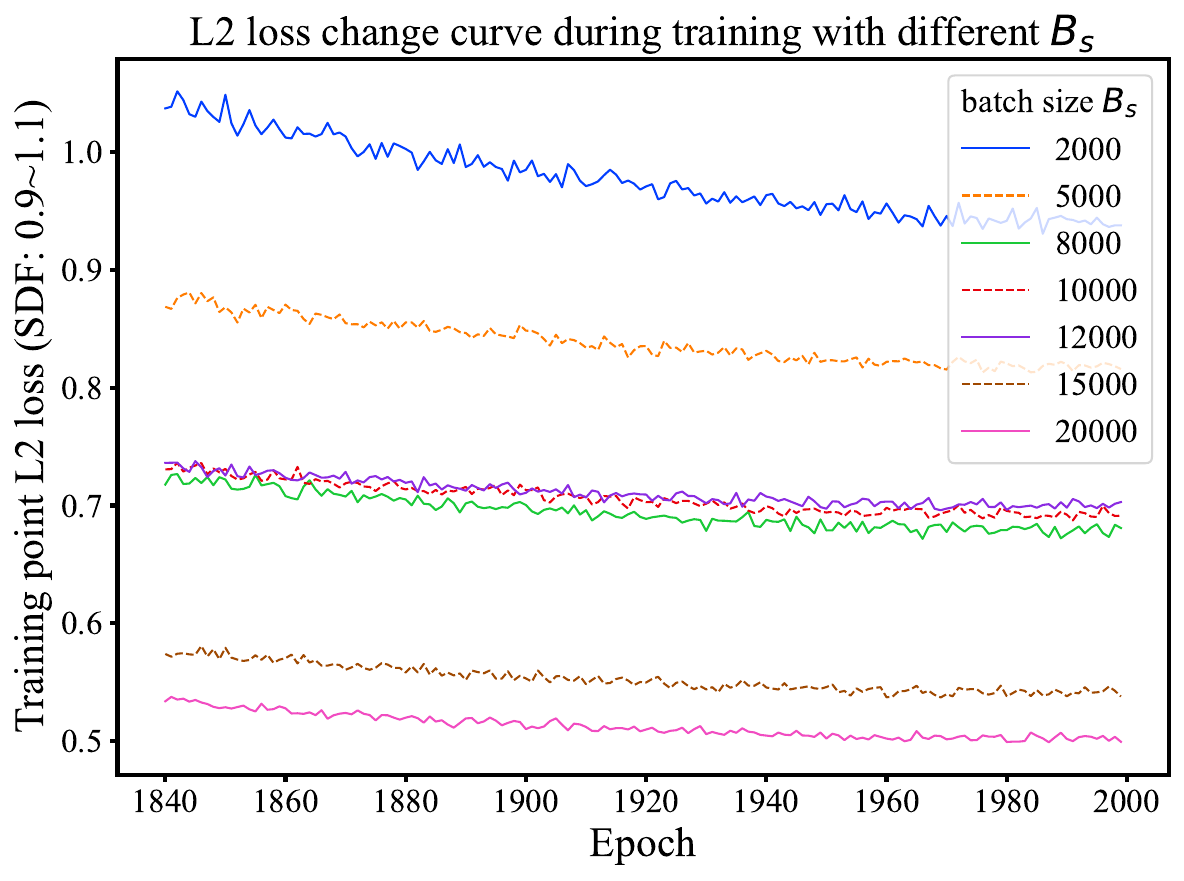}
        \includegraphics[width=0.24\textwidth]{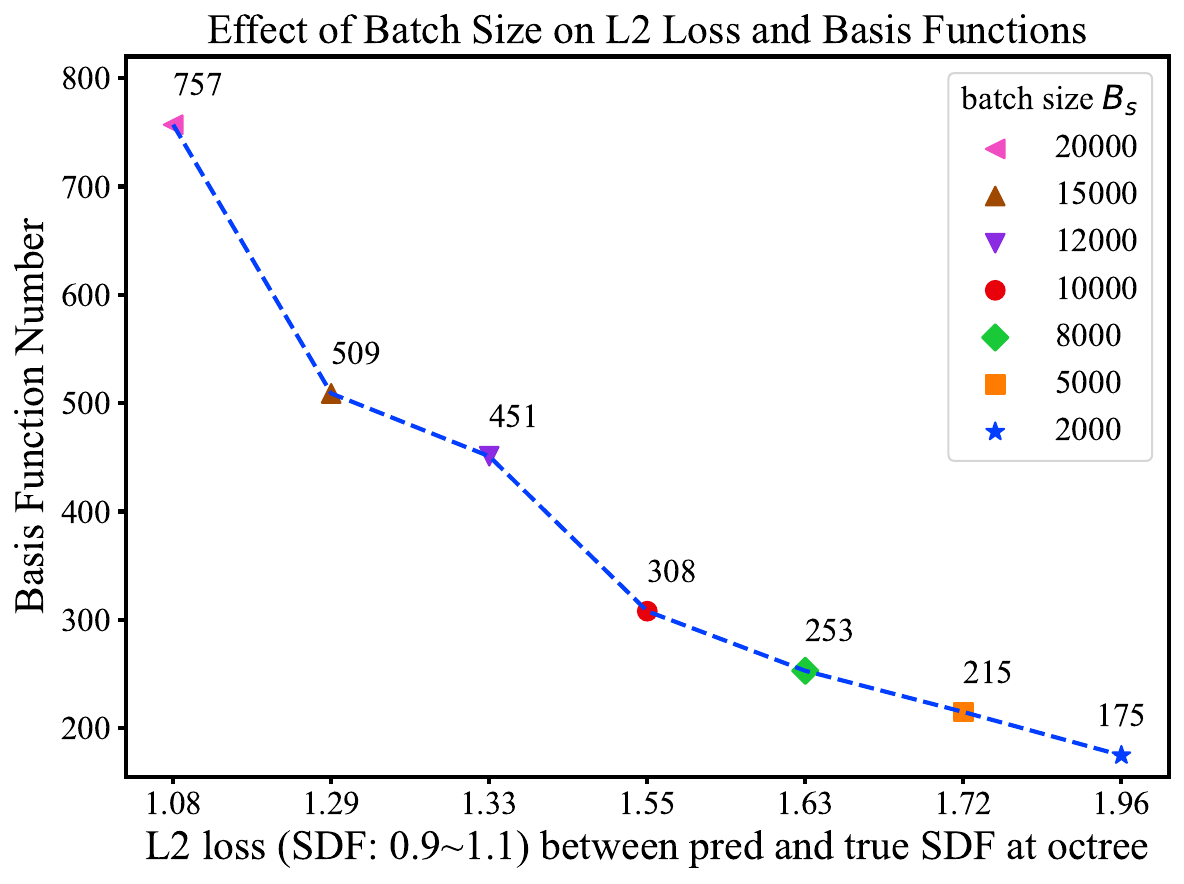}
    }
    \caption{Effect of batch size $B_s$ on implicit surface approximation results. (a) shows the implicit surface approximation results and the distribution of the basis function centers for the Utah\_teapot\_solid surface under different batch sizes; (b) shows the corresponding metric curves, from left to right: chamfer distance, training time, loss convergence, and the number of effective basis functions.}
    \label{fig:Bs_analyse}
\end{figure}

Batch size $B_s$ also influences the approximation quality. Figure \ref{fig:Bs_analyse} shows results on the Utah\_teapot\_solid model from the Famous dataset. Figure \ref{fig:Utah_bs} displays the approximated implicit surfaces and basis function center distributions for varying $B_s$. Increasing $B_s$ leads to more effective basis functions: as shown in the rightmost plot of Figure \ref{fig:Utah_bs_eval}, the number of basis functions increases from 175 at $B_s=2000$ to 757 at $B_s=20000$.
This is attributed to training stability: smaller batch sizes induce weight fluctuations under $\mathcal{L}_1$ regularization, causing basis functions to oscillate and be pruned, while larger batch sizes yield steadier updates and less pruning.
The leftmost and third plots of Figure \ref{fig:Utah_bs_eval} indicate that the CD decreases and loss converges lower as $B_s$ increases, reflecting improved fitting accuracy due to more basis functions. 
Furthermore, the second plot of Figure \ref{fig:Utah_bs_eval} shows that training time first decreases with $B_s$ owing to efficient GPU parallelization, but then increases again as excessively large batch sizes induce a greater number of basis functions, leading to longer training times.
Therefore, we select $B_s=10000$ as a compromise between approximation accuracy and computational cost for this test case.

\subsubsection{Impact of the Number of Sampled Points}

\begin{figure}[!t]
\centering
    \subfloat[\small CD with Different Numbers of Sampled Points (10K, 20K, 30K, 40K)]{
        \label{fig:cd_sample}
        \includegraphics[width=0.22\textwidth]{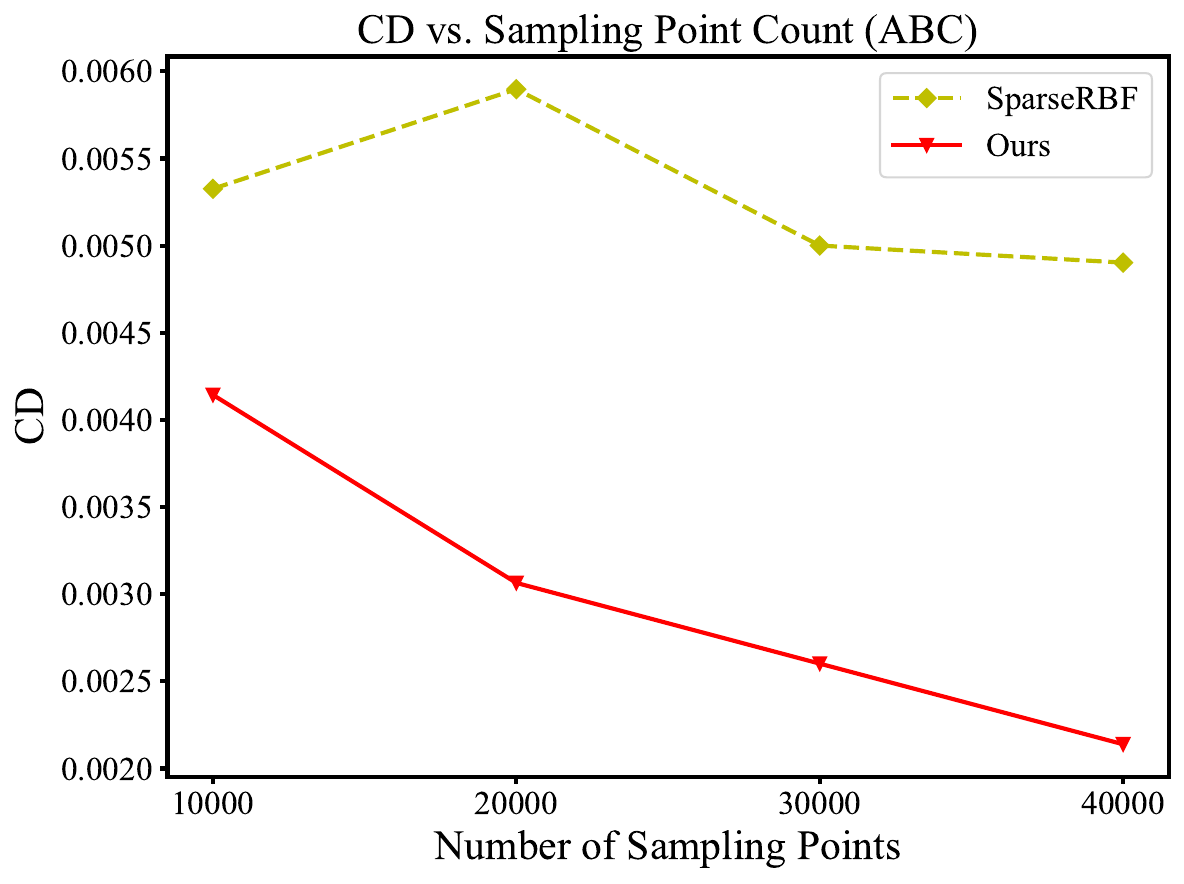}
        \includegraphics[width=0.22\textwidth]{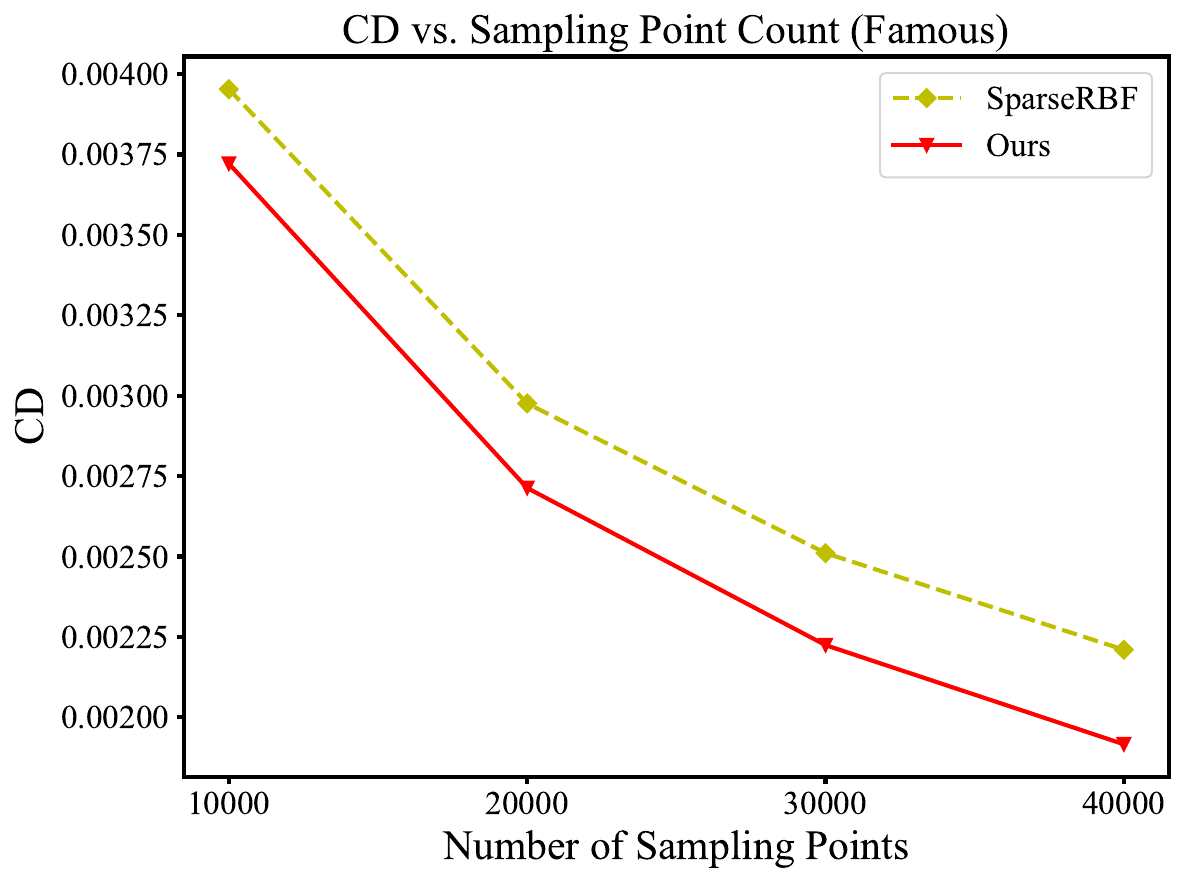}
        \includegraphics[width=0.22\textwidth]{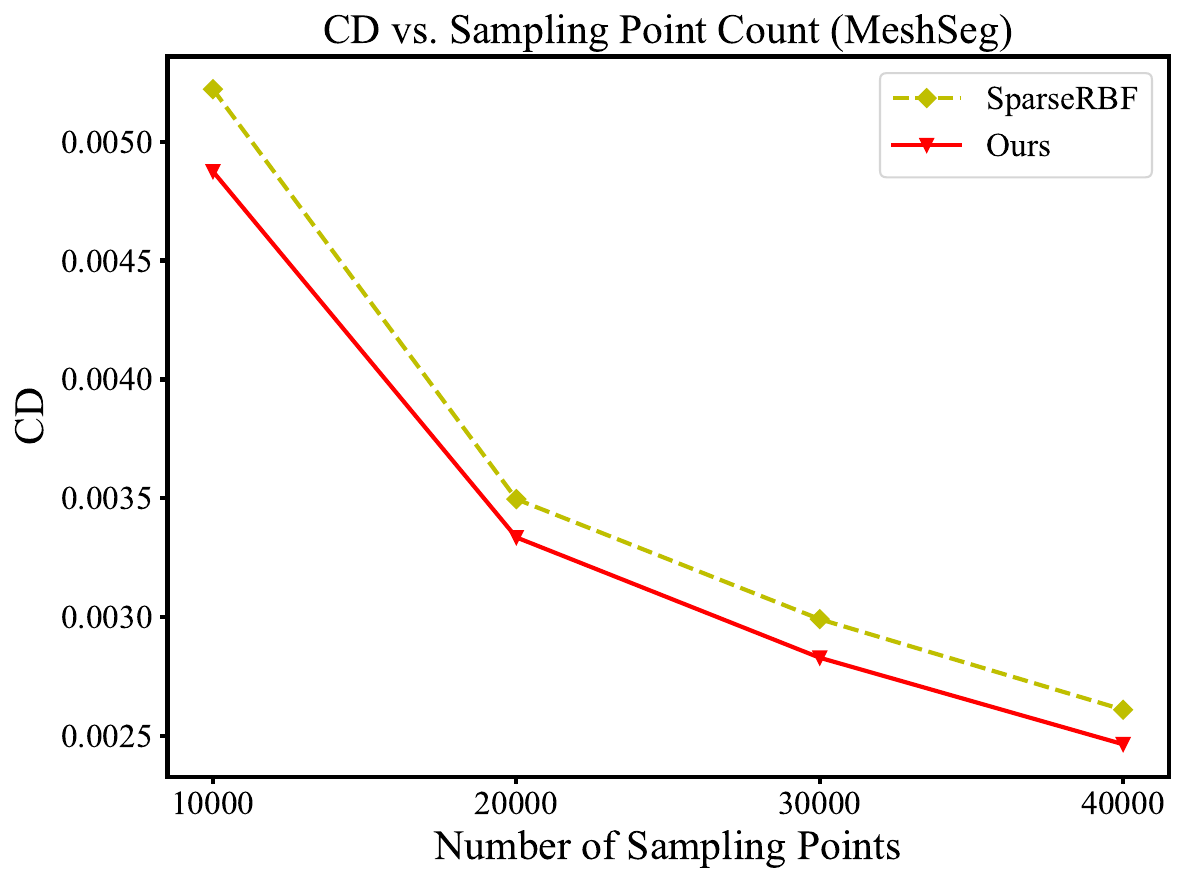}
        \includegraphics[width=0.22\textwidth]{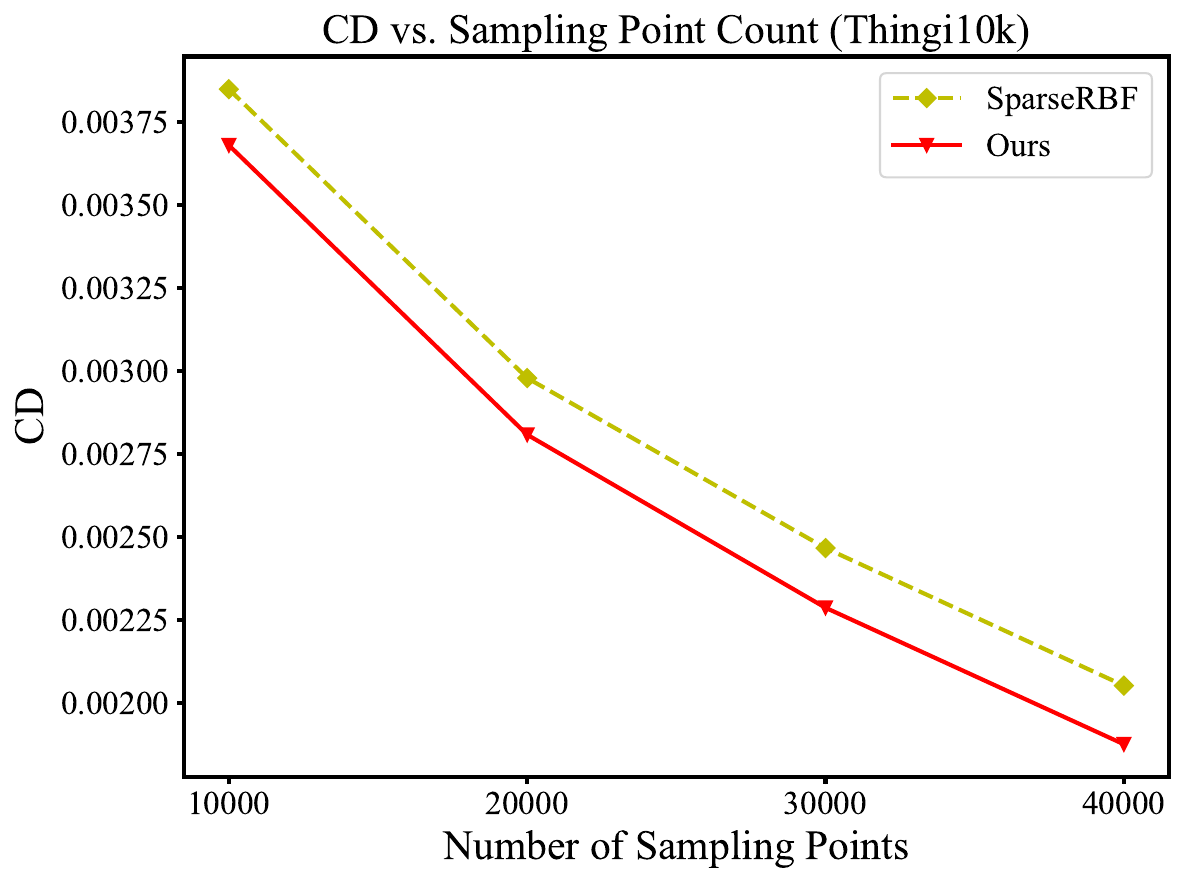}
    }
    \hfil
    \subfloat[\small CS with Different Numbers of Sampled Points (10K, 20K, 30K, 40K)]{
        \label{fig:cs_sample}
        \includegraphics[width=0.22\textwidth]{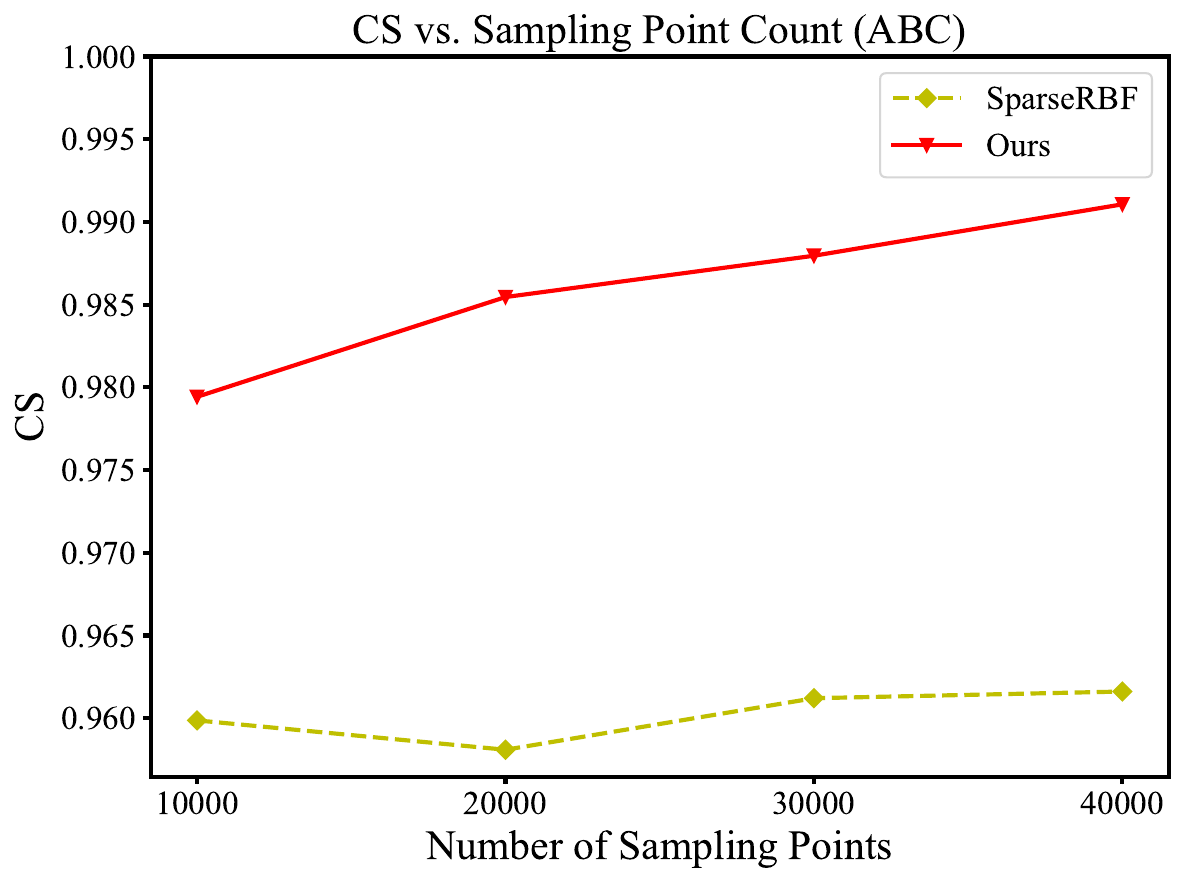}
        \includegraphics[width=0.22\textwidth]{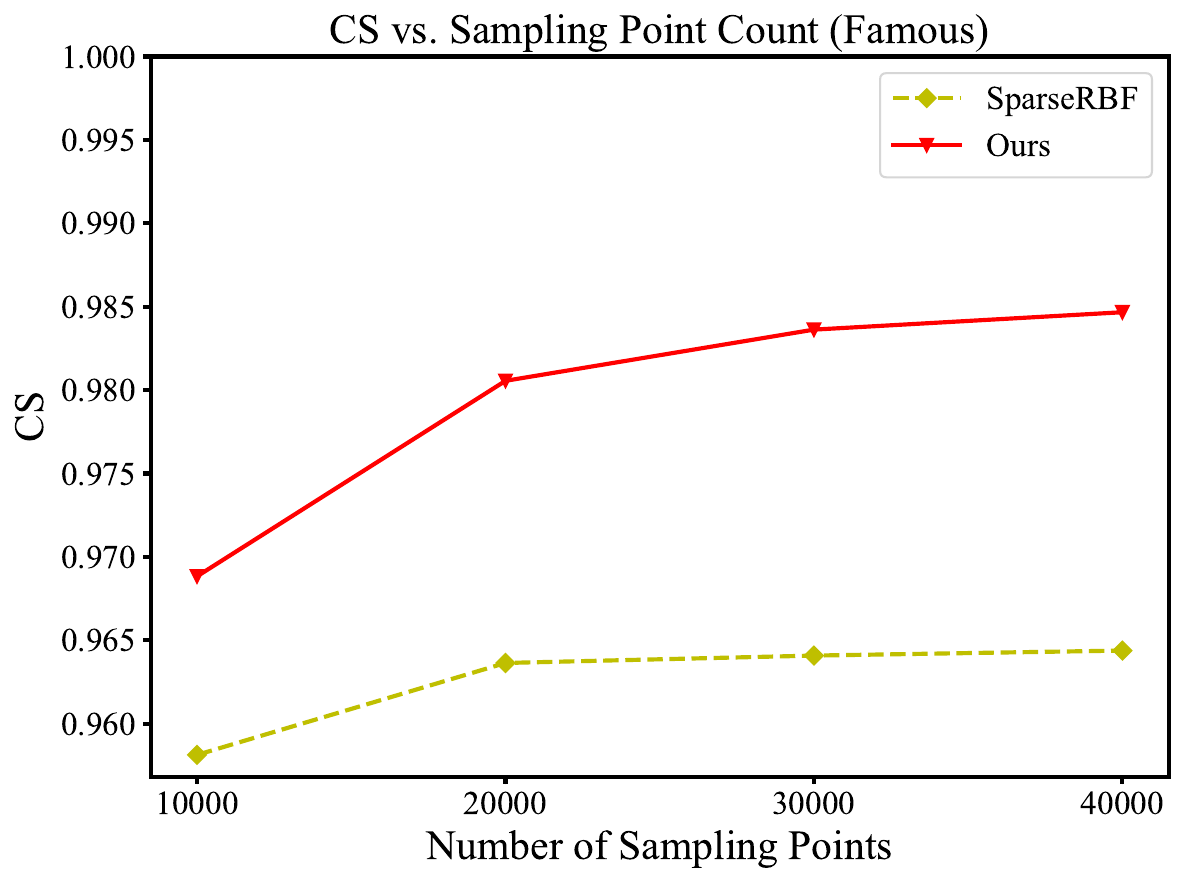}
        \includegraphics[width=0.22\textwidth]{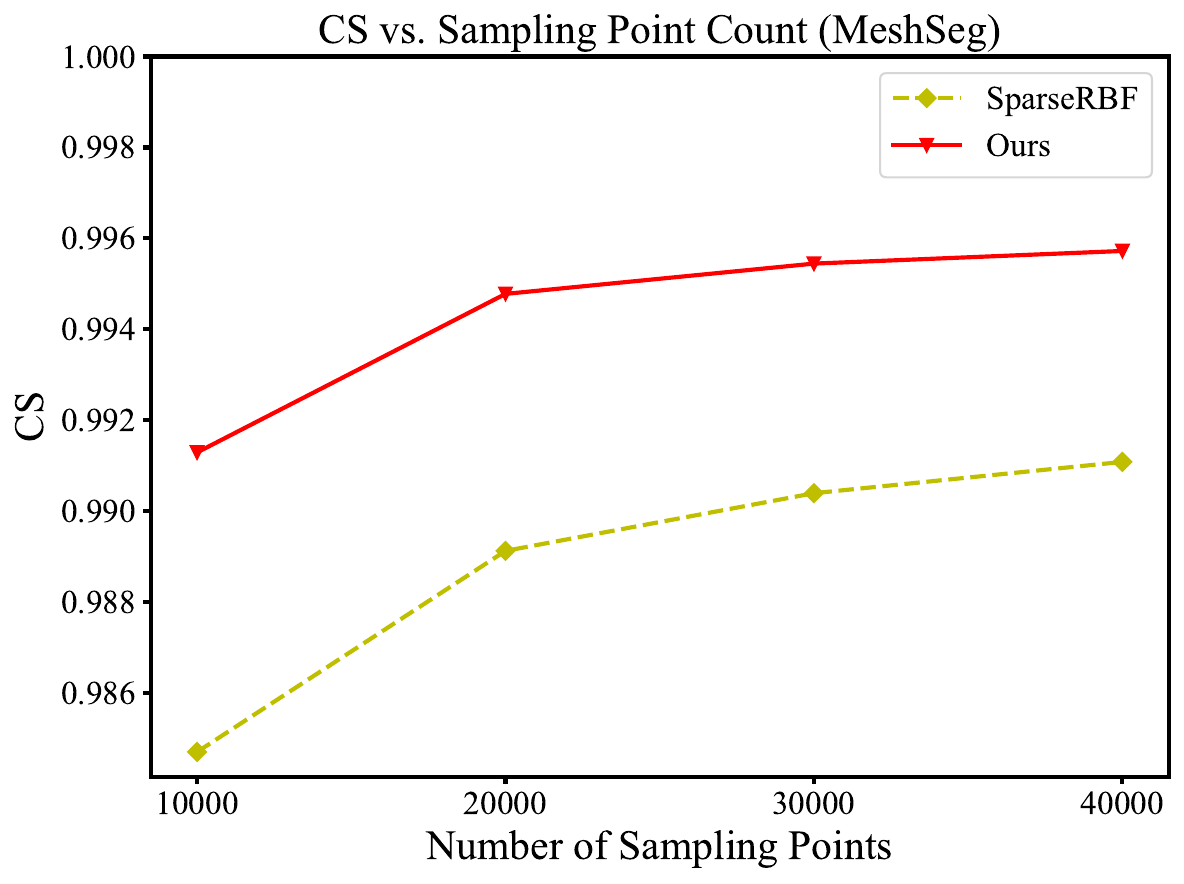}
        \includegraphics[width=0.22\textwidth]{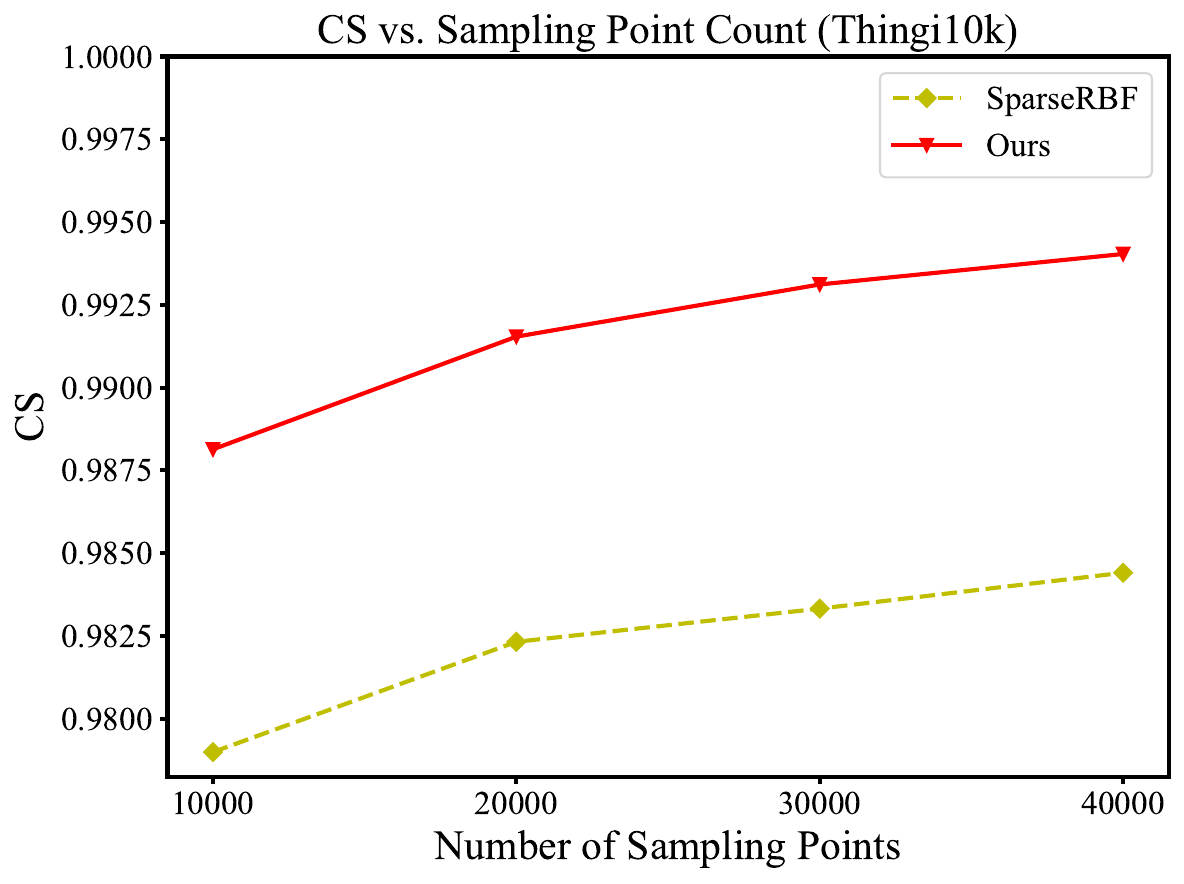}
    }
    \hfil
    \subfloat[\small Time with Different Numbers of Sampled Points (10K, 20K, 30K, 40K)]{
        \label{fig:time_sample}
        \includegraphics[width=0.22\textwidth]{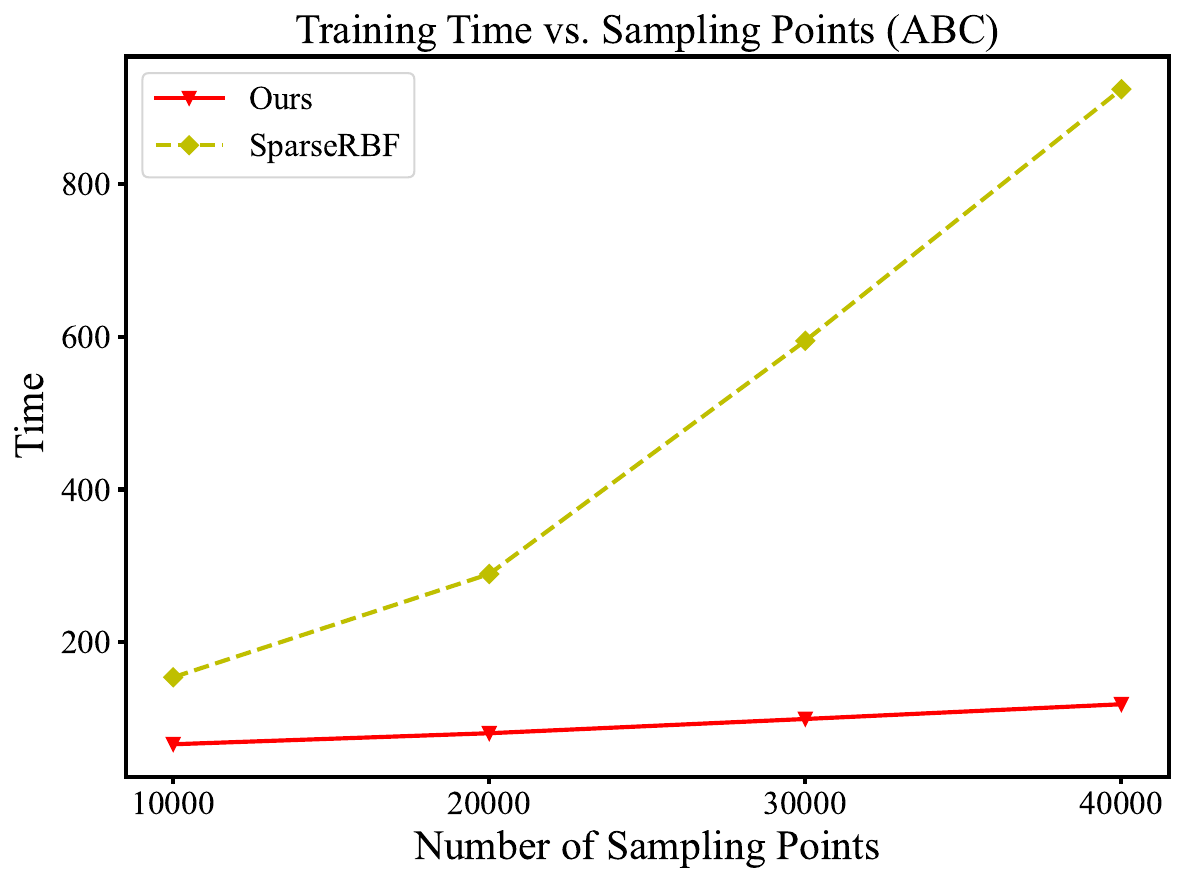}
        \includegraphics[width=0.22\textwidth]{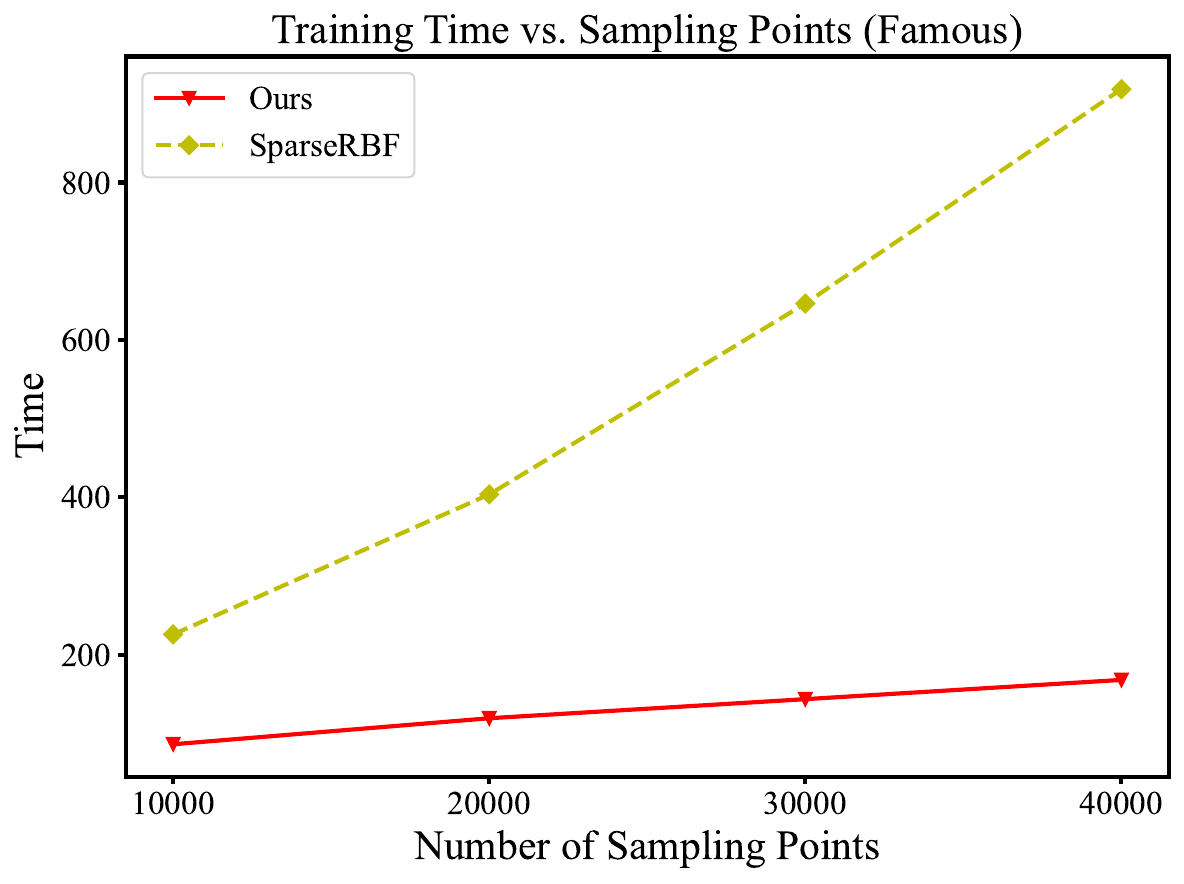}
        \includegraphics[width=0.22\textwidth]{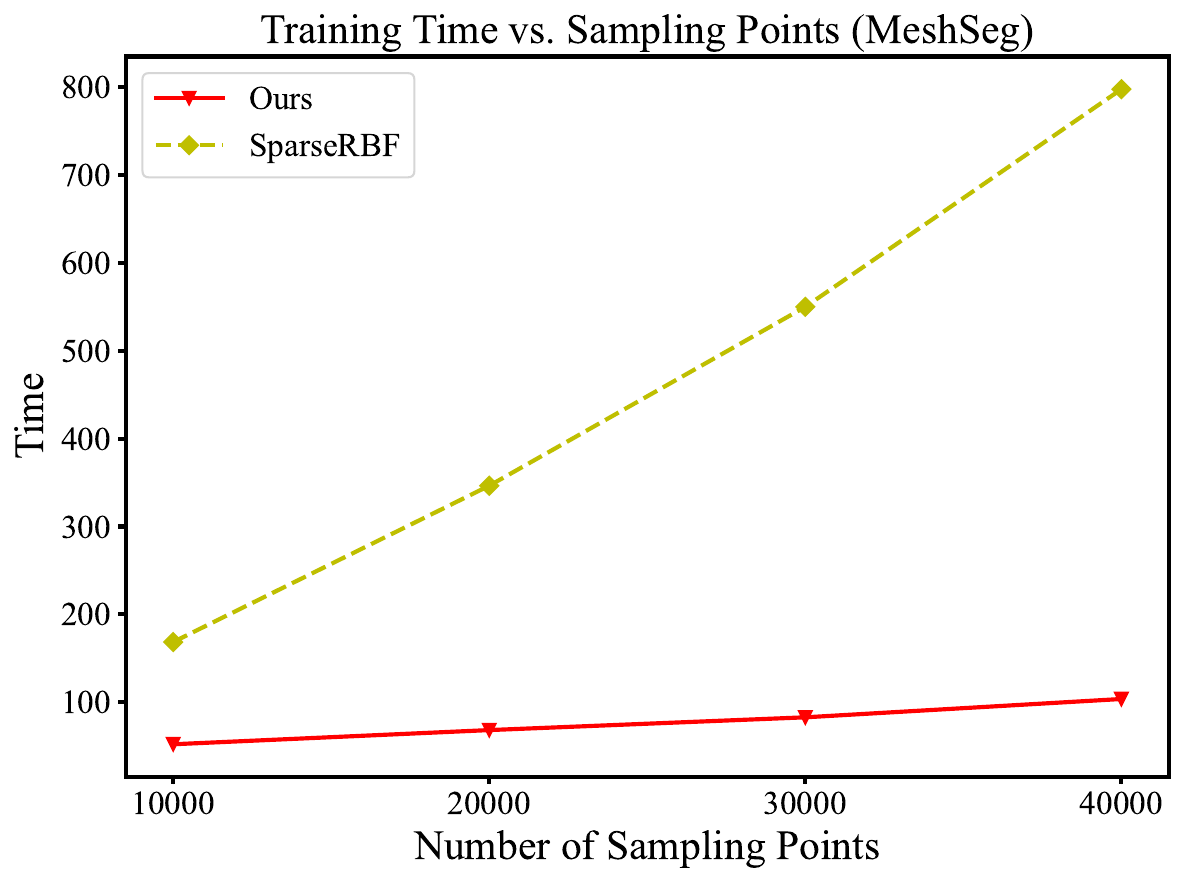}
        \includegraphics[width=0.22\textwidth]{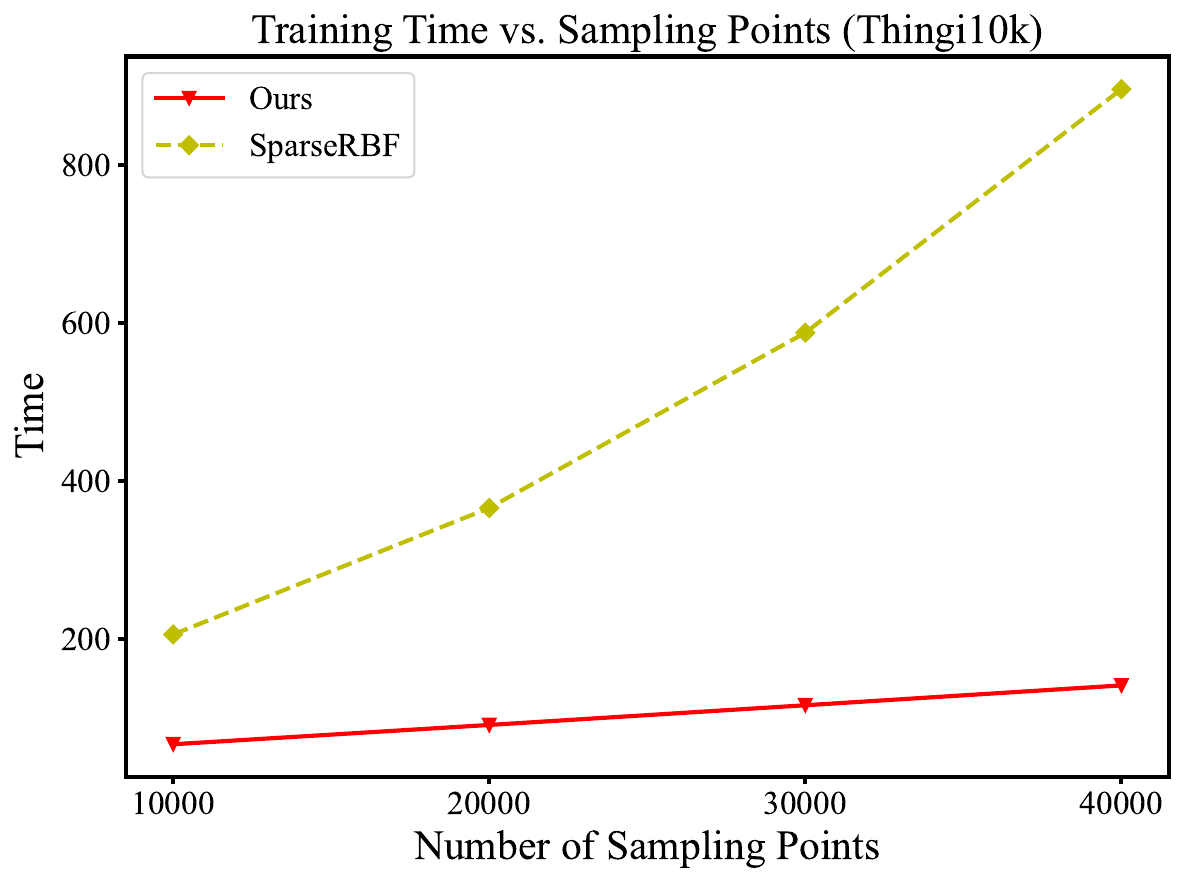}
    }
    \hfil
    \subfloat[\small Basis Functions with Different Numbers of Sampled Points (10K, 20K, 30K, 40K)]{
        \label{fig:basis_sample}
        \includegraphics[width=0.22\textwidth]{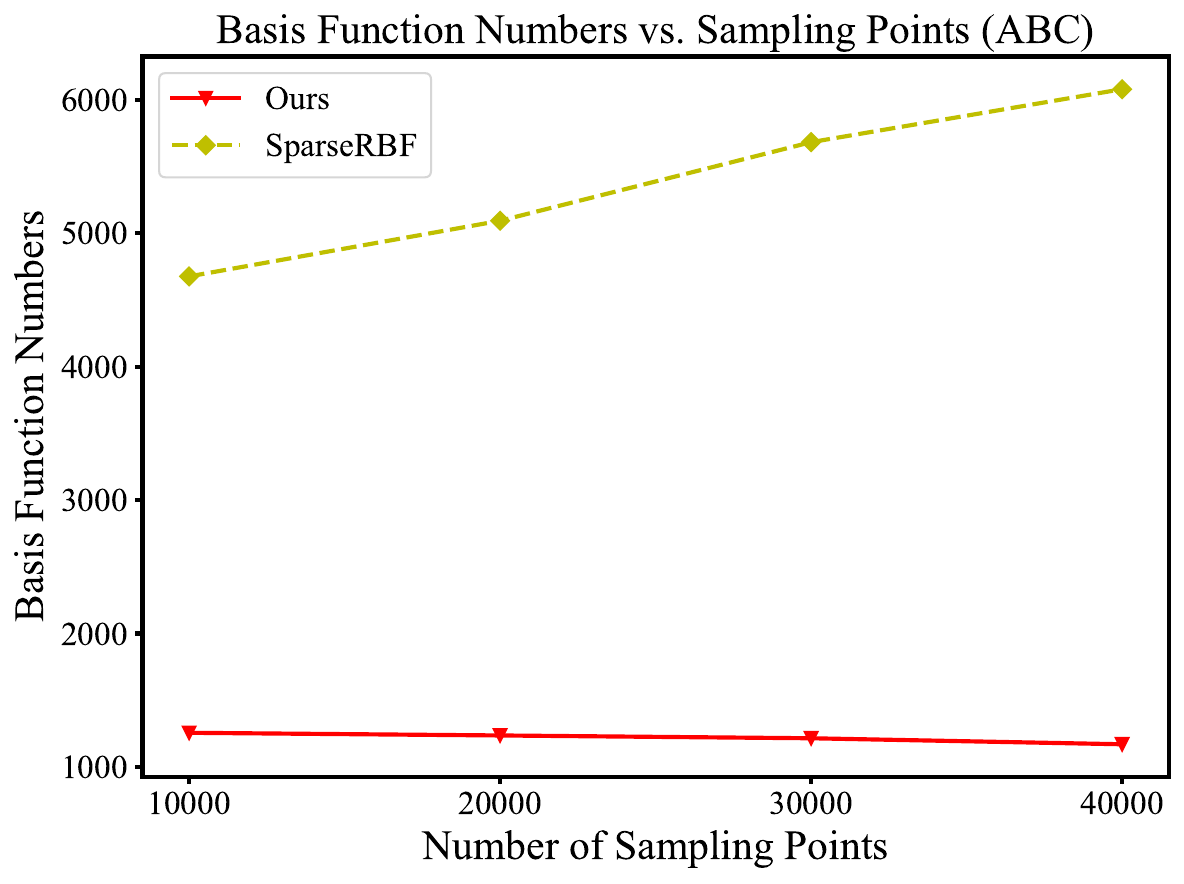}
        \includegraphics[width=0.22\textwidth]{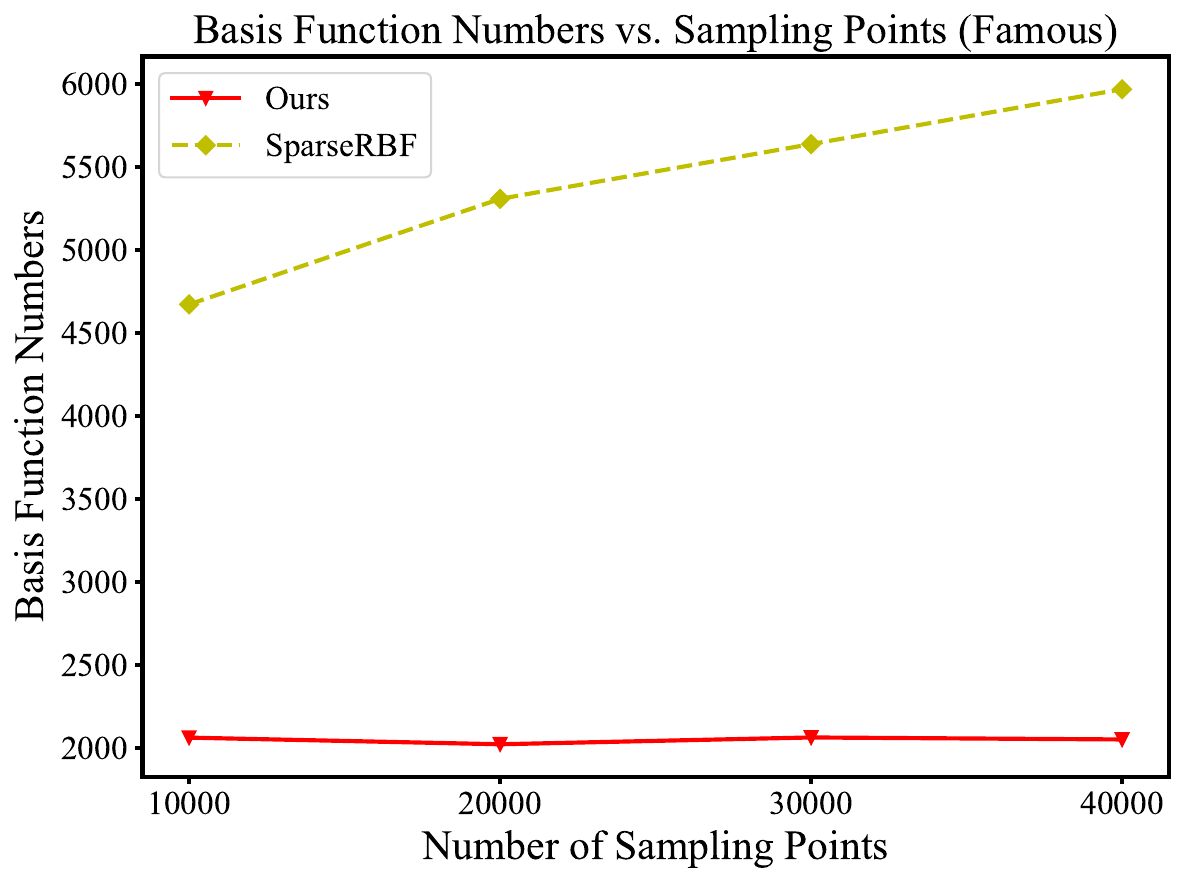}
        \includegraphics[width=0.22\textwidth]{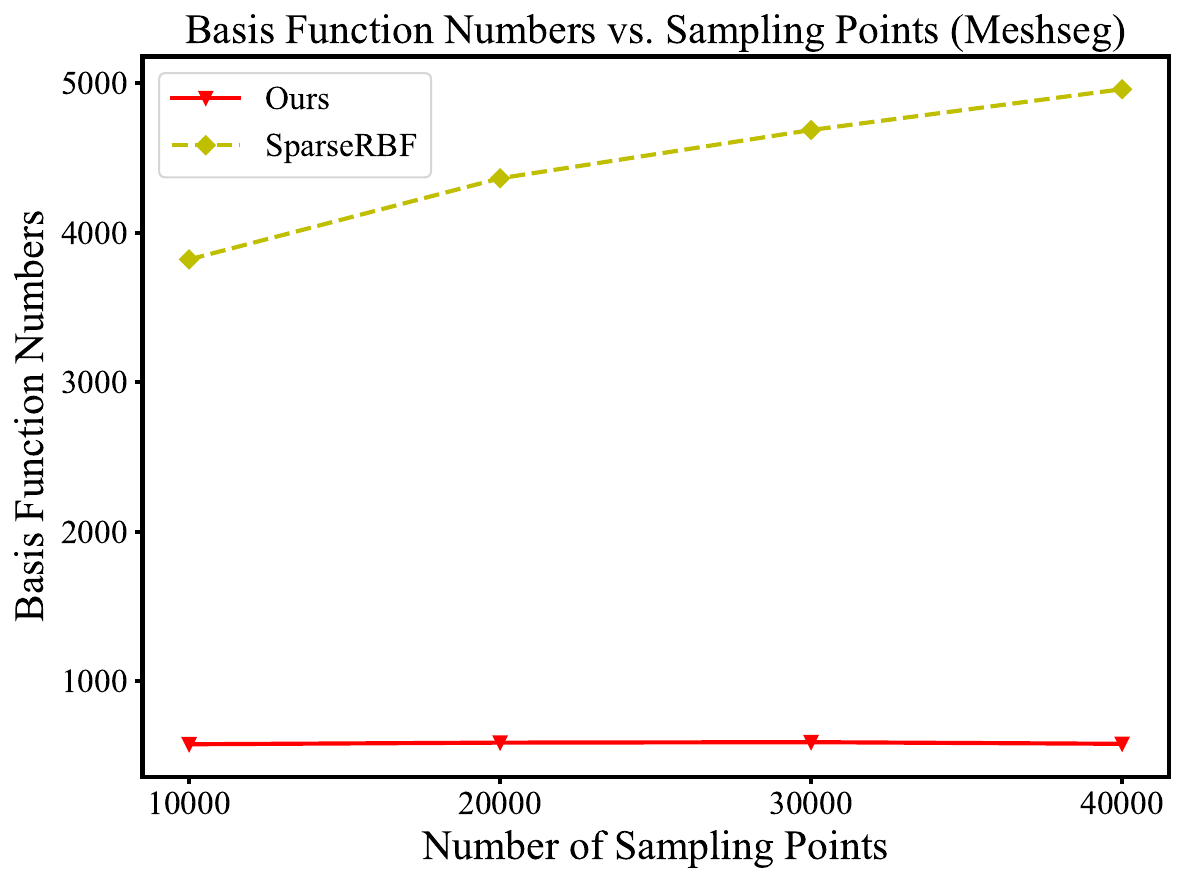}
        \includegraphics[width=0.22\textwidth]{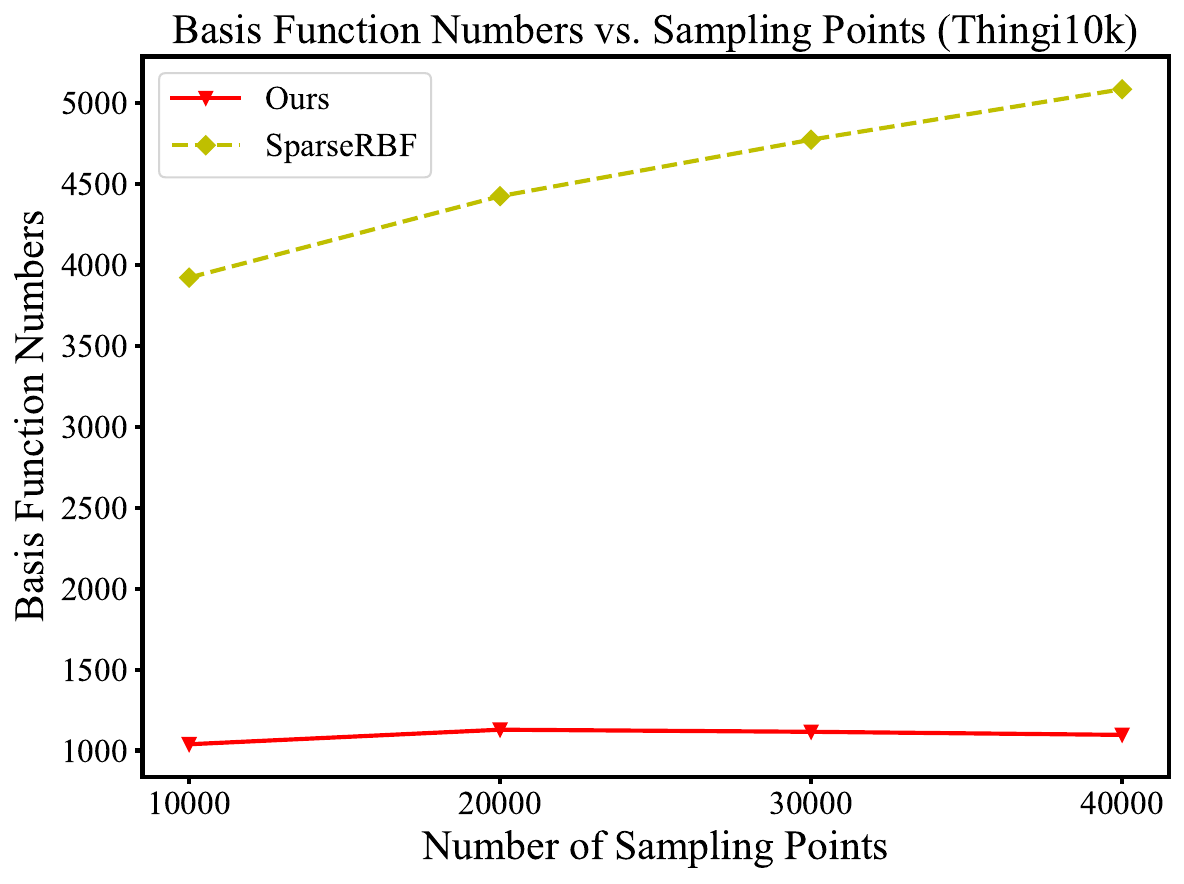}
    }
    \caption{Effect of the number of sampled points on implicit surface approximation results. Average results across the ABC, Famous, MeshSeg, and Thingi10K datasets for varying numbers of sampled points are shown in the four columns from left to right.} 
    \label{fig:sample_analyse}
\end{figure}

The number of sampled points directly affects the quality of SDF approximation, which in turn impacts the resulting implicit surface representation. More sampled points better cover the surface, capturing finer geometric details; sparser sampling increases interpolation difficulty and detail loss.
Figure \ref{fig:sample_analyse} summarizes the average performance across the ABC, Famous, MeshSeg, and Thingi10K datasets for different numbers of sampled points, with our method shown by the red curve.
As illustrated in Figures \ref{fig:cd_sample} and \ref{fig:cs_sample}, both our method and SparseRBF show that increasing the number of sampled points leads to lower CD and higher CS, indicating improved implicit surface quality. Nevertheless, our method consistently achieves better metric results across different numbers of sampled points.
Figure \ref{fig:time_sample} illustrates that our method scales more gracefully with the number of sampled points, incurring only moderate increases in approximation time, while SparseRBF's time grows sharply.
Figure \ref{fig:basis_sample} indicates that the number of basis functions in our method increases slightly with the number of sampled points, in contrast to SparseRBF, where it rises as the number of sampled points grows. Moreover, our method consistently requires significantly fewer basis functions, highlighting efficiency advantages in the storage and transmission of implicit surface representations.

\section{Conclusion and Future Work}
\label{section:conclusion}

This paper introduces SE-RBFNet, a sparse ellipsoidal radial basis function network designed for efficient and compact approximation of SDFs of implicit surfaces. Unlike methods that generate SDFs, our approach focuses on the sparse approximation of precomputed SDF samples, which can be derived from point clouds, triangular meshes, analytical SDFs, or neural implicit models. Specifically, the implicit surface representation problem is reformulated as a sparse nonlinear optimization task: given SDF values sampled on a spatial grid, SE-RBFNet approximates them using a small set of ellipsoidal RBFs, achieving a sparse representation that preserves the zero-level set of complex geometries while significantly reducing the number of parameters.
To balance accuracy and model complexity, we propose a dynamic multi-objective optimization framework, combined with a coarse-to-fine hierarchical strategy and a nearest-neighbor filtering mechanism, to accelerate training and reduce redundancy. Furthermore, an adaptive basis function adjustment mechanism is introduced, which iteratively selects new centers from high-error regions and dynamically updates the ellipsoidal parameters, enabling the network to represent complex surfaces with significantly fewer basis functions.
Extensive experiments demonstrate that SE-RBFNet can accurately approximate complex SDFs with substantially fewer parameters, achieving high-fidelity implicit surface representation while reducing storage and computational costs.

In future work, we will further explore adaptive strategies for basis function selection and parameter initialization, as well as investigate hybrid architectures that combine the advantages of sparse geometric modeling and neural implicit representations.
Additionally, we aim to explore the application of sparse representations to other challenges, such as surface matching and other related problems, where the compact and efficient nature of sparse models can offer significant benefits.

\appendix
\section{Complete Gradient Analysis for SE-RBFNet}
\label{section:gradients}

SE-RBFNet addresses the sparse approximation of SDFs for implicit surfaces by optimizing the following loss function:
\begin{equation}
\begin{aligned}
\mathcal{L}(\boldsymbol{\widetilde{\Theta}})=\alpha \cdot \sum_{i \in id_{all} }\left(\Psi\left(\boldsymbol{v}_i, \boldsymbol{\widetilde{\Theta}}\right)-t_i\right)^2+\beta \cdot\|\boldsymbol{W}\|_1 \text{,}
\end{aligned}
\end{equation}
where 
\begin{equation*}
\begin{aligned}
\Psi(\boldsymbol{v}_i, \boldsymbol{\widetilde{\Theta}}) &= \sum_{j=1}^M w_j \cdot |w_j| \hat{\Phi}(\boldsymbol{v}_i, \Theta_j) \\
&= \sum_{j=1}^M w_j \cdot |w_j| e^{-\left\|\boldsymbol{D}_j \boldsymbol{R}_j\left(\boldsymbol{v}_i-\boldsymbol{c}_j\right)\right\|_2^2} .
\end{aligned}
\end{equation*}
$\Psi(\boldsymbol{v}_i, \boldsymbol{\widetilde{\Theta}})$ is the SE-RBFNet output for the input $\boldsymbol{v}_i$, $t_i$ is the corresponding ground-truth SDF value, and $\boldsymbol{\widetilde{\Theta}}$ denotes the set of all SE-RBFNet parameters. Only the points corresponding to the indices in $id_{all}$ are used for loss calculation. 

The relevant parameters are defined as follows:
\begin{itemize}
    \item $V$: the total training points, which consists of $N$ points, i.e.,
    $V=\displaystyle\bigcup_{i=1}^{N}\boldsymbol{v}_i$, where $\boldsymbol{v}_i$ denotes the $i$-th training point.
    \item $\boldsymbol{F}$, denoted as the feature matrix, has dimensions $N\times M$, and each element represents the response value of the $i$-th point coordinate $\boldsymbol{v}_i$ under the influence of the $j$-th ERBF $\hat{\Phi}(\boldsymbol{v}_i, \Theta_j)$.
    \item $\boldsymbol{D}_j = \text{diag}(d_{j1}, d_{j2}, d_{j3})$: the lengths of the $j$-th ERBF along its principal axes. 
    \item $\boldsymbol{c}_j = (c_{j1}, c_{j2}, c_{j3})$: the center of the $j$-th ERBF.
    \item $w_j$: the weight of the $j$-th ERBF from the hidden layer to the output layer. 
    \item $\boldsymbol{a}_j = (\theta_{x_j}, \theta_{y_j}, \theta_{z_j})$: the rotation angles of the $j$-th ERBF.
    \item $\boldsymbol{R}\left(\theta_{x_j}, \theta_{y_j}, \theta_{z_j}\right)$: the rotation matrix of the $j$-th ERBF.
    \begin{equation*}
    \begin{aligned}
     \boldsymbol{R}\left(\theta_{x_j}, \theta_{y_j}, \theta_{z_j}\right) 
    = {\left[\begin{array}{ccc}
    r^{j}_{1} & r^{j}_{2} & r^{j}_{3} \\
    r^{j}_{4} & r^{j}_{5} & r^{j}_{6} \\
    r^{j}_{7} & r^{j}_{8} & r^{j}_{9}
    \end{array}\right], }
    \end{aligned}
    \end{equation*}
    where  \par
    $r^{j}_{1} = \cos \theta_{x_j} \cos \theta_{y_j}$;  \par 
    $r^{j}_{2} = -\cos \theta_{z_j} \sin \theta_{x_j}-\sin \theta_{y_j} \sin \theta_{z_j} \cos \theta_{x_j}$;  \par 
    $r^{j}_{3} = \sin \theta_{x_j} \sin \theta_{z_j}-\cos \theta_{x_j} \cos \theta_{z_j} \sin \theta_{y_j}$;  \par 
    $r^{j}_{4} = \cos \theta_{y_j} \sin \theta_{x_j}$;  \par
    $r^{j}_{5} = \cos \theta_{x_j} \cos \theta_{z_j}-\sin \theta_{x_j} \sin \theta_{y_j} \sin \theta_{z_j}$;  \par
    $r^{j}_{6} = -\sin \theta_{z_j} \cos \theta_{x_j}-\cos \theta_{z_j} \sin \theta_{x_j} \sin \theta_{y_j}$;  \par
    $r^{j}_{7} = \sin \theta_{y_j}$;  \par 
    $r^{j}_{8} = \cos \theta_{y_j} \sin \theta_{z_j} $;  \par
    $r^{j}_{9} = \cos \theta_{y_j} \cos \theta_{z_j}$;  \par
    
    \item $E_i = \Psi(\boldsymbol{v}_i, \boldsymbol{\widetilde{\Theta}}) - t_i$: the residual of the $i$-th training point $\boldsymbol{v}_i$.
\end{itemize}

In what follows, we derive the analytical gradients of the loss with respect to each parameter group: weights, axis lengths, centers, and rotation angles.
The gradient of each parameter is denoted as:
\begin{equation}
\label{equ:grad}
\begin{aligned}
\nabla{\mathcal{L}(\boldsymbol{\widetilde{\Theta}})}=\left\{\alpha \cdot \frac{\partial \mathcal{L}_2}{\partial \boldsymbol{C}}, \alpha \cdot \frac{\partial \mathcal{L}_2}{\partial \bar{\boldsymbol{D}}}, \alpha \cdot \frac{\partial \mathcal{L}_2}{\partial \boldsymbol{A}}, \alpha \cdot \frac{\partial \mathcal{L}_2}{\partial \boldsymbol{W}} + \beta \cdot \frac{\partial \mathcal{L}_1}{\partial \boldsymbol{W}}\right\} \text{.}
\end{aligned}
\end{equation}

\subsection{Weight Gradient}
The loss consists of a squared error term and an $\mathcal{L}_1$ regularization on the weights:
\begin{equation}
\begin{aligned}
\label{eq:grad_w_l1}
\frac{\partial \mathcal{L}_2}{\partial w_j} &= \sum_{i=1}^N 4 \cdot w_j \cdot \text{sign}(w_j) \cdot \boldsymbol{F}_{ij} \cdot E_i \text{,} \\
\frac{\partial \mathcal{L}_1}{\partial w_j} &=  \text{sign}(w_j)\text{,}
\end{aligned}
\end{equation}
where $\boldsymbol{F}_{ij}$ is the activation of the $j$-th ellipsoid on the $i$-th input.

\subsection{Axis Length Gradient}

Let $\boldsymbol{\Delta}_{\boldsymbol{x}_{i}} = \boldsymbol{x}_i - \boldsymbol{c}_j$ and $\widetilde{D}_{\boldsymbol{x}_{i}} = \boldsymbol{R}\left(\theta_{x_j}, \theta_{y_j}, \theta_{z_j}\right) \cdot \boldsymbol{\Delta}_{\boldsymbol{x}_{i}}$ denote the rotated input offset. Then, for each axis length $d_{jk}$:
\begin{equation}
\begin{aligned}
    \frac{\partial \mathcal{L}_2}{\partial d_{jk}} &= \sum_{i=1}^N -4 \cdot\text{sign}(w_j) \cdot w_j^2 \cdot d_{jk} \cdot \boldsymbol{F}_{ij} \cdot \widetilde{D}_{x_{ik}}^2 \cdot E_i, \\
    \quad k&=1,2,3. 
    \label{eq:grad_d}
\end{aligned}
\end{equation}

\subsection{Center Gradient}

Define the intermediate vector:
\begin{equation}
\begin{aligned}
    \widetilde{\boldsymbol{E}}_i &= 
    \begin{pmatrix}
        d_{j1}^2 \cdot w_j^2 \cdot \widetilde{D}_{x_{i1}} \cdot \boldsymbol{F}_{ij} \\
        d_{j2}^2 \cdot w_j^2 \cdot \widetilde{D}_{x_{i2}} \cdot \boldsymbol{F}_{ij} \\
        d_{j3}^2 \cdot w_j^2 \cdot \widetilde{D}_{x_{i3}} \cdot \boldsymbol{F}_{ij}
    \end{pmatrix},
    \\
    \widetilde{\boldsymbol{F}}_i &= \boldsymbol{R}\left(\theta_{x_j}, \theta_{y_j}, \theta_{z_j}\right)^\mathrm{T} \cdot \widetilde{\boldsymbol{E}}_i.
\end{aligned}
\end{equation}
Then, the gradient with respect to the center coordinate $c_{jk}$ is:
\begin{equation}
    \frac{\partial \mathcal{L}_2}{\partial c_{jk}} = \sum_{i=1}^N 4 \cdot \text{sign}(w_j) \cdot \widetilde{F}_{ik} \cdot E_i, \quad k=1,2,3. \label{eq:grad_c}
\end{equation}

\subsection{Rotation Angle Gradient}

The gradients with respect to each angle are as follows:

\subsubsection{\texorpdfstring{Gradient w.r.t. $\theta_{x_j}$}{Gradient w.r.t. theta\_xj}}
\begin{equation}
    \frac{\partial \mathcal{L}_2}{\partial \theta_{x_j}} = \sum_{i=1}^N -4 \cdot \text{sign}(w_j) \cdot E_i \cdot \left( \widetilde{E}_{i1} \cdot (-\widetilde{D}_{x_{i2}}) + \widetilde{E}_{i2} \cdot (-\widetilde{D}_{x_{i1}}) \right). \label{eq:grad_theta_x}
\end{equation}

\subsubsection{\texorpdfstring{Gradient w.r.t. $\theta_{y_j}$}{Gradient w.r.t. theta\_yj}}
Define:
\begin{align*}
    \widetilde{D}_X &= -\cos\theta_{x_j} \sin\theta_{y_j} \Delta_{x_{i1}} - r^j_1 \sin\theta_{z_j} \Delta_{x_{i2}} - r^j_1 \cos\theta_{z_j} \Delta_{x_{i3}}, \\
    \widetilde{D}_Y &= -\sin\theta_{x_j} \sin\theta_{y_j} \Delta_{x_{i1}} - r^j_4 \sin\theta_{z_j} \Delta_{x_{i2}} - r^j_4 \cos\theta_{z_j} \Delta_{x_{i3}}, \\
    \widetilde{D}_Z &= \cos\theta_{y_j} \Delta_{x_{i1}} - \sin\theta_{y_j} \sin\theta_{z_j} \Delta_{x_{i2}} - \sin\theta_{y_j} \cos\theta_{z_j} \Delta_{x_{i3}},
\end{align*}
and the gradient is:
\begin{equation}
    \frac{\partial \mathcal{L}_2}{\partial \theta_{y_j}} = \sum_{i=1}^N -4 \cdot \text{sign}(w_j) \cdot E_i \cdot \left( \widetilde{E}_{i1} \cdot \widetilde{D}_X + \widetilde{E}_{i2} \cdot \widetilde{D}_Y + \widetilde{E}_{i3} \cdot \widetilde{D}_Z \right). \label{eq:grad_theta_y}
\end{equation}

\subsubsection{\texorpdfstring{Gradient w.r.t. $\theta_{z_j}$}{Gradient w.r.t. theta\_zj}}
Using the partial derivative of $\boldsymbol{R}$ with respect to $\theta_{z_j}$:
\begin{align*}
    \hat{D}_X &= r^j_3 \Delta_{x_{i2}} - r^j_2 \Delta_{x_{i3}}, \\
    \hat{D}_Y &= r^j_6 \Delta_{x_{i2}} - r^j_5 \Delta_{x_{i3}}, \\
    \hat{D}_Z &= r^j_9 \Delta_{x_{i2}} - r^j_8 \Delta_{x_{i3}},
\end{align*}
the gradient becomes:
\begin{equation}
    \frac{\partial \mathcal{L}_2}{\partial \theta_{z_j}} = \sum_{i=1}^N -4 \cdot \text{sign}(w_j) \cdot E_i \cdot \left( \widetilde{E}_{i1} \cdot \hat{D}_X + \widetilde{E}_{i2} \cdot \hat{}{D}_Y + \hat{E}_{i3} \cdot \hat{D}_Z \right). \label{eq:grad_theta_z}
\end{equation}

\bibliographystyle{elsarticle-num}
\bibliography{ref}

@article{li2016sparse,
  title={Sparse RBF surface representations},
  author={Li, Manyi and Chen, Falai and Wang, Wenping and Tu, Changhe},
  journal={Computer Aided Geometric Design},
  volume={48},
  pages={49--59},
  year={2016},
  publisher={Elsevier}
}

@article{calakli2011ssd,
  title={SSD: Smooth signed distance surface reconstruction},
  author={Calakli, Fatih and Taubin, Gabriel},
  journal={Computer Graphics Forum},
  volume={30},
  number={7},
  pages={1993--2002},
  year={2011},
  organization={Wiley Online Library}
}

@inproceedings{park2019deepsdf,
  title={Deepsdf: Learning continuous signed distance functions for shape representation},
  author={Park, Jeong Joon and Florence, Peter and Straub, Julian and Newcombe, Richard and Lovegrove, Steven},
  booktitle={Proceedings of the IEEE/CVF conference on computer vision and pattern recognition},
  pages={165--174},
  year={2019}
}

@inproceedings{carr2001reconstruction,
  title={Reconstruction and representation of 3D objects with radial basis functions},
  author={Carr, Jonathan C and Beatson, Richard K and Cherrie, Jon B and Mitchell, Tim J and Fright, W Richard and McCallum, Bruce C and Evans, Tim R},
  booktitle={Proceedings of the 28th annual conference on Computer graphics and interactive techniques},
  pages={67--76},
  year={2001}
}

@article{dinh2002reconstructing,
  title={Reconstructing surfaces by volumetric regularization using radial basis functions},
  author={Dinh, Huong Quynh and Turk, Greg and Slabaugh, Greg},
  journal={IEEE transactions on pattern analysis and machine intelligence},
  volume={24},
  number={10},
  pages={1358--1371},
  year={2002},
  publisher={IEEE}
}

@article{turk2002modelling,
  title={Modelling with implicit surfaces that interpolate},
  author={Turk, Greg and O'brien, James F},
  journal={ACM Transactions on Graphics (TOG)},
  volume={21},
  number={4},
  pages={855--873},
  year={2002},
  publisher={ACM New York, NY, USA}
}

@article{ohtake2006sparse,
  title={Sparse surface reconstruction with adaptive partition of unity and radial basis functions},
  author={Ohtake, Yutaka and Belyaev, Alexander and Seidel, Hans-Peter},
  journal={Graphical Models},
  volume={68},
  number={1},
  pages={15--24},
  year={2006},
  publisher={Elsevier}
}

@inproceedings{samozino2006reconstruction,
author={Samozino, Marie and Alexa, Marc and Alliez, Pierre and Yvinec, Mariette},
title = {Reconstruction with Voronoi centered radial basis functions},
year = {2006},
publisher = {Eurographics Association},
booktitle = {Proceedings of the Fourth Eurographics Symposium on Geometry Processing},
pages = {51–60},
numpages = {10},
}

@inproceedings{kazhdan2006poisson,
  title={Poisson surface reconstruction},
  author={Kazhdan, Michael and Bolitho, Matthew and Hoppe, Hugues},
  booktitle={Proceedings of the fourth Eurographics symposium on Geometry processing},
  volume={7},
  pages = {61–70},
  numpages = {10},
  year={2006},
  publisher = {Eurographics Association},
}

@article{kazhdan2013screened,
  title={Screened poisson surface reconstruction},
  author={Kazhdan, Michael and Hoppe, Hugues},
  journal={ACM Transactions on Graphics (ToG)},
  volume={32},
  number={3},
  pages={1--13},
  year={2013},
  publisher={ACM New York, NY, USA}
}

@article{kazhdan2020poisson,
  title={Poisson surface reconstruction with envelope constraints},
  author={Kazhdan, Misha and Chuang, Ming and Rusinkiewicz, Szymon and Hoppe, Hugues},
  journal={Computer graphics forum},
  volume={39},
  number={5},
  pages={173--182},
  year={2020},
  organization={Wiley Online Library}
}

@article{guerrero2018pcpnet,
  title={Pcpnet learning local shape properties from raw point clouds},
  author={Guerrero, Paul and Kleiman, Yanir and Ovsjanikov, Maks and Mitra, Niloy J},
  journal={Computer graphics forum},
  volume={37},
  number={2},
  pages={75--85},
  year={2018},
  organization={Wiley Online Library}
}

@inproceedings{Erler_2020,
  title={Points2surf learning implicit surfaces from point clouds},
  author={Erler, Philipp and Guerrero, Paul and Ohrhallinger, Stefan and Mitra, Niloy J and Wimmer, Michael},
  booktitle={European Conference on Computer Vision},
  pages={108--124},
  year={2020},
  organization={Springer}
}

@article{gropp2020implicit,
  title={Implicit geometric regularization for learning shapes},
  author={Gropp, Amos and Yariv, Lior and Haim, Niv and Atzmon, Matan and Lipman, Yaron},
  journal={arXiv preprint arXiv:2002.10099},
  year={2020}
}

@article{erler2024ppsurf,
  title={PPSurf: Combining Patches and Point Convolutions for Detailed Surface Reconstruction},
  author={Erler, Philipp and Fuentes-Perez, Lizeth and Hermosilla, Pedro and Guerrero, Paul and Pajarola, Renato and Wimmer, Michael},
  journal={Computer Graphics Forum},
  volume={43},
  number={1},
  pages={e15000},
  year={2024},
  organization={Wiley Online Library}
}

@article{gui2020molecular,
  title={Molecular sparse representation by a 3d ellipsoid radial basis function neural network via l1 regularization},
  author={Gui, Sheng and Chen, Zhaodi and Lu, Benzhuo and Chen, Minxin},
  journal={Journal of Chemical Information and Modeling},
  volume={60},
  number={12},
  pages={6054--6064},
  year={2020},
  publisher={ACS Publications}
}

@article{kerbl20233d,
  title={3d gaussian splatting for real-time radiance field rendering.},
  author={Kerbl, Bernhard and Kopanas, Georgios and Leimk{\"u}hler, Thomas and Drettakis, George},
  journal={ACM Transactions on Graphics (ToG)},
  volume={42},
  number={4},
  pages={139--1},
  year={2023}
}

@article{tibshirani1996regression,
  title={Regression shrinkage and selection via the lasso},
  author={Tibshirani, Robert},
  journal={Journal of the Royal Statistical Society Series B: Statistical Methodology},
  volume={58},
  number={1},
  pages={267--288},
  year={1996},
  publisher={Oxford University Press}
}

@inproceedings{chen2020sparse,
  title={sparse representation of images based on RBF neural network},
  author={Chen, Zhaodi and Gui, Sheng and Chen, Hong and Wu, Chenjian and Chen, Minxin},
  booktitle={2020 IEEE 4th Information Technology, Networking, Electronic and Automation Control Conference (ITNEC)},
  volume={1},
  pages={830--835},
  year={2020},
  organization={IEEE}
}

@inproceedings{wang2021point,
  title={Point cloud surface reconstruction using sparse ellipsoid radial basis function neural network},
  author={Wang, Dandan and Chen, Hong and Wu, Chenjian and Chen, Minxin},
  booktitle={2021 IEEE 5th Information Technology, Networking, Electronic and Automation Control Conference (ITNEC)},
  volume={5},
  pages={1613--1618},
  year={2021},
  organization={IEEE}
}

@article{sener2018multi,
  title={Multi-task learning as multi-objective optimization},
  author={Sener, Ozan and Koltun, Vladlen},
  journal={Advances in neural information processing systems},
  volume={31},
  year={2018}
}

@article{wilhelms1992octrees,
  title={Octrees for faster isosurface generation},
  author={Wilhelms, Jane and Van Gelder, Allen},
  journal={ACM Transactions on Graphics (TOG)},
  volume={11},
  number={3},
  pages={201--227},
  year={1992},
  publisher={ACM New York, NY, USA}
}

@inproceedings{lorensen1998marching,
  title={Marching cubes: A high resolution 3D surface construction algorithm},
  author={Lorensen, William E and Cline, Harvey E},
  booktitle={Seminal graphics: pioneering efforts that shaped the field},
  pages={347--353},
  year={1998}
}

@inproceedings{curless1996volumetric,
  title={A volumetric method for building complex models from range images},
  author={Curless, Brian and Levoy, Marc},
  booktitle={Proceedings of the 23rd annual conference on Computer graphics and interactive techniques},
  pages={303--312},
  year={1996}
}

@article{pan2011continuous,
  title={Continuous global optimization in surface reconstruction from an oriented point cloud},
  author={Pan, Rongjiang and Skala, Vaclav},
  journal={Computer-Aided Design},
  volume={43},
  number={8},
  pages={896--901},
  year={2011},
  publisher={Elsevier}
}

@article{pan2012surface,
  title={Surface Reconstruction with higher-order smoothness},
  author={Pan, Rongjiang and Skala, Vaclav},
  journal={The Visual Computer},
  volume={28},
  pages={155--162},
  year={2012},
  publisher={Springer}
}

@article{sellan2022stochastic,
  title={Stochastic poisson surface reconstruction},
  author={Sell{\'a}n, Silvia and Jacobson, Alec},
  journal={ACM Transactions on Graphics (TOG)},
  volume={41},
  number={6},
  pages={1--12},
  year={2022},
  publisher={ACM New York, NY, USA}
}

@article{hou2022iterative,   
    title={Iterative Poisson Surface Reconstruction (iPSR) for Unoriented Points},  
    journal={ACM Transactions on Graphics},  
    author={Hou, Fei and Wang, Chiyu and Wang, Wencheng and Qin, Hong and Qian, Chen and He, Ying},  
    volume = {41},
    number = {4},
    year={2022},  
    pages={1--13}
}

@article{casciola2006shape,
  title={Shape preserving surface reconstruction using locally anisotropic radial basis function interpolants},
  author={Casciola, Giulio and Lazzaro, Damiana and Montefusco, Laura Bacchelli and Morigi, Serena},
  journal={Computers \& Mathematics with Applications},
  volume={51},
  number={8},
  pages={1185--1198},
  year={2006},
  publisher={Elsevier}
}

@article{walder2006implicit,
  title={Implicit surfaces with globally regularised and compactly supported basis functions},
  author={Walder, Christian and Chapelle, Olivier and Sch{\"o}lkopf, Bernhard},
  journal={Advances in Neural Information Processing Systems},
  volume={19},
  year={2006}
}

@article{pan2011two,
  title={A two-level approach to implicit surface modeling with compactly supported radial basis functions},
  author={Pan, Rongjiang and Skala, Vaclav},
  journal={Engineering with Computers},
  volume={27},
  pages={299--307},
  year={2011},
  publisher={Springer}
}

@inproceedings{xia2006orthogonal,
  title={Orthogonal least squares in partition of unity surface reconstruction with radial basis function},
  author={Xia, Qi and Wang, Michael Yu and Wu, Xiaojun},
  booktitle={Geometric Modeling and Imaging--New Trends (GMAI'06)},
  pages={28--33},
  year={2006},
  organization={IEEE}
}

@inproceedings{brazil2010sketching,
  title={Sketching variational hermite-rbf implicits},
  author={Brazil, E Vital and Macedo, Ives and Sousa, M Costa and de Figueiredo, Luiz Henrique and Velho, Luiz},
  booktitle={Proceedings of the Seventh Sketch-Based Interfaces and Modeling Symposium},
  pages={1--8},
  year={2010}
}

@article{macedo2011hermite,
  title={Hermite radial basis functions implicits},
  author={Mac{\^e}do, Ives and Gois, Joao Paulo and Velho, Luiz},
  journal={Computer graphics forum},
  volume={30},
  number={1},
  pages={27--42},
  year={2011},
  organization={Wiley Online Library}
}

@article{liu2016closed,
  title={A closed-form formulation of HRBF-based surface reconstruction by approximate solution},
  author={Liu, Shengjun and Wang, Charlie CL and Brunnett, Guido and Wang, Jun},
  journal={Computer-Aided Design},
  volume={78},
  pages={147--157},
  year={2016},
  publisher={Elsevier}
}

@article{majdisova2017big,
  title={Big geo data surface approximation using radial basis functions: A comparative study},
  author={Majdisova, Zuzana and Skala, Vaclav},
  journal={Computers \& Geosciences},
  volume={109},
  pages={51--58},
  year={2017},
  publisher={Elsevier}
}

@article{drake2022implicit,
  title={Implicit surface reconstruction with a curl-free radial basis function partition of unity method},
  author={Drake, Kathryn P and Fuselier, Edward J and Wright, Grady B},
  journal={SIAM Journal on Scientific Computing},
  volume={44},
  number={5},
  pages={A3018--A3040},
  year={2022},
  publisher={SIAM}
}

@article{zeng2022implicit,
  title={Implicit surface reconstruction based on a new interpolation/approximation radial basis function},
  author={Zeng, Yajun and Zhu, Yuanpeng},
  journal={Computer Aided Geometric Design},
  volume={92},
  pages={102062},
  year={2022},
  publisher={Elsevier}
}

@inproceedings{shen2004interpolating,
author = {Shen, Chen and O'Brien, James F. and Shewchuk, Jonathan R},
title = {Interpolating and approximating implicit surfaces from polygon soup},
year = {2004},
booktitle = {ACM SIGGRAPH 2004 Papers},
pages = {896–904},
}

@article{fuhrmann2014floating,
  title={Floating scale surface reconstruction},
  author={Fuhrmann, Simon and Goesele, Michael},
  journal={ACM Transactions on Graphics (ToG)},
  volume={33},
  number={4},
  pages={1--11},
  year={2014},
  publisher={ACM New York, NY, USA}
}

@inproceedings{Kazhdan_2005,
  title={Reconstruction of solid models from oriented point sets},
  author={Kazhdan, Michael},
  booktitle={Proceedings of the third Eurographics symposium on Geometry processing},
  pages={73--es},
  year={2005}
}

@article{lu2018surface,
  title={Surface reconstruction based on the modified Gauss formula},
  author={Lu, Wenjia and Shi, Zuoqiang and Sun, Jian and Wang, Bin},
  journal={ACM Transactions on Graphics (TOG)},
  volume={38},
  number={1},
  pages={1--18},
  year={2018},
  publisher={ACM New York, NY, USA}
}

@article{lin2022surface,
  title={Surface reconstruction from point clouds without normals by parametrizing the gauss formula},
  author={Lin, Siyou and Xiao, Dong and Shi, Zuoqiang and Wang, Bin},
  journal={ACM Transactions on Graphics},
  volume={42},
  number={2},
  pages={1--19},
  year={2022},
  publisher={ACM New York, NY}
}

@inproceedings{liu2021deep,
  title={Deep implicit moving least-squares functions for 3D reconstruction},
  author={Liu, Shi-Lin and Guo, Hao-Xiang and Pan, Hao and Wang, Peng-Shuai and Tong, Xin and Liu, Yang},
  booktitle={Proceedings of the IEEE/CVF Conference on Computer Vision and Pattern Recognition},
  pages={1788--1797},
  year={2021}
}

@inproceedings{mescheder2019occupancy,
  title={Occupancy networks: Learning 3d reconstruction in function space},
  author={Mescheder, Lars and Oechsle, Michael and Niemeyer, Michael and Nowozin, Sebastian and Geiger, Andreas},
  booktitle={Proceedings of the IEEE/CVF conference on computer vision and pattern recognition},
  pages={4460--4470},
  year={2019}
}

@inproceedings{peng2020convolutional,
  title={Convolutional occupancy networks},
  author={Peng, Songyou and Niemeyer, Michael and Mescheder, Lars and Pollefeys, Marc and Geiger, Andreas},
  booktitle={Computer Vision--ECCV 2020: 16th European Conference, Glasgow, UK, August 23--28, 2020, Proceedings, Part III 16},
  pages={523--540},
  year={2020},
  organization={Springer}
}

@inproceedings{ben2022digs,
  title={Digs: Divergence guided shape implicit neural representation for unoriented point clouds},
  author={Ben-Shabat, Yizhak and Koneputugodage, Chamin Hewa and Gould, Stephen},
  booktitle={Proceedings of the IEEE/CVF Conference on Computer Vision and Pattern Recognition},
  pages={19323--19332},
  year={2022}
}

@article{wang2023neural,
  title={Neural-singular-hessian: Implicit neural representation of unoriented point clouds by enforcing singular hessian},
  author={Wang, Zixiong and Zhang, Yunxiao and Xu, Rui and Zhang, Fan and Wang, Peng-Shuai and Chen, Shuangmin and Xin, Shiqing and Wang, Wenping and Tu, Changhe},
  journal={ACM Transactions on Graphics (TOG)},
  volume={42},
  number={6},
  pages={1--14},
  year={2023},
  publisher={ACM New York, NY, USA}
}

@article{ma2020neural,
  title={Neural-pull: Learning signed distance functions from point clouds by learning to pull space onto surfaces},
  author={Ma, Baorui and Han, Zhizhong and Liu, Yu-Shen and Zwicker, Matthias},
  journal={arXiv preprint arXiv:2011.13495},
  year={2020}
}

@article{Park-rbf1993,
  author={Park, Jooyoung and Sandberg, Irwin W.},
  journal={Neural Computation}, 
  title={Approximation and Radial-Basis-Function Networks}, 
  year={1993},
  volume={5},
  number={2},
  pages={305-316},
  doi={10.1162/neco.1993.5.2.305}
}

@article{majdisova2017radial,
  title={Radial basis function approximations: comparison and applications},
  author={Majdisova, Zuzana and Skala, Vaclav},
  journal={Applied Mathematical Modelling},
  volume={51},
  pages={728--743},
  year={2017},
  publisher={Elsevier}
}

@article{ismayilova2024universal,
  title={On the universal approximation property of radial basis function neural networks},
  author={Ismayilova, Aysu and Ismayilov, Muhammad},
  journal={Annals of Mathematics and Artificial Intelligence},
  volume={92},
  number={3},
  pages={691--701},
  year={2024},
  publisher={Springer}
}

@inproceedings{skala2020radial,
  title={Radial basis function approximation optimal shape parameters estimation},
  author={Skala, Vaclav and Karim, Samsul Ariffin Abdul and Zabran, Marek},
  booktitle={Computational Science--ICCS 2020: 20th International Conference, Amsterdam, The Netherlands, June 3--5, 2020, Proceedings, Part VI 20},
  pages={309--317},
  year={2020},
  organization={Springer}
}

@inproceedings{zwicker2001ewa,
  title={EWA volume splatting},
  author={Zwicker, Matthias and Pfister, Hanspeter and Van Baar, Jeroen and Gross, Markus},
  booktitle={Proceedings Visualization, 2001. VIS'01.},
  pages={29--538},
  year={2001},
  organization={IEEE}
}

@article{jatavallabhula2019kaolin,
  title={Kaolin: A pytorch library for accelerating 3d deep learning research},
  author={Jatavallabhula, Krishna Murthy and Smith, Edward and Lafleche, Jean-Francois and Tsang, Clement Fuji and Rozantsev, Artem and Chen, Wenzheng and Xiang, Tommy and Lebaredian, Rev and Fidler, Sanja},
  journal={arXiv preprint arXiv:1911.05063},
  year={2019}
}

@article{cornea2024curve,
  title={Curve-skeleton properties, applications, and algorithms},
  author={Cornea, Nicu D and Silver, Deborah and Min, Patrick},
  journal={IEEE Transactions on visualization and computer graphics},
  volume={13},
  number={3},
  pages={530--548},
  year={2024},
  publisher={IEEE}
}

@article{kingma2014adam,
  title={Adam: A method for stochastic optimization},
  author={Kingma, Diederik P and Ba, Jimmy},
  journal={arXiv preprint arXiv:1412.6980},
  year={2014}
}

@article{chen2009benchmark,
  title={A benchmark for 3D mesh segmentation},
  author={Chen, Xiaobai and Golovinskiy, Aleksey and Funkhouser, Thomas},
  journal={ACM Transactions on Graphic (TOG)},
  volume={28},
  number={3},
  pages={1--12},
  year={2009},
  publisher={ACM New York, NY, USA}
}

@inproceedings{koch2019abc,
  title={Abc: A big cad model dataset for geometric deep learning},
  author={Koch, Sebastian and Matveev, Albert and Jiang, Zhongshi and Williams, Francis and Artemov, Alexey and Burnaev, Evgeny and Alexa, Marc and Zorin, Denis and Panozzo, Daniele},
  booktitle={Proceedings of the IEEE/CVF conference on computer vision and pattern recognition},
  pages={9601--9611},
  year={2019}
}

@article{Thingi10K,
  title={Thingi10K: A Dataset of 10,000 3D-Printing Models},
  author={Zhou, Qingnan and Jacobson, Alec},
  journal={arXiv preprint arXiv:1605.04797},
  year={2016}
}

@misc{trimesh,
  title = {trimesh},
  author = {{Dawson-Haggerty et al.}},
  note = {https://trimesh.org/},
  year = {2019}
}

@inproceedings{amenta2001power,
  title={The power crust},
  author={Amenta, Nina and Choi, Sunghee and Kolluri, Ravi Krishna},
  booktitle={Proceedings of the sixth ACM symposium on Solid modeling and applications},
  pages={249--266},
  year={2001}
}

@misc{alglib,
  author       = {{Sergey Bochkanov}},
  title        = {{ALGLIB}},
  howpublished = {\url{http://www.alglib.net}},
}




\end{document}